\documentclass[traditabstract]{aa} 
\usepackage{txfonts}
\usepackage{graphicx}
\usepackage{natbib}

\newcommand{\hyeri}{\mbox{\object{HY\,Eri}}}
\newcommand{\rxjone}{\mbox{\object{RX\,J0154.0--5947}}}
\newcommand{\rxjsix}{\mbox{\object{RX\,J0600.5--2709}}}
\newcommand{\rxjeight}{\mbox{\object{RX\,J0859.1+0537}}}
\newcommand{\rxjnine}{\mbox{\object{RX\,J0953.1+1458}}}
\newcommand{\rxjten}{\mbox{\object{RX\,J1002.2--1925}}}
\newcommand{\jone}{\mbox{\object{J0154}}}
\newcommand{\jsix}{\mbox{\object{J0600}}}
\newcommand{\jeight}{\mbox{\object{J0859}}}
\newcommand{\jnine}{\mbox{\object{J0953}}}
\newcommand{\jten}{\mbox{\object{J1002}}}
\newcommand{\halp}{H$\alpha$}
\newcommand{\hbet}{H$\beta$}
\newcommand{\hgam}{H$\gamma$}
\newcommand{\hdel}{H$\delta$}
\newcommand{\heii}{\ion{He}{ii}$\lambda$4686}
\newcommand{\po}{$P_\mathrm{orb}$}
\newcommand{\kms}{km\,s$^{-1}$}
\newcommand{\ebv}{$E_\mathrm{B-V}$}
\newcommand{\nh}{$N_\mathrm{H}$}
\newcommand{\nhgal}{$N_\mathrm{H,gal}$}
\newcommand{\nhint}{$N_\mathrm{H,int}$}
\newcommand{\atoms}{H-atoms\,cm$^{-2}$}
\newcommand{\fpc}{$f_\mathrm{pc}$}

\newcommand{\tbb}{$T_\mathrm{bb}$}
\newcommand{\ktbb}{k$T_\mathrm{bb}$}
\newcommand{\ktbbi}{k$T_\mathrm{bb1}$}

\newcommand{\fbbibol}{$f_\mathrm{bb1,bol}$}

\newcommand{\fsxbol}{$f_\mathrm{sx,bol}$}

\newcommand{\fthbol}{$f_\mathrm{th,bol}$}
\newcommand{\lbol}{$L_\mathrm{bol}$}

\newcommand{\ergs}{erg\,cm$^{-2}$s$^{-1}$}
\newcommand{\ergsa}{erg\,cm$^{-2}$s$^{-1}$\AA$^{-1}$}
\newcommand{\cps}{cts\,s$^{-1}$}
\newcommand{\ten}[2]{#1\,\times\!10^{#2}}

\newcommand{\rsun}{$R_\odot$}
\newcommand{\msun}{$M_\odot$}
\newcommand{\msunyr}{$M_{\odot}$yr$^{-1}$}
\newcommand{\oc}{$O\!-\!C$}

\usepackage{amstext}

\begin{document}

 \title{Neglected X-ray discovered polars: III. \rxjone, \rxjsix,
 \rxjeight, \rxjnine, and \rxjten}

\author{
Beuermann, K. \inst{1} \and 
Burwitz, V. \inst{2} \and 
Reinsch, K. \inst{1} \and 
Schwope, A. \inst{3} \and
Thomas, H.-C. \inst{4}\thanks{Deceased 18 Jan 2012}
} 

\institute{ 
  Institut f\"ur Astrophysik, Georg-August-Universit\"at,
  Friedrich-Hund-Platz 1, D-37077 G\"ottingen, Germany \and
  MPI f\"ur extraterrestrische Physik, Giessenbachstr. 6, 85740
  Garching, Germany \and
  Leibniz-Institut f\"ur Astrophysik Potsdam (AIP), An der Sternwarte
  16, 14482 Potsdam, Germany \and
  MPI f\"ur Astrophysik, Karl-Schwarzschild-Str. 1, D-85740 Garching,
  Germany
}

\date{Received 8 June 2020; accepted 6 October 2020}

\authorrunning{K. Beuermann et al.} 
\titlerunning {Neglected X-ray discovered polars III.}

\abstract{We report results on the ROSAT-discovered noneclipsing
  short-period polars RX\,J0154.0-5947, RX\,J0600.5-2709,
  RX\,J0859.1+0537, RX\,J0953.1+1458, and RX\,J1002.2-1925 collected
  over 30 years. We present accurate linear orbital ephemerides that
  allow a correct phasing of data taken decades apart. Three of the
  systems show cyclotron and Zeeman lines that yield magnetic field
  strengths of 36\,MG, 19\,MG, and 33\,MG for the last three targets,
  respectively.  RX\,J0154.0-5947, RX\,J0859.1+0537, and
  RX\,J1002.2-1925 show evidence for part-time accretion at both
  magnetic poles, while RX\,J0953.1+1458 is a polar with a stable
  one-pole geometry. RX\,J1002.2-1925 shows large variations in the
  shapes of its light curves that we associate with an unstable
  accretion geometry. Nevertheless, it appears to be synchronized. We
  determined the bolometric soft and hard X-ray fluxes and the
  luminosities at the Gaia distances of the five stars. Combined with
  estimates of the cyclotron luminosities, we derived high-state
  accretion rates that range from $\dot M\!=\!2.9\!\times\!10^{-11}$\msunyr\
  to $9.7\!\times\!10^{-11}$\msunyr\ for white dwarf masses between 0.61
  and 0.82\,\msun, in agreement with predictions based on the observed
  effective temperatures of white dwarfs in polars and the theory of
  compressional heating.  Our analysis lends support to the hypothesis that different
  mean accretion rates appply for the subgroups of short-period polars and
  nonmagnetic cataclysmic variables.
}


\maketitle


\section{Introduction}

The discovery of hard and soft X-ray emission and of strong circular
polarization of the close binaries AM Her, VV\,Pup, and AN\,UMa
\citep{hearnrichardson77,tapia77} identified them as a very distinct
type of object, for which \citet{krzeminskiserkowski77} proposed the
name ``polar''. They contain a mass-losing late-type main-sequence
star and an accreting magnetic white dwarf (WD) in synchronous
rotation. Of the more than 1200 cataclysmic variables (CVs) in the
catalog of \citet[][final online version 7.24 of 2016,]{ritterkolb03}
114 are confirmed polars. Many of these were discovered by
their X-ray emission, which dominates the bolometric luminosity in
high accretion states. The basic physical mechanism underlying the
hard and soft X-ray emission was described by \citet{fabianetal76},
\citet{kinglasota79}, and \citet{lambmasters79}. Increasing the still
moderate number of well-studied polars will shed light on incompletely
understood aspects of the physics of accretion
\citep{bonnetbidaudetal15,busschaertetal15}, close binary evolution
\citep{kniggeetal11,bellonietal20}, the origin of magnetic fields in
CVs \citep{briggsetal18,bellonischreiber20}, and the complex magnetic
field structure of accreting WDs
\citep{beuermannetal07,wickramasingheetal14,ferrarioetal15}.

Our optical identification program of high-galactic latitude soft
ROSAT X-ray sources led to the discovery of 27 new polars
\citep{thomasetal98,beuermannetal99,schwopeetal02}. Twenty sources
have been described in previous publications. In this series of three
papers, we describe the remaining seven sources. In Papers I and II,
we presented comprehensive analyses of V358 \,Aqr
(=\,RX\,J2316.1--0527) \citep[][Paper~I]{beuermannetal17} and of the
eclipsing polar HY\,Eri (=\,RX J0501.7--0359)
\citep[][Paper~II]{beuermannetal20}.  Here, we present shorter
analyses of the remaining five sources RX\,J0154.0--5947,
RX\,J0600.5--2709, RX\,J0859+0536, RX\,J0953+1458, and RX\,J1002--1925
that are all noneclipsing. Our results include accurate long-term
orbital ephemerides that permit the correct phasing of observations
taken decades apart.

\begin{table}[b]
\begin{flushleft}
  \caption{Short names, epoch 2000 coordinates, high-state $V$-band
    magnitude, orbital period, and Gaia distance with 90\%
    confidence errors.}

\begin{tabular}{@{\hspace{1mm}}l@{\hspace{3mm}}l@{\hspace{2mm}}l@{\hspace{2mm}}c@{\hspace{1mm}}c@{\hspace{1mm}}c@{\hspace{2mm}}c}\\[-2ex]
\hline\hline \\[-1.5ex]
Short&\multicolumn{3}{c}{Optical position}                 & $V$    &  \po            & $d_\mathrm{\,Gaia,DR2}$\\
Name &\multicolumn{2}{c}{RA,DEC (2000)}      & l, b         & high   &  (min)              &    (pc)  \\[0.5ex]
\hline\\[-1ex]                                       
\jone & 01~54~00.9 & --59~47~49               & $289.0,-55.6$ & 15 & \hspace{1.5mm}88.95 & \hspace{-1.0mm}$320\,^{+4}_{-3}$\\[0.3ex]
\jsix & 06~00~33.3 & --27~09~19               & $233.0,-22.5$ & 19 & \hspace{1.5mm}78.69 &$1138\,^{+621}_{-325}$\\[0.3ex]
\jeight  & 08~59~09.2 &\hspace{-1.3mm} +05~36~54 & $223.2,+30.9$ & 17 &  143.93             & $437\,^{+32}_{-28}$ \\[0.3ex]
\jnine  & 09~53~08.2 &\hspace{-1.3mm} +14~58~36 & $220.0,+46.9$ & 17 &  103.71             & $448\,^{+61}_{-48}$ \\[0.3ex]
\jten & 10~02~11.7 & --19~25~37               & $257.0,+28.0$ & 17 &\hspace{1.5mm}99.98  & $797\,^{+93}_{-67}$ \\[1.0ex]
\hline\\
\end{tabular}\\[-1.0ex]
\label{tab:basic}
\end{flushleft}

\end{table}

\begin{table*}[t]
\begin{flushleft}
  \caption{Journal of X-ray observations (Cols. 1--6), interstellar
    extinction (Cols. 7--9), and the results of X-ray spectral fits
    with the sum of a multitemperature thermal component and a
    single-blackbody soft X-ray component with \ktbbi\ (Cols.
    10--12). The bolometric flux in Col. (11) is three times that of
    the single-blackbody fit (see Sect.~\ref{sec:31}). The cyclotron
    flux in Col. (13) is an estimate based on nonsimultaneous data.
    The luminosities in Cols. (14--16) are calculated with the Gaia
    distances in Table~\ref{tab:basic}. The luminosity ratio in Col.
    (17) is $R_\mathrm{L}\!=\!L_\mathrm{sx}/(L_\mathrm{hx}+L_\mathrm{cyc})$. 
    The blackbody temperatures in Col. (10) are quoted with
    errors. The errors are omitted for the derived parameters in
    Cols. (12-20).  The accretion rates in Cols. (18) and (19)
    were calculated for the WD mass in Col. (20).  Abbreviations:
    RASS = ROSAT All Sky Survey with the PSPC as detector, R = ROSAT
    pointed mode, X = XMM-Newton, P = PSPC as detector, H = HRI, pn =
    EPIC camera with pn as detector, M = MOS1\&2, hi = high state, in
    = intermediate state, and lo = low state. Numbers in brackets indicate
    errors in the last digits. A colon indicates an uncertain
    value.}

\begin{tabular}{@{\hspace{0.0mm}}c@{\hspace{2.0mm}}r@{\hspace{2.0mm}}l@{\hspace{1.0mm}}l@{\hspace{2.0mm}}r@{\hspace{1.0mm}}r@{\hspace{1.0mm}}c@{\hspace{1.0mm}}c@{\hspace{0.5mm}}c@{\hspace{0.0mm}}c@{\hspace{1.0mm}}r@{\hspace{1.0mm}}r@{\hspace{0.0mm}}r@{\hspace{1.0mm}}c@{\hspace{2.0mm}}c@{\hspace{0.6mm}}c@{\hspace{0.6mm}}c@{\hspace{1.0mm}}c@{\hspace{1.0mm}}c@{\hspace{1.0mm}}l@{\hspace{0.0mm}}l}\\[-4ex]
\hline\hline \\[-1.5ex]
(1)&(2)~~~~& \multicolumn{2}{l}{~~(3)} & (4)~ & (5)~~~& (6) & (7) & (8) & (9) & (10) & (11) & (12)~~ & (13) & (14) & (15) & (16) & (17)  & (18)  & (19) & (20) \\
Short &Start date~~~&\multicolumn{2}{l}{Instr.}&Exp.& $CR$~~~& $HR1$ & $N_\mathrm{H,gal}$ & $N_\mathrm{H,E_{B-V}}$ & $N_\mathrm{H,ad}$ & k$T_\mathrm{bb1}$ & ~$f_\mathrm{sx,bol}$ & ~~$f_\mathrm{th,bol}$  & $f_\mathrm{cyc}$ & $L_\mathrm{sx}~~~$ & $L_\mathrm{hx}~~~$  & $L_\mathrm{cyc}$ & $R_\mathrm{L}$ & $\dot M_\mathrm{x}$ & \hspace{-1mm}$\dot M_\mathrm{x+cyc}$ & $M_1$\\
Name   &      & \multicolumn{2}{l}{State~} & (s)~~ &(1/s)~~&&\multicolumn{3}{c}{\tiny \hspace{-2mm}($10^{20}$\,H-atoms/cm$^2$)}&\tiny \hspace{-1mm}(eV)& \multicolumn{3}{c}{\tiny \hspace{-2mm}$(10^{-11}$erg/cm$^2$s)} & \multicolumn{3}{c}{\tiny ($10^{32}$\,erg/s)} & &\multicolumn{2}{l}{\tiny  \hspace{-1mm}($10^{-11}\!M_\odot$/yr)}& \hspace{-1.0mm}(\msun)\\[0.5ex]
\hline\\[-1ex]
J0154 & 26~Nov~1990 & RASS & hi &   138 & 0.346~~ & $-0.78$ & 1.96 & 1.0\,(5) & ~~1.0 & $24^{+3~}_{-3}$  & $3.82$~~ & 0.25$^{1)}$ & 0.50 & 1.76~~     & 0.23~~   & 0.46~~ & 2.55 & $2.3$ & ~~2.9  & 0.75:                    \\[0.2ex]
        &  1~Jul~1992 & R~P & in &  5158 & 0.101~~ & $-0.57$ &      &          & ~~1.0 & $24^{+3~}_{-3}$  & $0.76$~~ & 0.20$^{1)}$ & 0.10 & 0.35~~     & 0.18~~   & 0.10~~ & 1.27 & $0.6$ & ~~0.7 &  0.75:\\
        &  3~Jan~1995 & R~H & hi & 11248 & 0.080~~ &         &      &          &       &                &          &            &      &            &          &      &                                &  \\[0.5ex]
   &1~May~2002 & X~pn& lo &  5040 & 0.005~~  &         &      &          &       &                &          &            &      &            &      &                                &  \\[0.5ex]
J0600 & 10~Sep~1990 & RASS & hi &   574 & 0.520~~ & $-0.92$ & 2.20 & 1.5\,(5) & ~~1.5 & $42^{+8~}_{-7}$  & $2.20^{2)}$ & 0.20$^{2)}$ & 0.08 & $6.53^{2)}$ &$1.19^{2)}$&  $0.47^{2)}$ & 3.93 & ~~$9.1^{2)}$ &~~$9.7^{2)}$& 0.75: \\
        & 12~Sep~1995 & R~H & lo & 13474 & 0.002~~ &         &      &          &       &                &          &            &      &            &          &                                &  \\[0.5ex]
J0859 & 28~Okt~1990 & RASS & hi &   368 & 0.300$^{3)}$ & $-0.86$ & 3.79 & 3.1\,(8) & ~~2.8 & $38^{+10}_{-8}$  & 3.26~~ &0.19$^{1)}$ & 0.50 & 2.79~~     & 0.33~~   & 0.86~~ & 2.36 & 5.4 & ~~6.9 & 0.61 \\
        & 24~Apr~1996 & R~H & in & 32846 & 0.022$^{3)}$ &         &      &          &       &                &          &            &      &            &          &                                &  \\[0.5ex]
J0953 & ~7~Nov~1990 & RASS & hi &   335 & 0.430$^{3)}$ & $-0.72$ & 3.11 & 2.2\,(6) & ~~2.2 & $46^{+9~}_{-8}$  & 1.99~~ & 0.56$^{1)}$ & 0.40 & 1.79~~     & 0.56~~   & 0.72~~ & 1.04 & 4.6 & ~~5.8 & 0.63 \\
        & 31~May~1996 & R~H & in &  4977 & 0.023$^{3)}$ &         &      &          &       &                &          &            &      &            &          &                                &  \\[0.5ex]
J1002 & 22~Nov~1990 & RASS & hi &   474 & 0.690$^{3)}$ & $-0.97$ &      &          &       &                &          &            &      &            &          &                                &  \\
        & 16~Nov~1992 & R~P & lo &  4192 &$-0.0002$ &       &     &          &       &                &          &            &      &            &          &                                &  \\ 
        & 30~Nov~1993 & R~P & hi &  1609 & 0.708$^{3)}$ & $-0.93$ & 3.96 & 2.9\,(9) &$\ga\!1.0$& $\la\!50$~~ &$1.78^{4)}$& 0.22$^{4)}$ & 0.43$^{4)}$ & $5.07^{4)}$ &$1.25^{4)}$ & 2.46$^{4)}$ & 1.37  & ~6.3$^{4)}$ &~~$8.7^{4)}$ & 0.82 \\
        &  4~Jun~1995 & R~H & hi & 12474 & 0.120~~ &         &      &          &       &                &          &            &      &            &      &                                &  \\[0.5ex]
        & 10~Dec~2001 & X~pn& in &  3160 &        &         & 3.96 & 2.9\,(9) &$\ga\!1.0$& $\la\!50$~~ &$0.86^{4)}$& 0.27$^{4)}$ & 0.13$^{4)}$ & $2.45^{4)}$ &$1.54^{4)}$ & 0.74$^{4)}$ & 1.08 & ~4.0$^{4)}$ & ~~$4.7^{4)}$ &  0.82\\
        &             & X~M & in &  5820 &        &         &      &          &       &                &          &            &      &            &      &                                &  \\[1.0ex]
\hline\\
\end{tabular}\\[-1.0ex]
\footnotesize{1) Fit with \nhint\,=\,0.~~ 2) Lower limit based on $d\,\ge\,813$\,pc. 3) Mean of bright phase.~~  4) Lower limits based on \nh$\,\ge\,1.0\!\times\!10^{20}$\,\atoms.  }
\label{tab:xray}
\end{flushleft}

\vspace{-6mm}
\end{table*}

\begin{table}[t]
\begin{flushleft}
\caption{Journal of time-resolved spectroscopic observations.}
\begin{tabular}{@{\hspace{0.0mm}}l@{\hspace{-4.0mm}}r@{\hspace{3.0mm}}l@{\hspace{1.0mm}}r@{\hspace{1.0mm}}r@{\hspace{0.0mm}}r@{\hspace{1.0mm}}r}\\[-3ex]
  \hline\hline \\[-1.5ex]
  Short Name& Date~~~~~~    & Wavelength & Res.  & Number  & Exp.  & Tel. \\
            &               &~~~~~(\AA)  &(\AA)~~& spectra & (s)~~ & \\[0.5ex]
  \hline\\[-1ex]                                             
  J0154--59 & 23\,Aug\,1993  & 3500--9500   & 10~~~ &   6~~~~~& 600~    & (1) \\
            & 17\,Dec\,1993  & 3500--9500   & 8~~~ &  18~~~~~& 480~    & (2) \\
         & 3--5\,Jul\,1995   & 3500--5400   &  6/8~~~ &  21~~~~~& 420~    & (1,4) \\
         & 24--25\,Nov\,1995 & 3800--9119   & 10~~~ &  35~~~~~& 300~    & (1) \\[0.5ex]
  J0600--27 & 16\,Nov\,1995  & 3600--10200  & 15~~~ &   1~~~~~& 1800~   & (1) \\
            & 4\,Mar\,1997   & 3600--10200  & 15~~~ &   9~~~~~& 900~    & (1) \\[0.5ex]
  J0859+05  & 13\,Dec\,1993  & 3500--9200   & 20~~~ &   1~~~~~& 1200~   & (1) \\
            & 5--7\,Feb\,1995& 3810--5556   &  6~~~ & 140~~~~~&trailed~ & (3) \\
            & 5--7\,Feb\,1995& 5560--9200   &  6~~~ & 150~~~~~&trailed~ & (3) \\
            & 7--8\,Feb\,1995& 4173--5071   & 1.6 &  54~~~~~&trailed~~ & (3) \\[0.5ex]
  J0953+14  & 13\,Dec\,1993  & 3500--9200   & 20~~~ &   1~~~~~& 1200~   & (1) \\
            & 5--7\,Feb\,1995& 3810--5556   &  6~~~ &  66~~~~~&trailed~ & (3) \\
            & 5--7\,Feb\,1995& 5560--9200   &  6~~~ &  78~~~~~&trailed~ & (3) \\[0.5ex]
  J1002--19 &  24\,Dec\,1992 & 3600--9120   & 15~~~ &   1~~~~~&1800~    & (1) \\
 & \hspace{-8mm}26\,Dec\,1992--1\,Jan\,1993 & 3500--5389   & 6~~~ &  21~~~~~& 600~    & (1) \\
            &   1\,Mar\,1997 & 3300--10500  & 15~~~ &  11~~~~~& 600~    & (1) \\
            &   2\,Mar\,1997 & 4070--7200   &  6~~~ &  10~~~~~& 600~    & (1) \\[1.0ex]
  \hline\\
\end{tabular}\\[-1.0ex]
\footnotesize{(1) ESO/MPI 2.2-m EFOSC2, (2) ESO 1.5m B and C
  spectrograph, (3) Calar Alto 3.5m TWIN, trailed spectra, exposure
  times $25\!-\!60$~min, (4) spectral resolution 6\,A for 1\farcs0
  slit width, and 8\,A for 1\farcs5.}
\label{tab:spec}
\end{flushleft}

\end{table}

\begin{table}[t]
\begin{flushleft}
\caption{Journal of time-resolved photometric observations.}
\begin{tabular}{@{\hspace{1.0mm}}l@{\hspace{-2.0mm}}r@{\hspace{1.0mm}}r@{\hspace{2.0mm}}r@{\hspace{2.0mm}}c@{\hspace{2.0mm}}r@{\hspace{3.0mm}}c}\\[-1ex]
  \hline\hline \\[-1.5ex]
  Short Name& Dates~~~~~~    & No. of & Total~ & Bands   & Exp. & Tel. \\
            &                & nights & hours  &         & (s)~~&      \\[0.5ex]
  \hline\\[-1ex]                                             
  J0154--59 & 16--17~Sep~1993 &  2~~ &  6.3~~ & V       & 150  &  (1) \\    
            &   8--9\,Jul~1995 &  2~~ &  6.4~~ & V       & 120  &  (2) \\
            & 29--31\,Aug~2015 &  2~~ &  5.5~~ & Sloan r & 120  &  (3) \\
            &   1--3\,Sep\,2015 &  2~~ &  5.6~~ & grizJHK &  90  &  (3) \\
        & May\,2016--Oct\,2018 & 22~~ & 39,9~~ & WL      &  60  &  (4) \\[0.5ex]
  J0600--27 &     5\,Feb\,1995 &   ~~ &  4.7~~ & V+WL    & 300  &  (1) \\
        & Sep\,2017--Jan\,2019 & 30~~ & 53.9~~ & WL      &  60  &  (4) \\[0.5ex]
  J0859+05  &    15\,Jan\,1996 &   ~~ &  9.0~~ & V       &  30  &  (1) \\
         &Feb\,2010--Jan\,2015 & 22~~ & 38.1~~ & WL      &  60  &  (5) \\
         &Feb\,2018--Jan\,2019 &  7~~ & 12.2~~ & WL      &  60  &  (4) \\[0.5ex]
  J0953+14  &     4\,Feb\,1995 &   ~~ &  7.5~~ & V       & 240  &  (1) \\
            &    19\,Mar\,2002 &   ~~ &  7.0~~ & WL      & 180  &  (6) \\
        & Jan\,2010--Feb\,2015 & 14~~ & 19.4~~ & WL      &  60  &  (5) \\
        & Feb\,2018--Jan\,2019 &  5~~ &  8.6~~ & WL      &  60  &  (4) \\[0.5ex]
  J1002--19 &     1\,Feb\,1995 &   ~~ &  2.0~~ & V       & 150  &  (1) \\
        & Feb\,2010--Feb\,2015 & 11~~ & 25.6~~ & WL      &  60  &  (5) \\
       &  Jun\,2016--Jan\,2019 & 16~~ & 42.3~~ & WL      &  60  &  (4) \\[1.0ex]
 \hline\\
\end{tabular}\\[-1.0ex]
\footnotesize{
  (1) ESO\,/\,Dutch 90 cm, (2) ESO\,/\,Danish 1.5 m, ~(3) MPI\,/\,ESO
  2.2 m, GROND, (4) MONET\,/\,S 1.2 m, (5) MONET\,/\,N 1.2 m, and (6)
  Observatorio Astron{\'o}mico de Mallorca, 30 cm; WL = white light}.
\label{tab:phot}
\end{flushleft}

\end{table}

\section{Observations}
\label{sec:obs}

In Table~\ref{tab:basic} we summarize characteristic parameters of
our targets such as the positions of the optical counterparts, the
high-state $V$-band magnitude, the orbital periods derived in this
paper, and trigonometric distances from the Gaia DR2
\citep{gaia18,bailerjonesetal18}\footnote{https://vizier.u-strasbg.fr/viz-bin/VizieR?-source=I/347}.
Finding charts were acquired from the PanSTARRS data
archive\footnote{https://ps1images.stsci.edu/cgi-bin/ps1cutouts}
\citep{chambersetal16} and are provided in Appendix~B.

\subsection{X-ray observations}

All targets discussed in this paper were discovered as variable very
soft high galactic latitude X-ray sources in the ROSAT All Sky Survey
(RASS) and observed subsequently with ROSAT in the pointed phase.  The
RASS covered the entire sky within one half year, starting in July
1990. Any target within its actual viewing strip was visited every 96
min for an exposure time of up to 30\,s. All X-ray observations of our
targets are summarized in Cols. 1--6 of Table\,\ref{tab:xray}. Columns
7--9 contain information on the interstellar extinction, and
Cols. \mbox{10--12} list results of the X-ray spectral fits that are
discussed in turn below. The hardness ratio $HR1$ in Col.~7 refers to
the ROSAT observations with the Position-Sensitive Proportional
Counter (PSPC) as detector. The PSPC was sensitive from
$0.10\!-\!2.4$\,keV, with two windows from $0.1\!-\!0.28$\,keV (pulse
height channels 11-41) and $\sim\!0.5\!-\!2.0$\,keV (pulse height
channels 51-201), which define the count rates $S$ and $H$ in the soft
and hard bands, respectively. The hardness ratio was defined as
$HR1\!=\!(H\!-\!S)/(H\!+\!S)$. The PSPC spectra of the polars in this
paper are dominated by soft X-rays and have large negative
$HR1$. Subsequent pointed ROSAT observations were made with the PSPC
or, after its shutdown due to the small amount of counter gas left,
with the High Resolution Imager (HRI). The HRI lacked energy
resolution and was less sensitive than the PSPC, but had a higher
spatial resolution and an even lower background. Some information on
the relative response of PSPC and HRI and on the energy flux per PSPC
unit count rate are given in Appendix~\ref{sec:A}. An additional
observation with XMM-Newton and the pn and MOS detectors of the EPIC
camera is available for \jten\ \citep{ramsaycropper03}.

\subsection{Optical spectrophotometry}

Time-resolved spectrophotometry was acquired between 1992 and 1997,
using the ESO 1.5 m telescope at La Silla in Chile with the Boller and Chivens

spectrograph and the ESO/MPI 2.2 m telescope with EFOSC2. The spectra
were taken with different spectral resolutions and wavelength
coverage. We refer to spectra covering the entire optical range with a
full width at half maximum (FWHM) of $10\!-\!20\AA$ as low
resolution, spectra with reduced wavelength coverage and
FWHM$\,\simeq\!6\AA$ as medium resolution, and with a
FWHM$\,=\!1.6$\AA\ as high resolution.  Trailed simultaneous blue and
red medium-resolution spectra of \jeight\ and \jnine\ were acquired
with the TWIN spectrograph on the 3.5 m telescope of the Centro
Astronomico Hispano Alem{\'a}n Calar Alto, Spain. Table~\ref{tab:spec}
summarizes the observations.

\subsection{Optical photometry}

Time-resolved photometry was acquired between 1993 and 2019. Details
are provided in Table~\ref{tab:phot}. Most of the data were taken in
white light (WL), using the two robotic \mbox{1.2 m} MONET
telescopes\footnote{https://monet.uni-goettingen.de}, MONET/N at the
University of Texas McDonald Observatory and MONET/S at the South
African Astronomical Observatory. All images were corrected for dark
current and flat-fielded in the usual way.  All times were measured in
UTC and converted into barycentric dynamical time (TDB), using the tool
provided by
\citet{eastmanetal10}\,\footnote{http://astroutils.astronomy.ohio-state.edu/time/},
which also accounts for the leap seconds.  All times that enter the
calculation of the ephemerides are listed in Appendix~\ref{sec:C}.

As described in Paper\,I, we performed synthetic photometry in order
to tie the WL measurements into the standard $ugriz$ system. We
defined a MONET-specific WL AB magnitude $w$, which has its pivot
wavelength $\lambda_\mathrm{piv}\!=\!6379$\,\AA\ in the Sloan $r$
band. For a wide range of incident spectra, the synthetic color
$|(w\!-\!r)_\mathrm{syn}|\!\lesssim\!0.1$.  Setting $w\!\simeq\!r$ is
correct within 0.1 mag, except for very red stars.

\section{General approach}
\label{sec:general}

All five targets were identified as polars by one or more of the
following properties: (i) variable soft X-ray emission, (ii) optical
spectroscopic and photometric variability, (iii) optical emission
lines with skewed profiles caused by streaming motions, (iv) strong
\heii\ line emission indicating the presence of an XUV source, (v)
cyclotron emission lines, and (vi) Zeeman absorption lines, both
indicative of magnetic field strengths in the tens of MG regime, and
(vii) long-term variations in the form of high and low states.

Observationally, a high state is characterized by intense soft X-ray
emission and strong He\,II lines produced by photoionization.
Physically, ``high'' refers to accretion rates adequate to drive the
standard secular evolution of CVs. ``Low'' refers to low-level or to
switched-off accretion. Living long-term $V$-band light curves of
\jeight, \jnine, and \jten\ are available from the monitoring program
of Ritter CVs in the Catalina Sky Survey
\citep{drakeetal09}\,\footnote{http://crts.caltech.edu/}.

\subsection{Analyzing the X-ray data}
\label{sec:31}

In a high state, the bolometric luminosity \lbol\ of a polar
is dominated by soft and hard X-ray emission.
Our targets emit intense quasi-blackbody soft X-rays and thermal hard
X-rays. The hard component originates from the post-shock cooling flow
in competition with cyclotron emission. The peak temperature in the flow
is about 30\,keV (20\,keV) for a WD of 0.75\,\msun\ (0.60\,\msun) in a
pure bremsstrahlung model with a shock near the WD surface, but lower
for a tall shock or a strong magnetic field
\citep{woelkbeuermann96,fischerbeuermann01}. Soft X-rays arise from
the complex reprocessing of the energy carried into the WD atmosphere
in the spot and its vicinity
\citep{lambmasters79,kinglasota79,kuijperspringle82}.  The bolometric
fluxes of both spectral components are difficult to measure because
the XUV flux is severely degraded by interstellar extinction and
effectively inaccessible below about 0.1\,keV.  Of the hard component,
the ROSAT PSPC catches only a glimpse and the detectors of the EPIC
camera on board XMM-Newton cover it still incompletely.

The soft component was conventionally modeled with a single blackbody
with a temperature \ktbbi. A model like this is a gross simplification,
however, as shown by the highly resolved optically thick XUV spectrum
of the prototype polar AM~Her, taken with the Low Energy Transmission
Grating Spectrometer (LETGS) on board Chandra. An improved model
involves temperatures \ktbb, ranging from about 0.5 to
$2.0\,\times\,$\ktbbi.  In the case of AM~Her, the single-blackbody
fit underestimated the bolometric energy flux by a factor of
$3.7\,\pm\,0.7$ \citep[][see their Table\,1]{beuermannetal12}. This
factor may well differ for individual polars, but in the absence of
further information, we corrected the bolometric energy fluxes
\fbbibol\ of our single-blackbody PSPC fits upward by a factor
$c_\mathrm{sx}\!=\!3$ and treated the EPIC pn-spectrum in the same
way. A separate problem in measuring $f_\mathrm{bb1,bol}$ from PSPC
spectra is the tradeoff between the blackbody temperature \ktbbi\ and
the interstellar absorbing column density \nh\ in front of the source,
which leads to large correlated errors in the fit parameters. This
problem is relaxed by recent progress in the construction of 3D models
of the galactic extinction \ebv\
\citep[e.g.,][]{schlaflyfinkbeiner11,lallementetal18}\footnote{https://irsa.ipac.caltech.edu/applications/DUST/\newline
  \hspace*{5.1mm}http://stilism.obspm.fr/}, the conversion of \ebv\ into
\nh\ \citep{nguyenetal18}, and the availability of Gaia distances
\citep{gaia18,bailerjonesetal18}. Combined, they allow an
educated guess of \nh\ that may be more trustworthy than the result of
a free PSPC fit, except for particularly well-exposed PSPC \mbox{spectra}.

In addition to absorption by interstellar gas, the emerging X-rays may suffer
internal absorption by atmospheric and infalling matter, which we
assume only affects the hard X-ray component. The internal absorber
probably fluctuates in space and time, a situation that is only
approximately described by the concept of a partial absorber with a
covering fraction \fpc\ and an unabsorbed fraction $1-$\fpc.  Because
the quality of the PSPC spectra is only moderate, we opted for a simple model
that includes (i) a multitemperature thermal hard X-ray component
with an added partial-absorber feature and (ii) a single-blackbody
soft X-ray component with the energy flux corrected upward by
$c_\mathrm{sx}\!=\!3$. The thermal component approximates a
cooling-flow model by including a Mekal or bremsstrahlung component
with a fixed high temperature of 20\,keV and one or two fitted
low-temperature Mekal components with temperatures between 0.2 and
2\,keV. For consistency reasons, we adopted the same simple model for
the single XMM EPIC pn spectrum we considered. Our
own more complex cooling-flow model \citep{beuermannetal12} did not
give significantly different integrated energy fluxes. The fits to the
ROSAT PSPC spectra are not sensitive to different levels of internal
absorption, while the fit to the XMM-Newton EPIC pn spectrum of \jten\
improves substantially when internal absorption is added, as noted
already by \citet{ramsaycropper03}.

\begin{figure}[t]
\includegraphics[height=89.0mm,angle=270,clip]{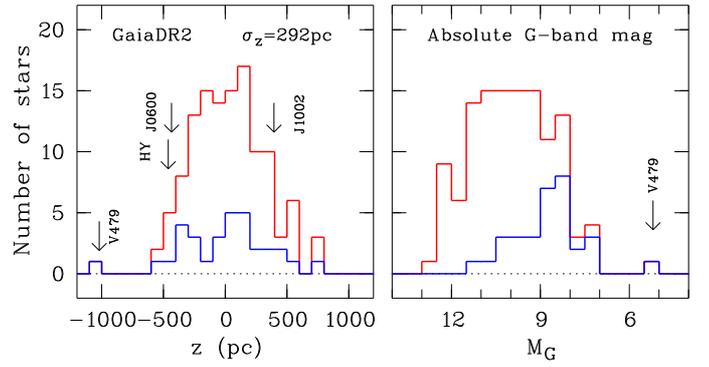}

\caption[chart] {\emph{Left: } Frequency distribution of polars
    perpendicular to the galactic plane based on the distances in the
    Gaia DR2 for the stars listed in the final 2016 version 7.24 of
    the catalog of \citet{ritterkolb03}. The red histogram shows all
    polars, and the blue histogram shows long-period systems with
    \po$\,>\!0.12$\,d. \emph{Right: } Distribution of the absolute
    G-band magnitude. Arrows indicate individual objects discussed in
    the text.}
\label{fig:z}

\end{figure}

\subsection{Distances, scale height, and luminosities}
\label{sec:2}

Part of this study was facilitated by the availability of Gaia DR2 distances
for practically all known polars \citep{gaia18,bailerjonesetal18}.
We show in Fig.~\ref{fig:z}, the frequency distribution of polars as a
function of the separation $z\!=\!d\,\mathrm{sin}\,b$ from the
Galactic plane, with $d$ the distance and $b$ the galactic latitude,
for systems listed in the final 2016 online version 7.24 of the
catalog of \citet{ritterkolb03}\footnote{The unusual magnetic CVs
  discovered by \citet{hongetal12} in a window 1\fdg4 from the
  galactic center were excluded because they distort the local
  sample, as were two polars with obviously incorrectly assigned very
  large distances.}. In Paper II and in the present paper, we found
that \hyeri, \jsix, and \jten reside at
$\mid\!z\!\mid\,\simeq\!400$\,pc, raising the question whether they
might be halo objects. The three objects are marked by arrows in the
left panel of Fig.~\ref{fig:z}, and we conclude that their distances
from the plane are entirely compatible with their being members of what may
be a single population of polars.  Its standard deviation is
$\sigma_\mathrm{z}\!=\!292$\,pc; the possible long-period polar
V479\,And \citep{gonzalezetal13} was excluded. When the
distribution is split into short-period and long-period polars, the
standard deviations become 288\,pc and 311\,pc,
respectively. These distributions may obviously be affected by an
increasing incompleteness at higher $d$ and $\mid\!z\!\mid$.
Nevertheless, the value of 288\,pc is compatible with the scale
heights of 260\,pc and 280\,pc for short-period magnetic CVs (mCVs)
advocated by \citet{pretoriusetal13} and \citet{palaetal20},
respectively, but the value of 311\,pc disagrees with the 120\,pc
assigned to young (i.e., long-period) systems by
\citet{pretoriusetal13}. The right-hand panel shows the distribution
of the Gaia DR2 absolute G-band magnitudes, which cluster between
$M_\mathrm{G}\!=\!8$ and 12 and reflect the spread in in the
luminosities caused by differences in the system parameters, notably
the instantaneous accretion rate.  Again, V479\,And deviates from the
well-defined sample of polars.

\subsection{X-ray luminosity, accretion rate, and WD temperature}
\label{sec:33}

We calculated the bolometric luminosity of component $x$ as
$L_\mathrm{x,bol}\!=\!\eta\,\pi d^2f_\mathrm{x,bol}$, where $d$ is the
distance, $f_\mathrm{x,bol}$ the respective bolometric energy flux,
and $\eta$ is a geometry factor, which is $\eta\,=\,4$ for isotropic
emission. For a plane surface element that is viewed at an angle
$\theta$, the geometry factor is $\eta\!=\!1/\mathrm{cos}\,\theta$.
Because $\theta$ is only approximately known, we used a conservative
mean $\eta\!=\!1.5$, although \citet{heiseetal85} argued for an
'emitting mound' with $\eta\,\simeq\,2$ (see also the $\eta$-values in
Table~2 of Beuermann et al. 2012). We used $\eta\!=\!3$ for the
thermal component, which accounts approximately for the X-ray albedo
of the WD atmosphere, and $\eta\!=\!3$ for the beamed
cyclotron radiation.  With $c_\mathrm{sx}\!=\!3$ from above, we
calculated the bolometric accretion-induced luminosity as 
\begin{equation}
  L_\mathrm{bol}\!\simeq\!\pi d^2(1.5\,c_\mathrm{sx}f_\mathrm{bb1,bol}+3f_\mathrm{th,bol}+3f_\mathrm{cyc,bol})\!\simeq\!\mathrm{G}M_1\dot M/R_1
\label{eq:lum}
\end{equation}
and equated it to the gravitational energy released by matter accreted
from infinity at a rate $\dot M$ by a WD of mass $M_1$ and radius
$R_1$. We added the X-ray luminosity and an estimate of the cyclotron
luminosity in an attempt to describe the accretion-induced luminosity.
We disregarded the stream emission.

The effective temperature $T_1$ of a sufficiently old accreting WD is
thought to be largely determined by compressional heating
\citep{townsleygaensicke09}. This theory relates the equilibrium
temperature $T_\mathrm{eq}$
to the long-term mean accretion rate $\langle\dot
M\rangle_\mathrm{10}$ in units of $10^{-10}$\,\msunyr\ by
\begin{equation}
T_\mathrm{eq}\!=\!14200\,\langle\dot M\rangle_\mathrm{10}^{1/4}\,(M_1/0.75M_\mathrm{\odot})~~\mathrm{K}.
\label{eq:temp}
\end{equation}
We here derived the accretion rate $\dot M$ for a measured or
adopted $M_1$ and quote the temperature $T_1$ the WD would have if
$\dot M$ were identified with $\langle\dot M\rangle_\mathrm{10}$.  We
discuss to which extent the accretion rates derived by us fit into the
general picture of short-period polars.

\begin{figure}[t]
\includegraphics[height=74.0mm,angle=0,clip]{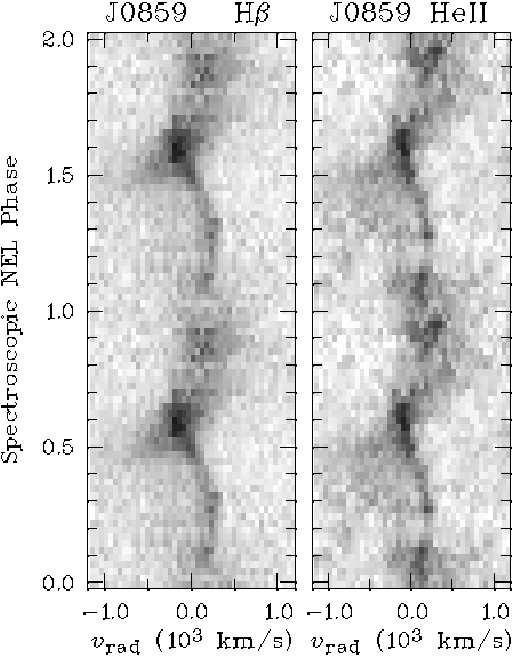}
\includegraphics[height=74.0mm,angle=0,clip]{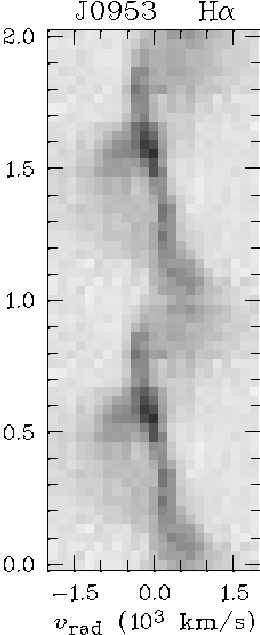}
\caption[chart] {2D representations of the phase-resolved \heii\
    and \hbet\ spectra of \jeight\ (FWHM 1.6\AA) and the \halp\
    spectra of \jnine\ (FWHM 6\AA); the flux increases from
    white to black. The NEL component is prominent over one half of
    the orbit centered on the superior conjunction of the secondary
    star. The data are shown twice for better visibility of the
    orbital structure. The ordinate is spectroscopic phase with
    $\phi\!=\!0$ at the blue-red crossing of the NEL (see
    Sects.~\ref{sec:0859} and \ref{sec:0953}). } 
\label{fig:vrad}
\end{figure}

\subsection{Ultraviolet and optical luminosity}

No simultaneous X-ray and UV or optical observations are
available for the present targets. We therefore constructed the
UV-optical-IR spectral energy distributions (SED) from all available
(nonsimultaneous) data in order to obtain an overview that would enable
us to pick appropriate pairs of X-ray and optical flux levels.  The
upper envelope to the SED is taken as a measure of the
UV-optical-IR flux in a high state of accretion. In
favorable cases, the lower envelope provides information on the
contributions by the stellar components.  We collected the data using
the Vizier SED tool provided by the Centre de Donn\'ees astronomiques
de Strasbourg\footnote{http://vizier.unistra.fr/vizier/sed/}. We
searched the Galaxy Evolution Explorer (GALEX, Bianchi et al. 2017),
the Sloan Digital Sky Survey (SDSS, Aguado et al. 2018)\footnote{Data
  Release 15, http://www.sdss.org/dr15}, the Pan-STARRS Data Release~1
\citep{chambersetal17}, the SkyMapper catalog \citep{wolfetal19}, the
Gaia catalog \citep{gaia18}, the Two Micron All Sky Survey (2MASS,
Skrutskie et al. 2006), the UKIRT Infrared Deep Sky Survey (UKIDSS,
Lawrence et al. 2007), the VISTA Catalog \citep{mcmahonetal13}, the
Wide-field Infrared Survey (WISE, Cutri et al. 2012,2014), the PPMXL
catalog \citep{roeseretal10}, and the NOMAD catalog
\citep{zachariasetal05}. \citet{harrisoncampbell15} studied the light
curves of our targets in the WISE W1 and W2 bands. SPITZER Space
Telescope data of \jone\ were discussed by \citep{howelletal06}.

\subsection{System parameters}
\label{sec:NEL}

Polars display complex Balmer and helium emission lines.
\citet{schwopeetal97} distinguished three components, a narrow
emission line (NEL) from the heated face of the secondary star, a
broad base component (BBC) from the magnetically guided part of the
accretion stream, and a high-velocity component (HVC) from the
ballistic part of the accretion stream in the orbital plane. The small
velocity dispersion of the NEL represents the distribution of emission
from the static stellar atmosphere, while the widths of the two other
components reflect the internal velocity variation of the accelerating
stream. Fig.~\ref{fig:vrad} shows the trailed spectra of the emission
lines of \heii\ and \hbet\ in \jeight\ and of \halp\ in \jnine, taken
on 7-8 and 5-7 February 1995, respectively. The lines are shown in the NEL
(binary) phase with red-to-blue crossing at $\phi\!=\!0.5$. As
expected for a sufficiently high inclination, the NEL is visible
approximately from quadrature over superior conjunction to quadrature.
The HVC crosses the NEL near spectroscopic phase 0.50. The fuzzy
excursions to high positive and negative velocities result from the
combined action of HVC and BBC. The NEL for the other three targets in our
sample is well resolved in \jten, is perhaps marginally
detected in \jone, and remains undetected in \jsix. We measured the radial velocity for systems with resolved NEL by fitting a single
Gaussian. For the combined BBC and HVC with its complex profile, we
used the centroid of the cursor-defined full extent of the line near
zero intensity, which suffices for our purpose because no
quantitative argument is based on the measured broad-line
velocities. For the two systems in which the NEL was not resolved, we
measured radial velocities by fitting single Gaussians to the total
line profiles.
 
The observed velocity amplitude $K_2'$ of the NEL represents the
centroid of the emission from the secondary star and its FWHM the
distribution of the emission over the star. These need not be the same
for the NEL of different species \citep[e.g.,][]{schwopeetal00}.
Transforming the observed $K_2'$ into the velocity amplitude $K_2$ of
the center of mass of the secondary star therefore requires a model
of the emission in the respective line. There is evidence that metal
lines with low-ionization potentials are best suited to trace the
secondary star \citep{schwopeetal00,beuermannreinsch08,beuermannetal20}.  
For this pilot study, we disregarded these complications and
applied the irradiation model BR08 \citep{beuermannreinsch08}, which was
devised for CaII$\lambda8498$, also to the Balmer lines and to
\heii. Using the inferred value of $K_2$ and a mass-radius relation
$R_2(M_2)$ of the Roche-lobe-filling secondary star, we calculated the
system parameters as a function of the unknown inclination $i$.  We
used the radii of main-sequence stars of solar composition and an age
of 1\,Gyr of \citet[][henceforth BHAC]{baraffeetal15} for masses
between 0.072 and 0.200\msun, represented by a power law
$R_\mathrm{BHAC}/R_\odot\!=\!A\,(M_2/M_\odot)^B$ with $A\!=\!0.831$
and $B\!=\!0.827$. Because the secondary stars in CVs are known to be
more or less bloated, we set $R_2\!=\!R_\mathrm{BHAC}f_{123}$, where
$f_{123}\!=\!f_1f_2f_3$ and $f_1\!=\!1.020$ accounts for expansion by
magnetic activity and spot coverage, $f_2\!=\!1.045$ for tidal and
rotational deformation of the Roche-lobe-filling star, and
$f_3\!\ge\!1.0$ for the increased radius of a star driven out of
thermal equilibrium \citep[][and discussion in
Paper~II]{kniggeetal11}.  Application of Kepler's third law and Roche
geometry \citep{kopal59} yields the component masses as functions of
$f_{3}$ and the inclination $i$ for a given $K_2'$. With $R_2$ and $d$
known, the $i$-band magnitude of the secondary star is obtained, using
the calibration of the $i$-band surface brightness
$S_\mathrm{i}\!=\!i_\mathrm{AB}+5\,\mathrm{log}(R_2/R_\odot/d_\mathrm{pc}+1)$
that we established as a function of color or spectral type in
Paper~II. For spectral types dM4 to dM8 in steps of one subclass,
$S_\mathrm{i}\!\simeq\!7.6$, 8.2, 8.7, 9.2, and $\sim\!10$.  The
expected spectral types of the secondaries are quoted by
\citet{kniggeetal11} in their Table 2. For the two of our objects that
show the secondary star in their spectra, the observed $i$-band
magnitudes and those predicted at the Gaia distance agree well.

None of the systems discussed in this paper is eclipsing, and obtaining
information on $i$ is often based on circumstantial evidence.  If the
primary accretion spot at colatitude $\beta$ suffers a self-eclipse
for a phase interval $\Delta \phi$, the angles $i$ and $\beta$ \mbox{are
related by}
\begin{equation}
  \mathrm{tan}\,i\,\mathrm{tan}\,\beta = 1/\mathrm{cos}(\pi \Delta \,\phi).
\label{eq:visi}
\end{equation}
The colatitude $\zeta$ of the field vector in the spot usually exceeds
$\beta$ somewhat, but accounting quantitatively for the difference
requires a closer study. The emitted cyclotron radiation is most
intense perpendicular to and minimum along the field direction
(cyclotron beaming). The cyclotron minimum is more than 10\degr\ wide,
and we did not differentiate explicitely between $\zeta$ and
$\beta$. The shape of the minimum may be modified by ff- and
bf-absorption in the infalling matter. For $i\!>\!\beta$, a narrow
absorption dip may occur when the line of sight crosses the
magnetically guided part of the accretion stream, or a wide
depression, when it is formed by an extended accretion curtain. The
X-ray light curves are similarly shaped by geometric effects and
photoabsorption.

\begin{figure}[t]
\includegraphics[height=89.0mm,angle=270,clip]{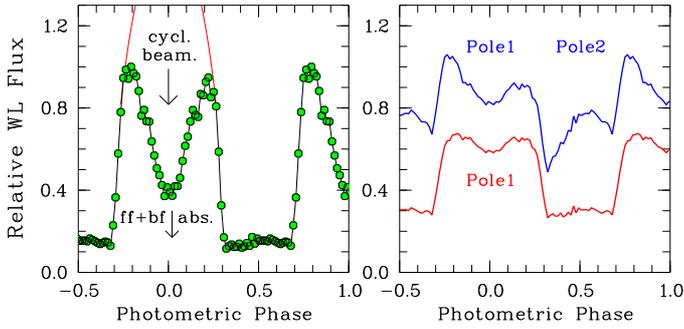}

\caption[chart]{\emph{Left: } WL light curve of the persistent
    one-pole system \jnine\ from \citet{kanbachetal08}. The accretion
    spot is visible for 60\% of the orbital period, and the central
    depression is created preferentially by cyclotron
    beaming. \emph{Right: } Derived light curves depicting the effect
    of reduced beaming (red) and of adding a second accretion spot
    (blue). The latter mimics the light curves observed at times in
    \jeight\ and \jten.}
\label{fig:GK}

\end{figure}

\subsection{Orbital light curves and ephemerides}

The interpretation of the optical light curves of noneclipsing polars
varies from simple to complex. As a simple case, we show in
Fig.~\ref{fig:GK} (left panel) the binned WL light curve of the stable
one-pole emitter \jnine\ observed by \citet{kanbachetal08}. The
accreting pole is visible between orbital phases $-0.30$ and $+0.30,$
and the light curve is shaped primarily by cyclotron beaming with a
central minimum at the instant of closest approach of the line of
sight to the accretion funnel, chosen here to define orbital phase
$\phi\!=\!0$. The solid red curve shows the expected light curve
formed by the varying aspect of the accretion spot without the beaming
effect. Using the observed light curve (with its fluctuations) as a
model, we constructed more complex cases. Increasing the minimum value
of $\mid\!\beta-i\!\mid$ or the optical depth of the emission region
reduces cyclotron beaming. An example is shown by the red curve in the
right panel. Adding a second emission region in the lower hemisphere
that is less influenced by beaming and is, for example, phase-shifted
by 200\degr\ produces the blue curve, in which the three orbital
minima can no longer uniquely be assigned to the two poles without
independent information. Circular spectropolarimetry can provide the
required information, as demonstrated for the case of \hyeri\ in Paper
II, but is not available in the present study. The blue curve mimics
the light curve 1996~V of \jeight\ in Fig.~\ref{fig:0859} and the
light curves of 25 February 2018 and 13-14 December 2010 of \jten\ in
Fig.~\ref{fig:1002}.

For each of our targets, we searched for an orbital feature that
reliably marks the orbital period. A detected period was accepted as
the orbital one if it agreed with the period of the radial-velocity
variation, preferentially of the narrow component. The spectroscopic
and photometric periods agreed in all cases within the uncertainties
and all objects were accepted as synchronized polars, in part with
tight margins for a possible remaining asynchronism. The errors of the
derived orbital periods are sufficiently small to exclude alias
periods over the time span of 30~yr, except for the faint object
\jsix, which exhibits a remaining uncertainty of one orbit in 150,000
cycles between 1995 and 2017.

We applied several methods to determine the timings of orbital
features. These included sinusoidal fits to light curves, Gaussian
fits to the fluxes around minima, and graphical methods. For instance,
in double-humped light curves as that of \jnine, we measured (i) the
ingress to and egress from the bright phase individually, (ii) its
center as the mean of the two timings, or (iii) the position of the
central minimum and chose the one that displayed the smallest
long-term scatter. We determined times of minima or maxima
preferentially by the bisected-chord technique, marking the center
between fall-off and rise at various intensity levels, and measuring
the desired time and its error from the mean and the scatter of the
markings. Our approach minimizes timing errors for objects whose light
curves are distorted by flickering or noise.

\begin{figure*}[t]
\includegraphics[height=89.0mm,angle=270,clip]{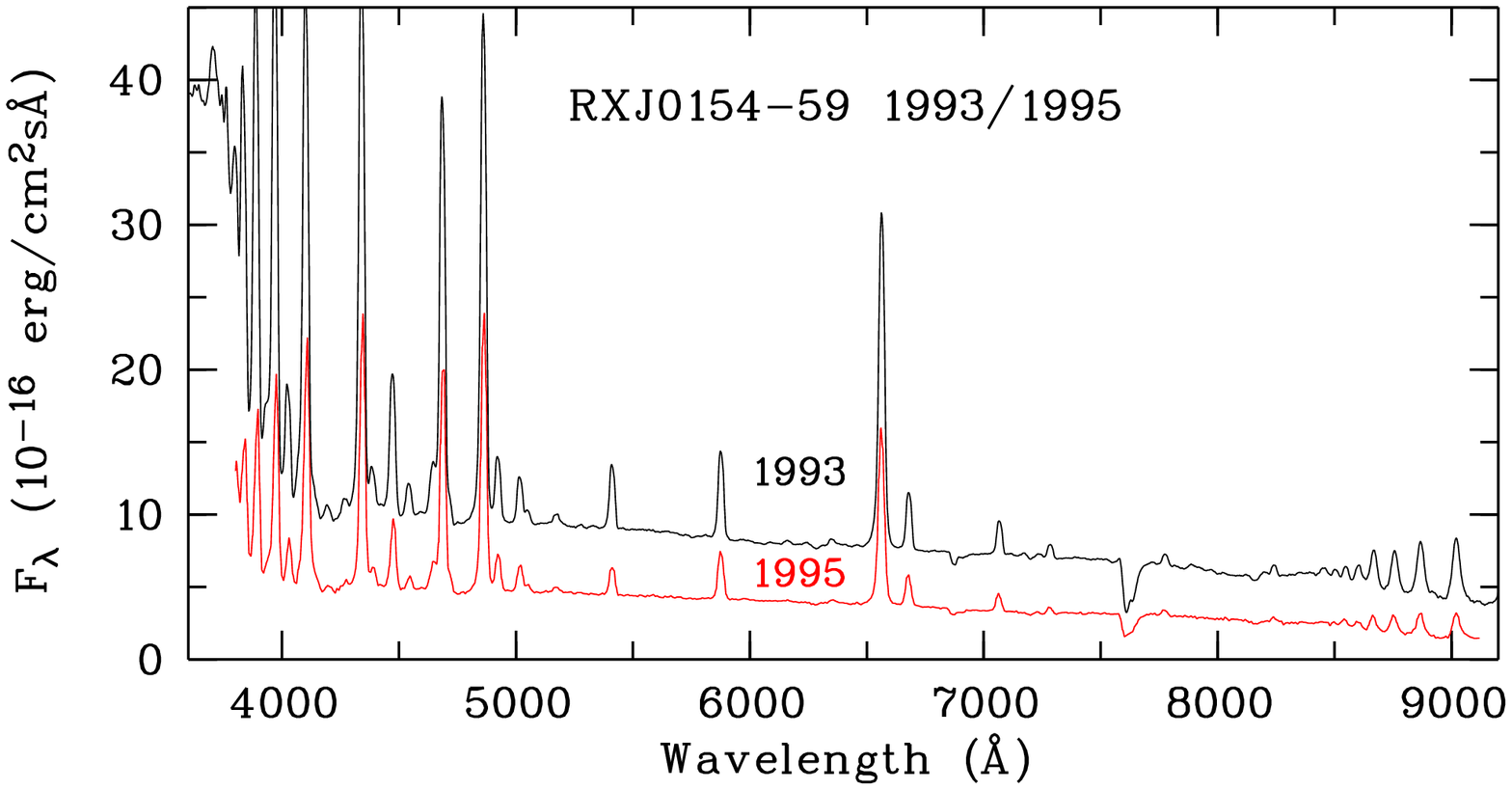}
\hfill
\includegraphics[height=89.0mm,angle=270,clip]{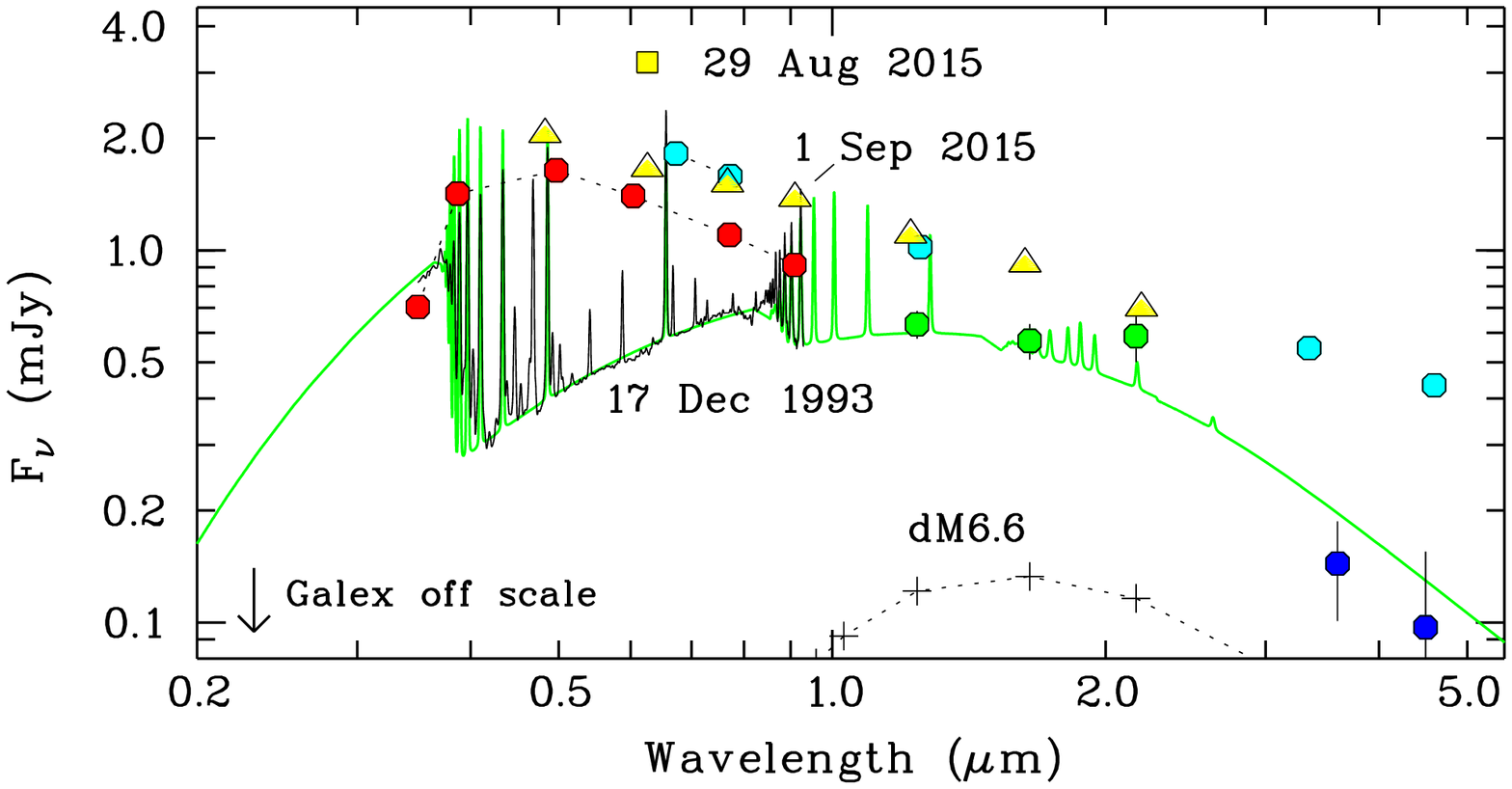}

\medskip
\begin{minipage}[t]{90mm}
\includegraphics[height=89.0mm,angle=270,clip]{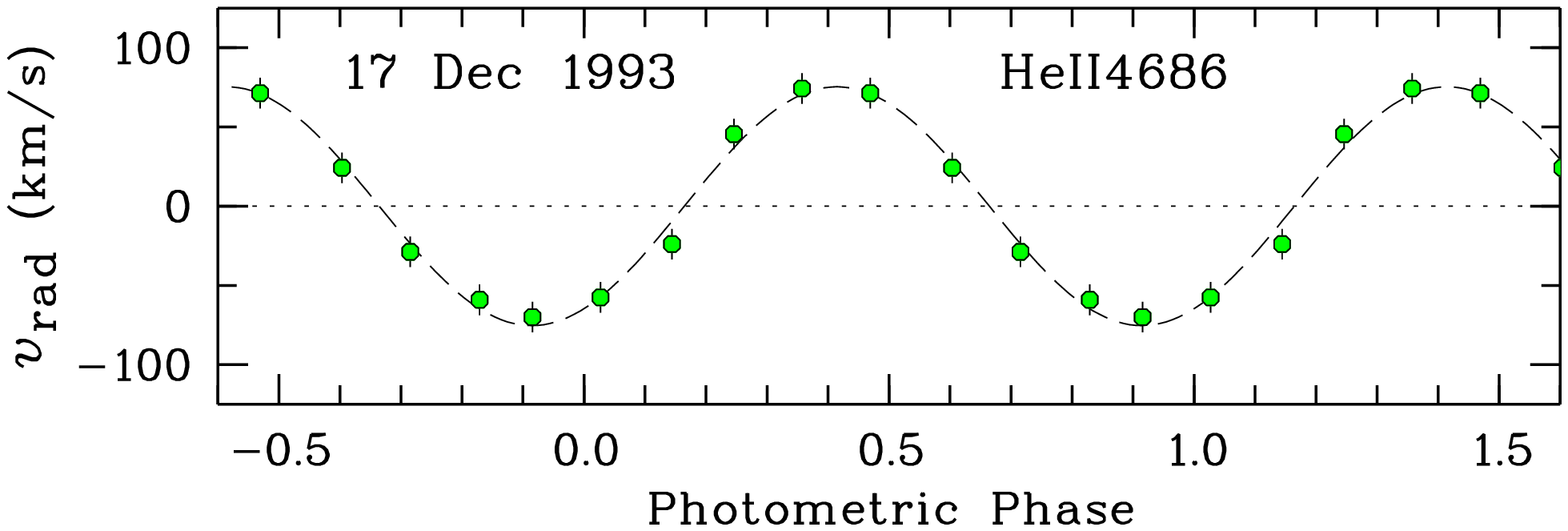}

\medskip
\includegraphics[height=89.0mm,angle=270,clip]{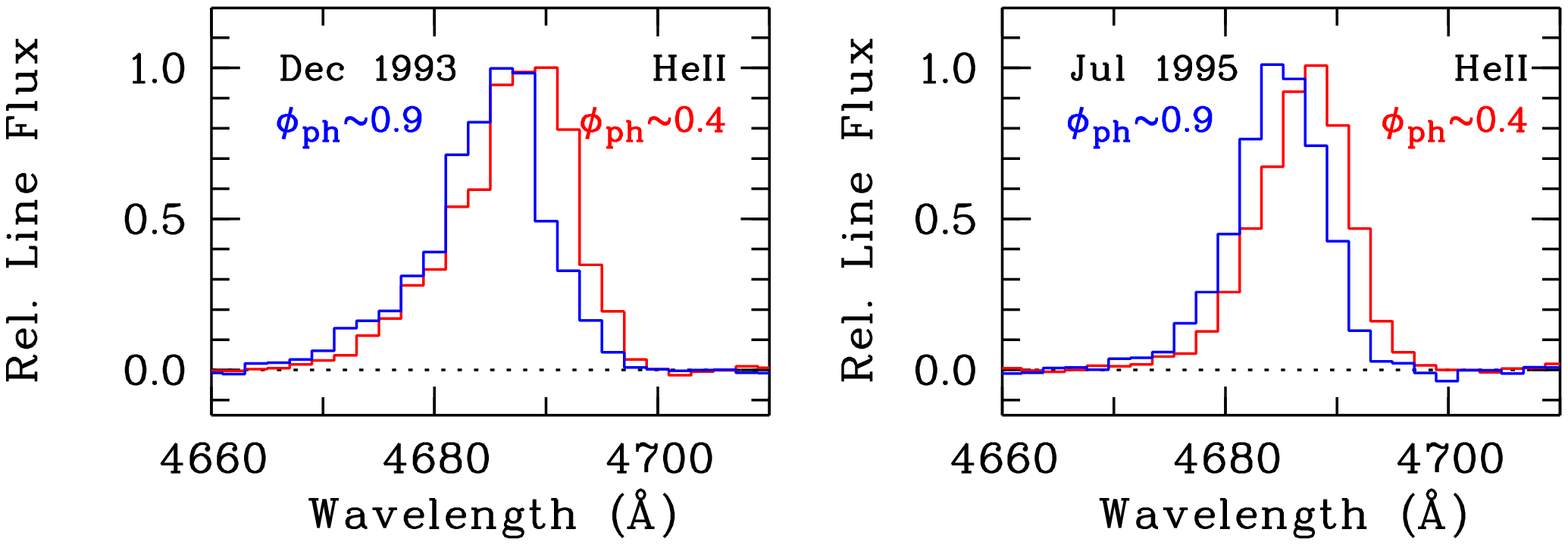}

\medskip
\includegraphics[height=89.0mm,angle=270,clip]{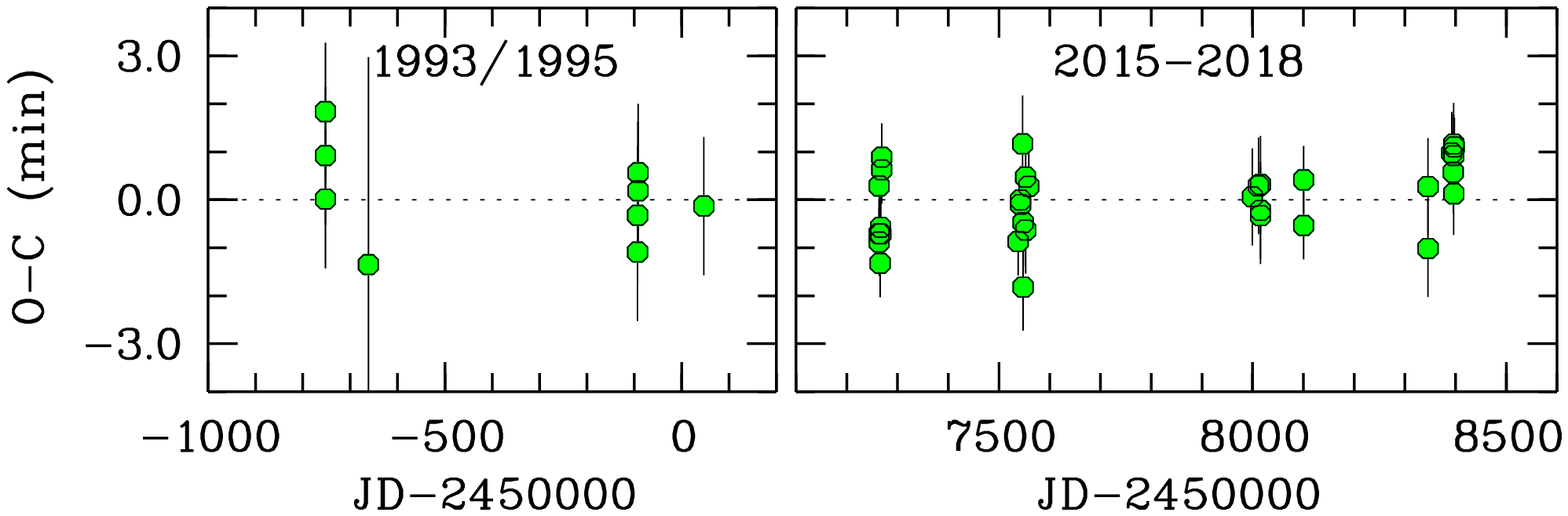}
\end{minipage}

\hspace{94mm}
\begin{minipage}[t]{90mm}
\vspace*{-94.3mm}
\includegraphics[height=89.0mm,angle=270,clip]{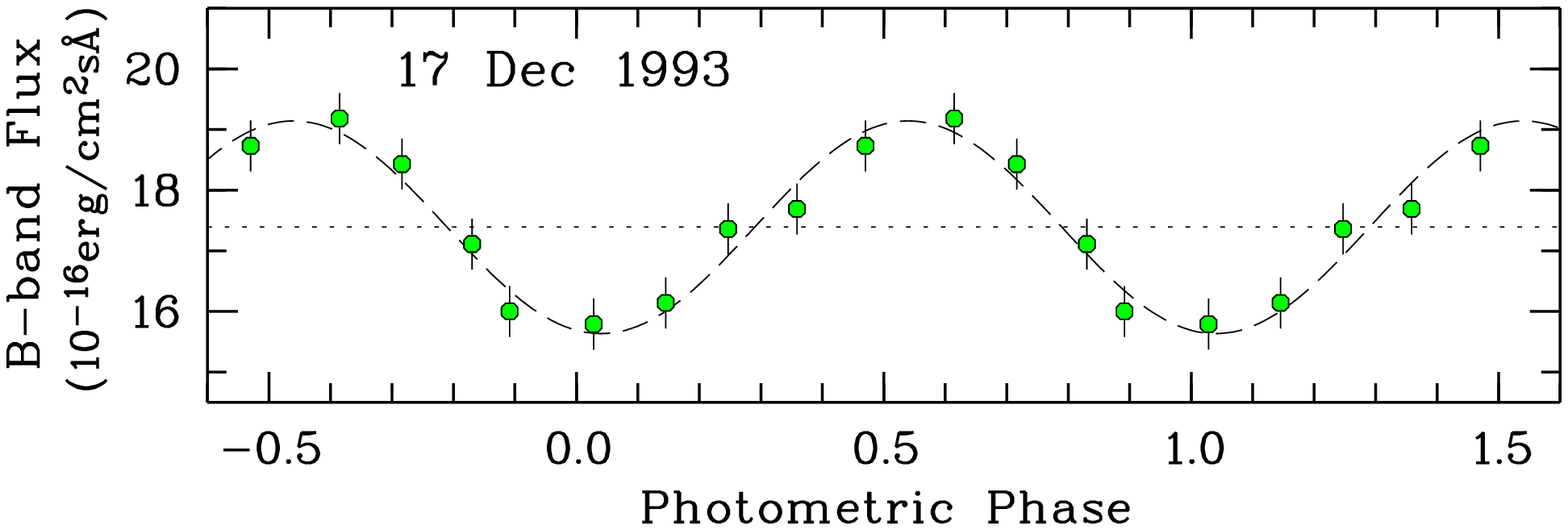}

\vspace{-1mm}
\includegraphics[height=89.0mm,angle=270,clip]{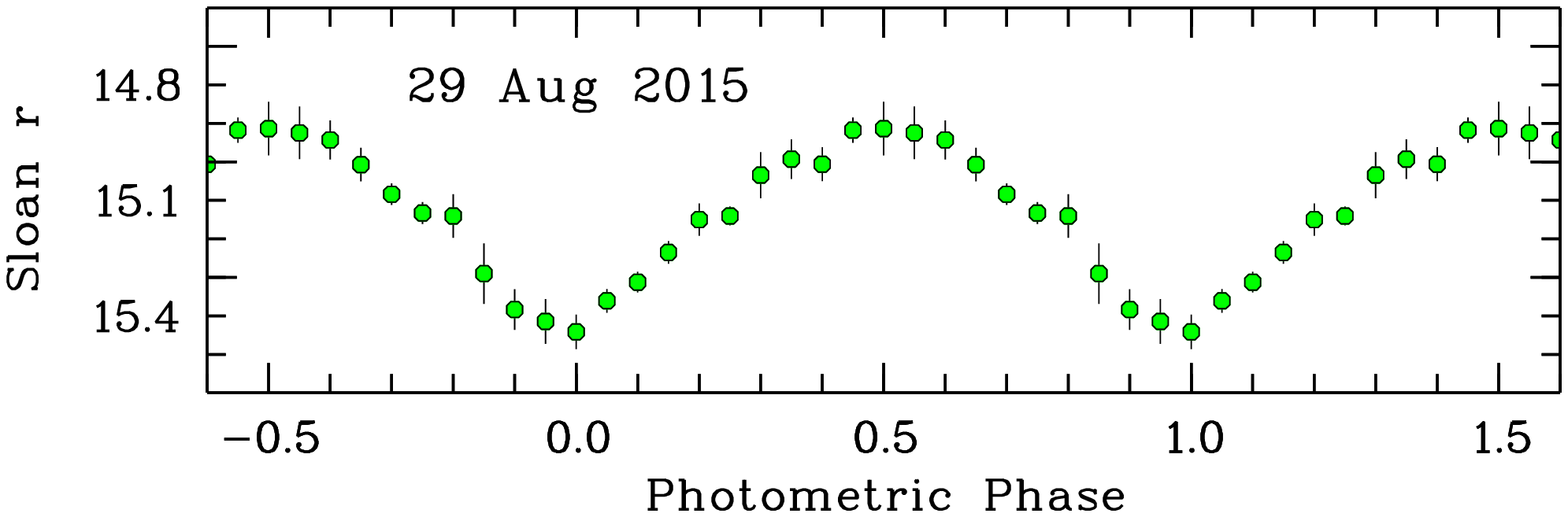}

\vspace{-1mm}
\includegraphics[height=89.0mm,angle=270,clip]{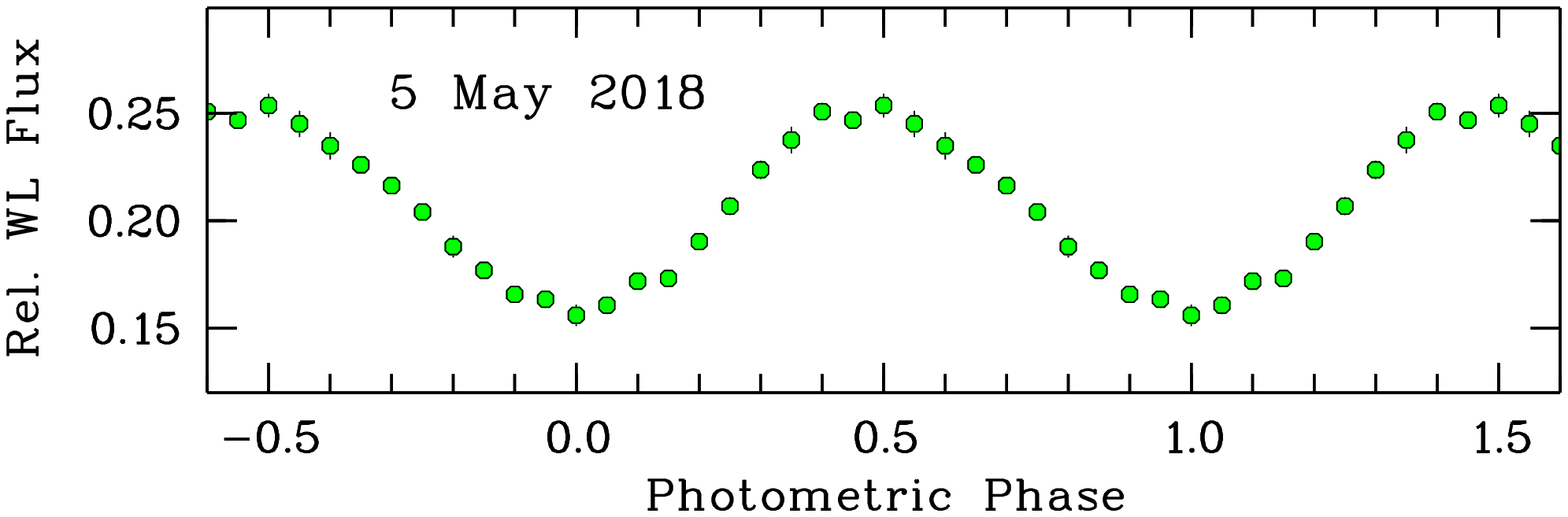}

\vspace{-1mm}
\includegraphics[height=89.0mm,angle=270,clip]{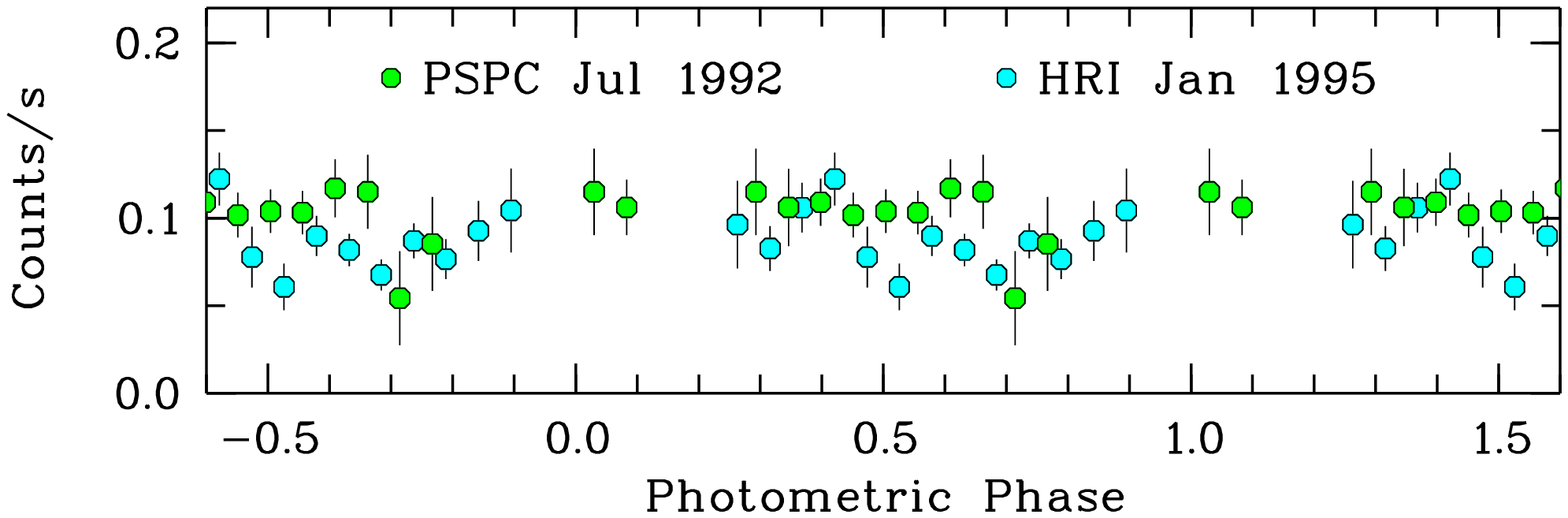}
\end{minipage}

\caption[chart]{\rxjone. \emph{Left, top: } Mean spectra of of the
  observations on 23 August 1993 and 24 November 1995. \emph{Second from top:}
  \heii\ radial velocities obtained from the 17 December 1993
  spectrophotometry with fitted sinusoid superimposed. \emph{Third
    from top:} \heii\ line profiles at maximum blueshift (blue curve)
  and redshift (red curve). \emph{Bottom:} $O\!-\!C$ diagram for
  the times of orbital minimum. \emph{Right, top:} Overall spectral
  energy distribution (see text). \emph{Bottom set of four panels:}
  Spectral flux in the $B$ band of 17 December 1993 with peak magnitude
  $B\!=\!16.3$, Sloan $r$ light curve of 29 August 2015 with peak
  brightness $r\!=\!14.9$, WL relative flux on 5 May 2018, reaching
  magnitude $w\!=\!15.2$, and ROSAT X-ray light curves taken with the
  PSPC in the intermediate state of 1--2 July 1992 and with the HRI in
  the high state of 3--7 January 1995.  All light curves are phased on the
  ephemeris of Eq.~\ref{eq:0154ephem}. }
\label{fig:0154}

\end{figure*}

\section{\rxjone\ (= \jone)~in Hydrus}
\label{sec:0154}

\jone\ was discovered 1990 in the RASS as a moderately bright soft
X-ray source, spectroscopically identified by us as a polar and listed
as such in \citet{beuermannthomas93}, \citet{beuermannburwitz95}, and
\citet{beuermannetal99}. With up to $r\!=\!14.9$, it is the brightest
star in our sample and with a Gaia distance of $d\!=\!320\,\pm\,4$\,pc
(Table~\ref{tab:basic}) also the nearest.

\subsection{X-ray observations}

In the RASS, the star was detected with a PSPC count rate of
$0.35\!\pm\!0.05$\,\cps\ and a hardness ratio
$HR1\!=\!-0.78\!\pm\!0.08$ (Table~\ref{tab:xray}). In a pointed
5.1\,ks PSPC observation of 1992, it was found at a reduced mean count
rate of 0.10\,PSPC\,\cps\ and an increased hardness ratio of $-0.57$,
suggesting that only the soft component had weakened. In an 11.2\,ks
HRI observation on 3--7 January 1995, the system was found at a mean count
rate of 0.080\,HRI\,cts/s, which translates into about 0.64\,PSPC\,\cps\
(Table~\ref{tab:crrat}), suggesting that the RASS observation
represents only a moderate high state. The count rates of both
observations in the bottom right-hand panel of Fig.~\ref{fig:0154}
show little orbital variation. Phases are from
Eq.~\ref{eq:0154ephem}. The visible pole of \jone\ had nearly stopped
accreting, when XMM-Newton barely detected it with 0.005(3) EPIC pn
\cps\ on 1 May 2002 \citep{ramsayetal04}.

We fit the 1992 PSPC spectrum (not shown) with the X-ray spectral
model described in Sect.~\ref{sec:31}. The fit gave
\ktbbi$\,=\!24$\,eV, \nh$\,=\!\ten{1.0}{20}$\,\atoms, and a blackbody
flux \fbbibol\ that translates to
$f_\mathrm{sx,bol}\!=\!c_\mathrm{sx}\,f_\mathrm{bb1,bol}\!=\!\ten{7.6}{-12}$
(Table~\ref{tab:xray}), with $c_\mathrm{sx}\!=\!3$ from
Sect.~\ref{sec:31}. The reddening in front of the target is
\ebv$\,\simeq\!0.011$ \citep{lallementetal18} or about half the
galactic value of 0.0195 \citep{schlaflyfinkbeiner11}, which
corresponds to \nh$\,=\!\ten{1.0}{20}$\,\atoms\ \citep{nguyenetal18},
supporting the PSPC fit.  In the brighter but statistically inferior
RASS spectrum, the count rate of the soft component is higher by a
factor of five, and the hard component remains about the same.  For
the same blackbody temperature of 24\,eV, the soft X-ray flux in
Table~\ref{tab:xray} is raised by a factor of five as well. A soft X-ray
variability, exceeding that of the hard X-ray component, was seen also
in other polars and taken as evidence for the concept of blobby
accretion \citep{kuijperspringle82}, which carries the kinetic
free-fall energy into subphotospheric layers and releases the
reprocessed energy as soft X-rays, independent of more tenuous
sections of the flow that pass through a free-standing shock,
radiating hard X-rays.

\subsection{Orbital ephemeris}

Time-resolved optical photometry of \jone\ was performed over a time
span of 25 yr. Its brightness was measured relative to a comparison
star located at RA(2000)$\,=\!01^\mathrm{h}54^\mathrm{m}00\fs9$,
DEC(2000)$\,=\!-59\degr43\arcmin33\arcsec$, or 0\arcsec\ W and
256\arcsec\ N of the target, which has $V\!=\!13.85$,
$B\,-\,V\!=\!0.62$, Sloan $r\!=\!13.61$, and
$w\,-\,r\!\simeq\!0.08$. All light curves of \jone\ are characterized
by a quasi-sinusoidal modulation with a period of 89~min
(Fig.~\ref{fig:0154}, lower right-hand panels). The star reached
$r\!=\!14.9$ on 29 August 2015 and $w\!=\!15.2$ on 5 May 2018. Radial
velocities measured from low-resolution spectra taken on 23 August and 17
December 1993 confirmed the photometric period as the orbital one (second
left-hand panel from the top).
>From 39 photometric minima of 1993, 1995, 2015, 2016, 2017, and
2018 (Table~\ref{tab:0154}), we obtained the alias-free linear ephemeris
\begin{equation}
T_\mathrm{min}\!=\!\mathrm{TDB}~2457263.76473(10) + 0.061767741(3)\,E.~~~
\label{eq:0154ephem}
\end{equation}
The $O\!-\!C$ diagram is shown in the lower left panel of
Fig.~\ref{fig:0154}. 
The tentative period $P_\mathrm{orb}\!=\!0\,\fd0556$
cited in the catalog of \citet[][final version 7.24 of
2016,]{ritterkolb03} is not confirmed.

\subsection{Spectrophotometry}
\label{sec:0154field}

Time-resolved low-resolution spectroscopy was collected on 23 August
1993, 17 December 1993, and on 24--25 November 1995. The top left panel shows
the slightly smoothed mean low-resolution spectra of August 1993 and November
1995. They correspond to mean AB magnitudes $r\!=\!16.1$ and 16.9,
respectively. The source displayed emission lines of H\,I, He\,I,
He\,II, a strong Balmer jump in emission, and weak metal lines. No
spectral features of the secondary star were detected. The large
\heii\ equivalent width with a ratio
$W$(\heii)/$W$(\hbet)$\,\simeq\!0.79$ (Table~\ref{tab:w}) is typical
of the high-state emission-line spectrum of a polar.  The line
profiles extend to $-1500$ and $+1000$\,\kms, more in the Balmer than
in the helium lines, but they are all peculiar in showing very little
orbital variation.  Single-Gaussian fits to lines observed in December
1993, July 1995, and November 1995 gave velocity amplitudes between 60 and
80\,\kms, with very little variation in the phasing. The radial-velocity curve of the \heii\ line in December 1993 is shown in the second
left panel from the top in Fig.~\ref{fig:0154}. The blue-to-red zero
crossing for this line occurred at photometric phase
$\phi_\mathrm{br}\!=\!0.164(15)$ in December 1993, at ~0.17(2) in July 1995,
and at 0.16(2) in November 1995. The \hbet\ results are very similar. We
show examples of the \heii\ lines at maximum positive and negative
excursion in Fig.~\ref{fig:0154} (third left panel from the top). The
spectral resolutions between 6 and 10\,\AA\ did not suffice to
identify the NEL component, although the variation in the line peak in
1993 may be an indication of its presence. Assuming that the line
nevertheless relates to a fixed structure in the binary system (as the
illuminated face of the secondary), these events define the
spectroscopic or binary orbital period,
$P_\mathrm{sp}\!=\!0\,\fd061767740(123)$. The period $P$ in
Eq.~\ref{eq:0154ephem}, on the other hand, represents the rotational
period of the WD. The difference is consistent with zero and limits
any asynchronism to a level of $\ten{2}{-6}$. Because identification of
the NEL is required to locate the secondary star, it may be rewarding
\mbox{to study this bright system at higher spectral resolution}.

\subsection{Spectral energy distribution}

Further insight into the properties of the system is obtained from the
nonsimultaneous overall SED in the upper right-hand panel of
Fig.~\ref{fig:0154}, which shows the mean spectrum of August 1993 (black
curve) along with a model spectrum (green curve) for an isothermal
slab of hydrogen at a temperature of 17000\,K, a pressure of
$10^3$\,dyne\,cm$^{-1}$, and a slab thickness of $10^8$\,cm.
The model fits the observed 1993 spectrum and is consistent with the
2MASS near-IR fluxes (green dots) and the Spitzer IRAC fluxes at
3.6 and 4.5\,$\mu$m reported by \citet{howelletal06} (blue dots).  We
interprete this spectrum as the signature of a luminous accretion
stream. There is no evidence for the associated accretion spot,
however, which is evidently located on the far side of the white dwarf
and is permanently out of view. On the other hand, our grizJHK
photometry of August and September 2015 (yellow triangles and yellow square),
photometry of SkyMapper (red), 2MASS (green), and Gaia\,2, VISTA, and
WISE (cyan blue) show higher fluxes that are probably associated with
a spot on the near hemisphere that accretes only temporarily. Our
optical and X-ray observations suggest that it was active in November 1990,
July 1992, January 1995, and August and September 2015, but not in August and December 1993, July
1995, and May 2002.  The persistent stream emission when the near spot
is inactive explains the preponderance of negative radial velocities
in the 1993 line profiles in Fig.~\ref{fig:0154} by the plasma motion
toward the unseen pole.

The variability of the source is also indicated by the GALEX far-UV and
near-UV fluxes of 0.050\,mJy at 0.153 and 0.231\,$\mu$m that belong to a
low state, consistent with representing the WD. No spectral signature
of the secondary star is detected.

\subsection{System parameters}
\label{sec:0154system}

For \po=89\,min, the evolutionary sequence of \citet{kniggeetal11}
with $M_1\!=\!0.75$\,\msun\ predicts a Roche-lobe-filling secondary
star with $M_2\!\simeq\!0.085$\,\msun, $R_2\!\simeq\!0.132$\,\rsun,
and spectral type dM6.6. For this spectral type, the $i$-band surface
brightness of the secondary star is $S_\mathrm{i}\!\simeq\!9.0$ and
its $i$-band magnitude becomes $i\!=\!20.9$\ mag (0.016\,mJy) for
$d\!=\!320$\,pc (Table~\ref{tab:basic}). Its $K$-band magnitude would
be 16.9 mag (0.116\,mJy). The expected flux distribution of the
secondary star is shown in the upper right-hand panel of
Fig.~\ref{fig:0154} (crosses and dotted line). For the Knigge et
al. component masses, the orbital velocity of the secondary star is
$\upsilon_2\!=\!458$\,\kms. For an assumed NEL amplitude of up to
$\sim\!100$\,\kms, interpreted by our irradiation model BR08
($K_2/K_2'\!=\!1.21$), the inclination does not exceed 15\degr. For a
WD of 0.75 \,\msun, the secondary star is moderately bloated with
$f_3\!=\!1.12$.  As long as the inclination cannot be tightly
constrained, similar models can be constructed with primary masses
from $M_1\!=\!0.50$ up to the Chandrasekhar mass and inclinations
between 20\degr\ and 12\degr. A promising path for progress involves a
spectroscopic measurement of the WD radius, and thereby its mass, when
\jone\ lapses into a low state or a measurement of the inclination by
the identification of the NEL.

The RASS soft and hard X-ray fluxes of Table~\ref{tab:xray} with the
geometry factors of Sect.~\ref{sec:general} give a high-state
bolometric X-ray luminosity of $2.0\,\times\!10^{32}$\,\ergs\ and an
X-ray based accretion rate of $\dot
M_\mathrm{x}\!=\!2.3\,\times\!10^{-11}$\,\msunyr\ for an adopted WD
mass of 0.75\,\msun\ (Table~\ref{tab:xray}). The optical level of 1
September 2015 probably represents a high state as well, and we estimate that
about $5.0\times\!10^{-12}$\,\ergsa\ arises from cyclotron
radiation. When it is included in the energy balance, the accretion
rate rises to $\dot M_\mathrm{x+cyc}\!=\!2.9\,\times\!10^{-11}$\,\msunyr.
If this rate equals the long-term mean, the expected WD temperature
due to compressional heating would be 10400\,K, near the lower end of
the observed temperature range \citep{townsleygaensicke09}. Its
4600\AA\ flux would be 0.040\,mJy, close to the observed GALEX
fluxes. In the 1995 HRI observation, the accretion rate could have
reached $\dot M_\mathrm{x+cyc}\!\sim\!5\,\times\!10^{-11}$\,\msunyr.

\begin{figure*}[t]
\includegraphics[height=89.0mm,angle=270,clip]{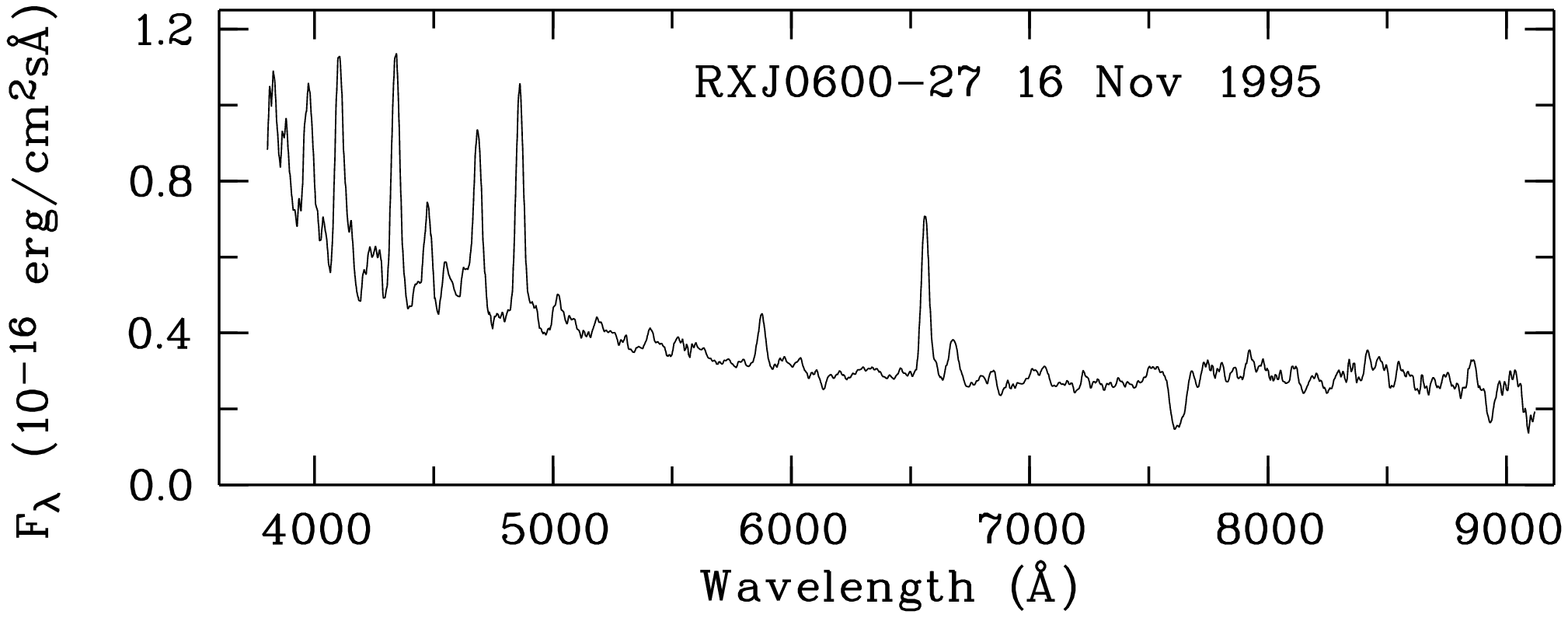}
\hfill
\includegraphics[height=89.0mm,angle=270,clip]{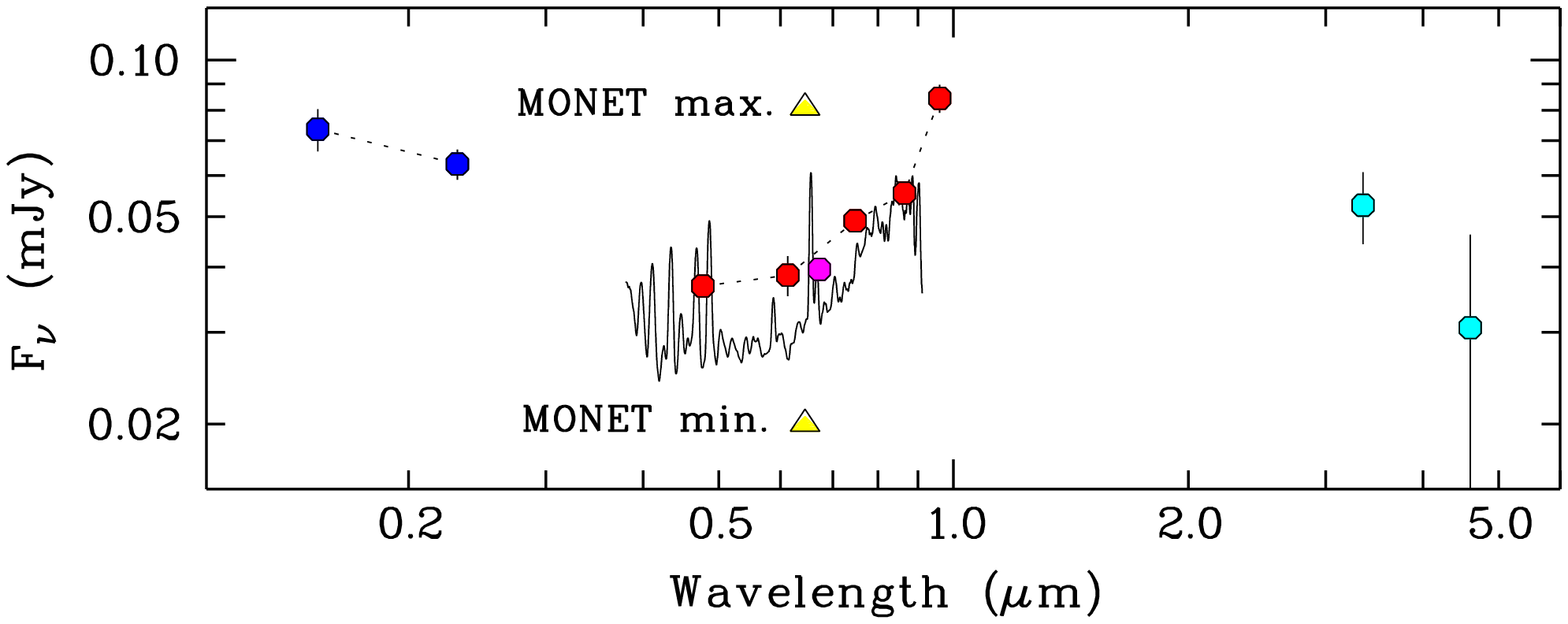}

\medskip
\includegraphics[height=89.0mm,angle=270,clip]{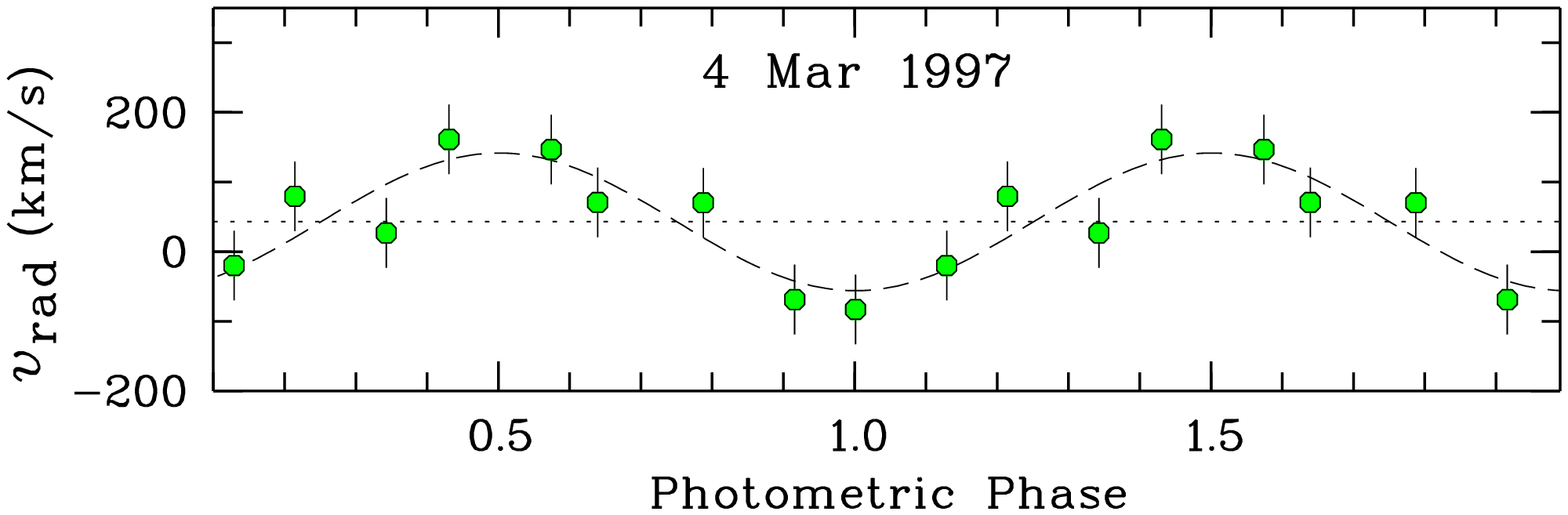}
\hfill
\includegraphics[height=89.0mm,angle=270,clip]{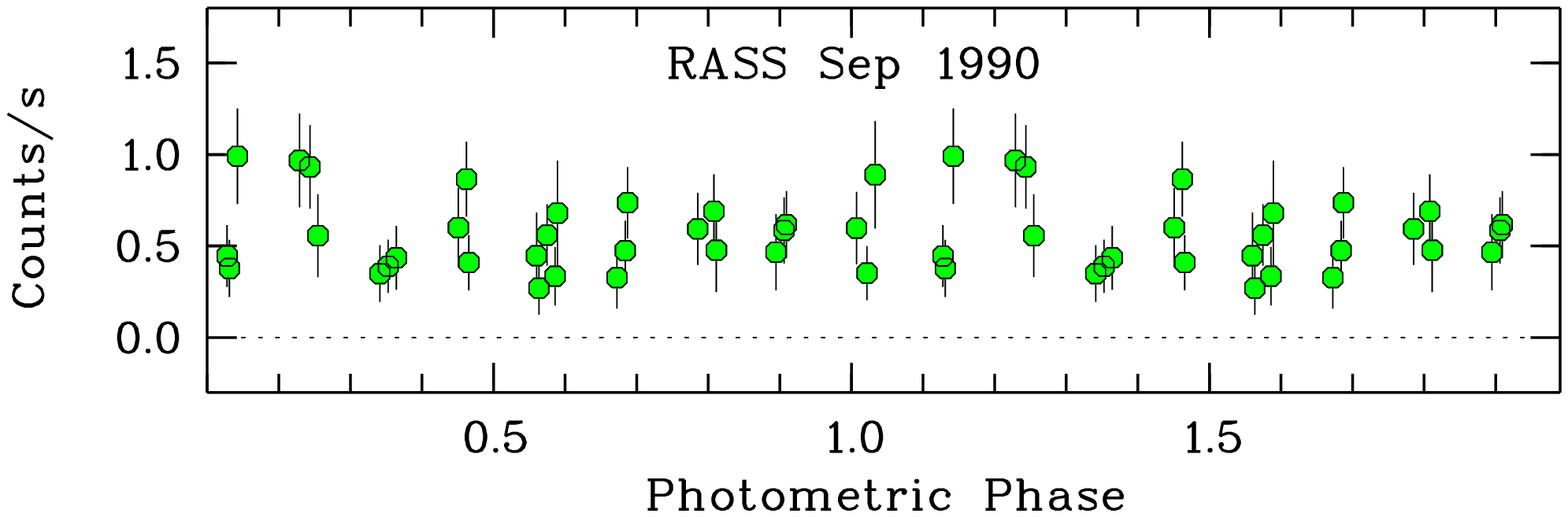}

\medskip
\includegraphics[height=89.0mm,angle=270,clip]{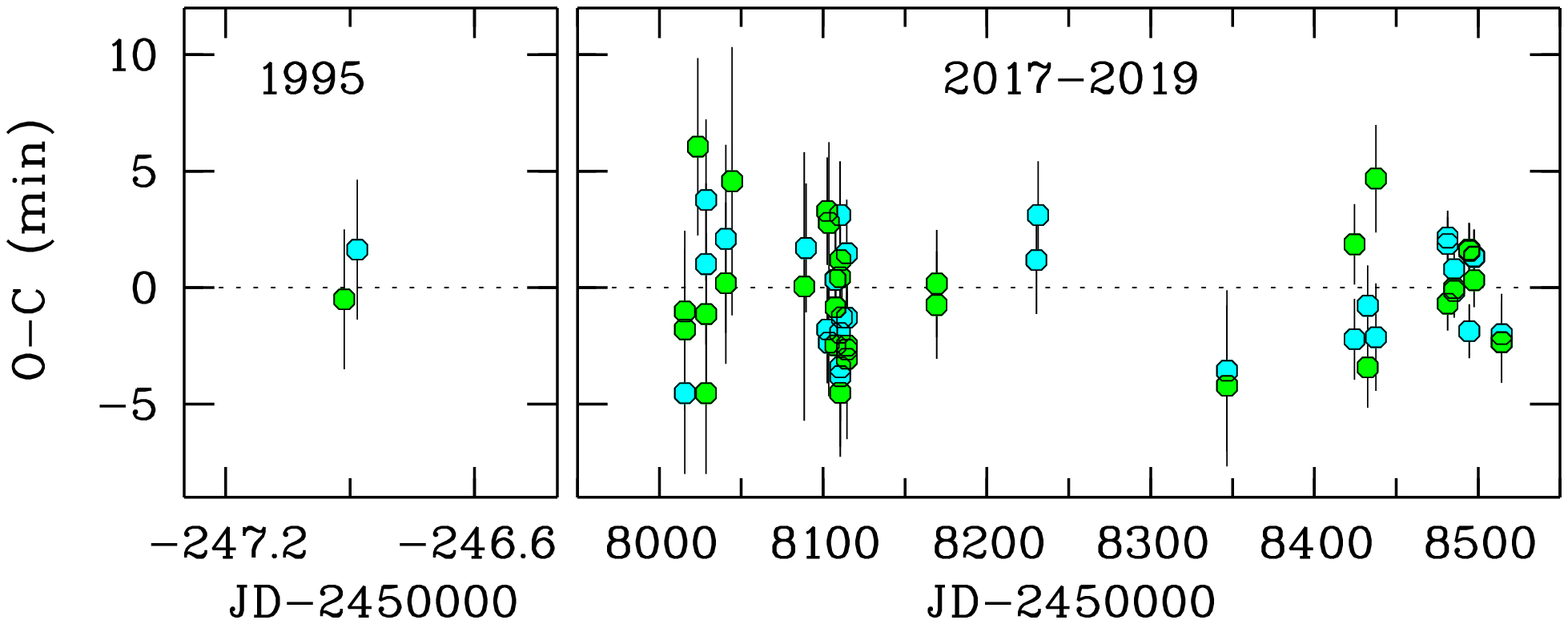}
\hfill
\includegraphics[height=89.0mm,angle=270,clip]{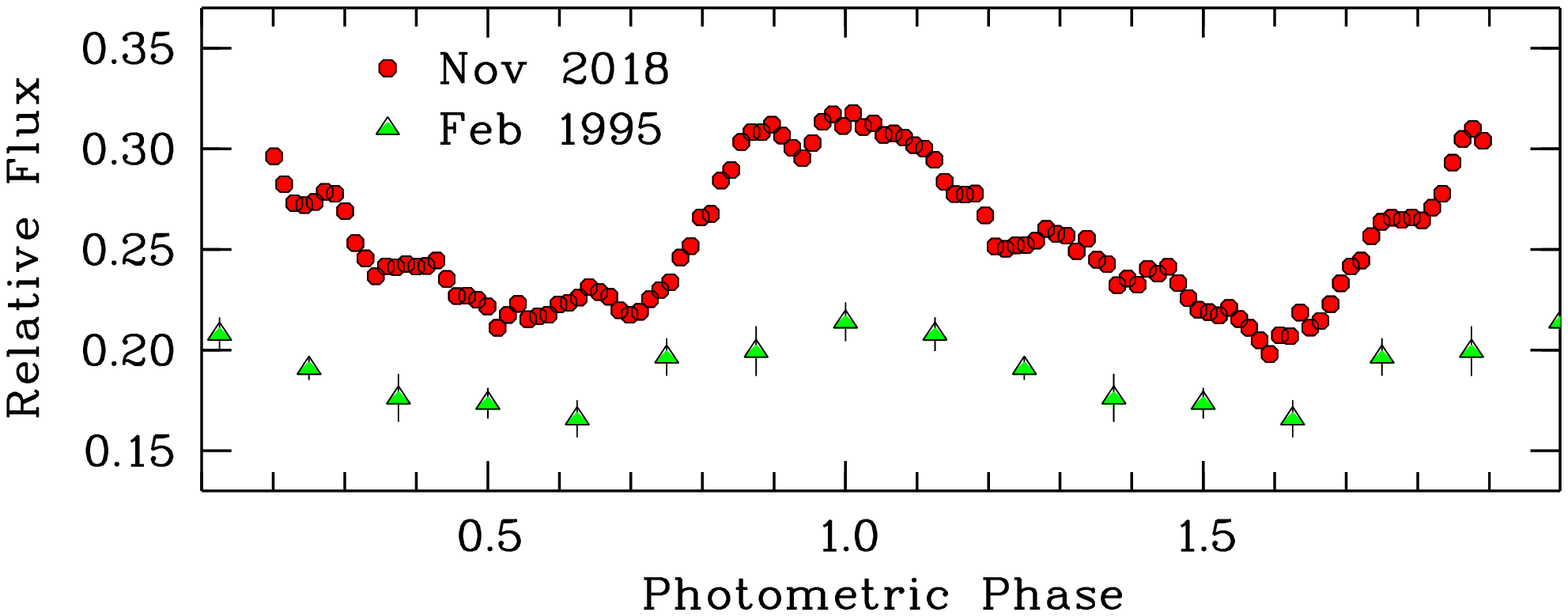}
\caption[chart]{\rxjsix. \emph{Left, top: } Identification spectrum
  taken on 16 November 1995. \emph{Left, center:} Balmer line radial-velocity curve from phase-resolved spectroscopy on 4 March
  1997. \emph{Left, bottom:} $O\!-\!C$ diagram for orbital maxima
  (green) and minima (cyan). \emph{Right, top: } Spectral energy
  distribution, showing the identification spectrum, and a summary of
  nonsimultaneous photometry (see text). \emph{Right, center:} X-ray
  light curve taken in the RASS between 10 and 13 September 1990.
  \emph{Right, bottom:} Light curves taken in WL on 5~February 1995 and 28
  December 2018. \mbox{The photometric phase is from Eq.~\ref{eq:0600ephem}.}}
\label{fig:0600}

\end{figure*}

\section{\rxjsix\ (= \jsix) in Lepus}
\label{sec:0600}

\jsix\ was discovered 1990 in the RASS as a bright and very soft X-ray
source. The most distant and optically faintest object in our sample
(Table~\ref{tab:basic}) was spectroscopically identified by us as a
polar and is listed as such in \citet{beuermannetal99}. Its orbital
period is close to the bounce period of CV evolution and
its secondary is therefore close to substellar.

\subsection{X-ray observations}

In the RASS, \jsix\ was detected with a mean PSPC count rate of
$0.52\!\pm\!0.03$\,\cps\ and a hardness ratio of $HR1\!=\!-0.92\!\pm\!0.03$ 
(Table\,\ref{tab:basic}). The rather long exposure time of 568\,s in
35 satellite visits allowed the construction of an orbital light
curve, which shows no evidence for a periodicity. The center right
panel of Fig.~\ref{fig:0600} shows the light curve folded over the
orbital period of Eq.~\ref{eq:0600ephem}. Judged by the X-ray
luminosity (Table~\ref{tab:xray}), \jsix\ was in a high state during
the RASS, and the lack of orbital modulation suggests that the
accretion spot was permanently in view. In a 13.5\,ks ROSAT HRI
observation between 12 and 28 September 1995, it was found in a low state
with a count rate of 0.0023 cts/s or a soft X-ray flux roughly a
factor of 30 below that of the RASS.

The RASS PSPC spectrum (not shown) is dominated by soft X-rays and
appears only moderately absorbed despite the large distance of
\jsix. No spectrally resolved follow-up X-ray observation is
available.  The unconstrained blackbody fit to the RASS spectrum
prefers an unrealistic \nh$\,\simeq\!0$ and
k$T_\mathrm{bb}\!\simeq\!75$\,eV with large errors
\citep[2RXS,][]{bolleretal16}. At the position of \jsix, the total
extinction is \ebv$\,=\!0.0215$ \citep{schlaflyfinkbeiner11} and the
\mbox{total} neutral hydrogen column density is
\nh$\,=\!2.2\!\times\!10^{20}$\,\atoms\ \citep{hi4pi16}. The
extinction in front of \jsix\ is \ebv\,$=\!0.016\!\pm\!0.006$
\citep{lallementetal18} and \nh$\,=\!(1.5\!\pm\!0.5)\!\times\!10^{20}$
\mbox{\atoms} \citep{nguyenetal18}. Adopting this value of \nh, the
PSPC fit with the X-ray spectral model of Sect.~\ref{sec:31} yields
\tbb$\,=\!42$\,eV and the X-ray fluxes and the accretion rate listed
in Table~\ref{tab:xray}. The Gaia distance for \jsix\
(Table~\ref{tab:basic}) has a large error, and the luminosity and
accretion rate are quoted for the 90\% confidence lower limit of
813\,pc and marked as lower limits.

\subsection{Spectroscopy, photometry, and ephemeris}

The optical counterpart of \jsix\ was identified as a 19--20 mag
periodically variable star in 4.7~h of WL photometry taken on 5 February
1995 with the ESO-Dutch 90 cm telescope. Its brightness was measured
relative to a comparison star C1 that has AB magnitude $r\!=\!17.89$
and is located at RA(2000)\,=\,$06^\mathrm{h}00^\mathrm{m}36\farcs2$,
DEC(2000)\,= $-27\degr09\arcmin27\arcsec$, 40\arcsec\,E and
8\arcsec\,S of the target (Fig.~\ref{fig:fc0600}).  The best period
of the quasi-sinusoidal variation was 0\,\fd\,0545(7)
(Fig.~\ref{fig:0600}, bottom right panel, green triangles). A
low-resolution optical spectrum taken on 16 November 1995 with the ESO/MPI
2.2 m telescope at La Silla, Chile, revealed \jsix\ as a polar (top
left panel). Strong \heii\ emission with an equivalent-width ratio
$W$(\heii)/$W$(\hbet)$\,=\!0.86$ (Table~\ref{tab:w}) indicated a high
state of accretion. Further nine low-resolution spectra of 15 min
exposure were taken on 4~March 1997, when the source was at a similar
brightness level of $19\!-\!20$\,mag. The emission-line radial
velocities were measured by cross-correlating the individual spectra
with the mean spectrum on a log\,$\lambda$ scale. The line profiles
display the asymmetries characteristic of polars, but are not
sufficiently well resolved to allow the separation of individual line
components. The radial velocities in the left center panel of
Fig.~\ref{fig:0600} yielded a period of $0\,\fd\,0535(35)$, revealing
the photometric period as the orbital one and placing the star near
the bounce period of CVs. The peak positive radial velocity occurs near
photometric minimum.  Except for the orbital modulation and the
day-to-day variability by factors of two each, the optical counterpart
to \jsix\ has shown little variability over the years.
No optical equivalent to the 1995 \mbox{X-ray} low state was found.
We derived a long-term ephemeris from the photometry of 5 February 1995
and WL photometry of 25 nights between September 2017 and January 2019 that
resulted in 31 additional maximum and minimum times each.  All times
are reported in Table~\ref{tab:0600} in Appendix~\ref{sec:C}. We fit the data
by a linear ephemeris for the orbital maxima,
\begin{equation}
T_\mathrm{max}\!=\!\mathrm{TDB}~2458015.55336(17) + 0.054644792(10)\,E,~~~
\label{eq:0600ephem}
\end{equation}
allowing for a shift of the minima relative to the maxima. The skewed
light curves have their minima on average at $\phi\!=\!0.55$.
There is no evidence for a deviation from linearity. Because of the
22-year hiatus between the two groups of timings, we cannot entirely
exclude the nearest alias period on each side, which corresponds to an
uncertainty of one in 151,194 cycles. These periods are separated by
$+29.7$\,ms and $-29.7$\,ms from the period of Eq.~\ref{eq:0600ephem},
reaching a reduced $\chi^2_{\nu}\!=\!1.4$ and 2.0, respectively,
relative to unity for the best period. 
The preliminary period $P_\mathrm{orb}\!=\!0\,\fd0757$ cited in the
catalog of \citet[][final version 7.24 of 2016,]{ritterkolb03} turned
out to be incorrect.

\subsection{Spectral energy distribution}

The top right panel in Fig.~\ref{fig:0600} combines all available data
into an SED that includes the low-resolution spectrum of the top left
panel. The two yellow triangles indicate the full range of the MONET/S
WL photometry, the blue dots the GALEX UV fluxes, the red dots fluxes
from the Pan-STARRS DR1, and the cyan-blue ones the WISE W1 and W2
fluxes, all obtained via the VizieR photometry viewer.
\citet{haakonsenrutledge09} misidentified the 2MASS image of a bright
star 26\arcsec\ WNW with \jsix\ (Fig.\ref{fig:fc0600}). The WISE
images of the region show several faint sources near the target
position\footnote{https://irsa.ipac.caltech.edu/applications/wise}, of
which the brightest is located 7\arcsec\
SW. \citet{harrisoncampbell15} may have mistaken this 15 mag object
for \jsix. The 2MASS
images\footnote{https://irsa.ipac.caltech.edu/applications/2MASS/IM/interactive.html}
show a faint equivalent to the WISE object, but no source at the
position of \jsix.  The 3$\sigma$ upper limits are 0.15, 0.22, and
0.28\,mJy in the $J,H$, and $K_\mathrm{s}$ bands, respectively, still
permitting a broad hump that extends over the entire near-IR
band. The hump looks suspiciously like the SED of a late M-dwarf, but
the secondary star would be much fainter at the Gaia distance.  A
possible interpretation involves optically thick cyclotron emission in
a magnetic field of $B\,\la\!20$\,MG.  The identification of the GALEX
source with \jsix\ seems trustworthy because of the close positional
coincidence. The origin of the UV emission remains uncertain.
Phase-resolved observations in the different wavelength bands could
resolve the open questions.

\subsection{System parameters}

Of all polars, only CV~Hyi and V4738~Sgr \citep{burwitzetal97} have
orbital periods shorter than \jsix. With less than 79\,min, all three
binaries fall below the bounce period 
$P_\mathrm{bounce}\!=\!81.8\pm0.9$\,min calculated by
\citet{kniggeetal11} for the evolution of the bulk of the (mostly
nonmagnetic) CVs. A lower value would be obtained for a reduced
braking efficiency, as may be appropriate for polars.  The
secondary mass at $P_\mathrm{bounce}$ is about 0.06\,\msun\
\citep{kniggeetal11}, and as a given system evolves through this point,
the accretion luminosity drops rapidly. The high observed X-ray
luminosity of \jsix\ suggests that it is still approaching the
minimum.  The secondary is then expected to be a very late star or a
brown dwarf. For example, a Roche-lobe-filling secondary of
$M_2\!\simeq\!0.07$\,\msun\ with $f_3\!=\!1.15$ would have
$R_2\!\simeq\!0.11$\,\rsun, a spectral type of dM8, and an $i$-band
surface brightness $S_\mathrm{i}\!\simeq\!10$. For the 90\% confidence
lower limit to the distance of 813\,pc, it would have
$i\!\simeq\!24.3$ ($0.7\,\mu $Jy), and $K\!=\!19.6$ ($100\,\mu $Jy),
which is far below the observed fluxes.

The narrow emission-line component could not be identified in \jsix.
When $K_2'$ would tentatively be equated to the observed velocity amplitude of
99\,\kms , this would imply an inclination of 14\degr\ for
$M_1\!=\!0.75$\msun.  Any primary mass between 0.5\,\msun\ and the
Chandrasekhar limit can be accommodated.

The bolometric fluxes and luminosities of the X-ray components are
listed in Table~\ref{tab:xray}. When the cyclotron flux is included in the
$\dot M$ calculation, the
accretion rate rises for an 0.75\,\msun\ WD only minimally to $\dot
M_\mathrm{x+cyc}\!\ga\!9.7\!\times\!10^{-11}$\,\msunyr,
placing the star at the upper end of the range of accretion rates
found in short-period polars \citep{townsleygaensicke09}.  The
equivalent equilibrium temperature of the WD would be 14100\,K,
implying a 4600\,\AA\ flux of the compressionally heated WD of
0.014\,mJy, somewhat below the observed minimum MONET WL
flux. Measuring the WD temperature and radius spectroscopically in a
low state appears feasible.

\begin{table}[b]
\begin{flushleft}
  \caption{Equivalent widths of prominent emission lines in \AA\ for
    the high-state spectra of Figs. 1 to 5. The letters b and f refer
    to the bright and faint orbital phase intervals, respectively.}

\medskip
\begin{tabular}{@{\hspace{1mm}}l@{\hspace{1.5mm}}r@{\hspace{1.5mm}}r@{\hspace{1.5mm}}r@{\hspace{1.5mm}}r@{\hspace{1.5mm}}r@{\hspace{1.5mm}}r@{\hspace{1.5mm}}r@{\hspace{1.5mm}}r@{\hspace{1.5mm}}c}\\[-5ex]
\hline\hline \\[-1.5ex]
Name    &  Date~~~~~ & &\hdel ~~&\hgam~~ &  HeII &\hbet ~~& HeII & HeI~  &\halp~~\\        
        &            & &        &        &  4686 &        & 5411 & 5876 &        \\        
\hline\\[-1ex]                                                                             
0154-59 &  23 Aug 93 & &  103.1 &   86.5 &  81.6 &  103.9 & 11.9 & 17.2 &   87.8 \\        
        &  17 Dec 93 & &  111.6 &  102.5 &  78.5 &   99.6 & 15.1 & 20.6 &   89.4 \\        
        &  24 Nov 95 & &  105.2 &  100.5 &  89.7 &  113.7 & 12.7 & 23.5 &  103.7 \\[0.5ex] 
0600-27 &  16 Nov 95 & &  45.7  &   57.8 &  47.7 &   55.6 &  6.7 & 17.7 &   62.3 \\[0.5ex] 
0859+05 &  10 Jan 03 & &  27.1  &   23.3 &  20.2 &   33.9 &  4.3 &  9.5 &   33.2 \\[0.5ex] 
0953+05 & 6/7 Feb 95 &b&  11.2  &   11.4 &   3.4 &   13.0 &  1.0 &  3.8 &   12.7 \\[0.5ex] 
        &            &f&  48.1  &   50.3 &  20.3 &   69.4 &  2.9 & 31.2 &   80.0 \\        
1002-19 &  24 Dec 92 & &  12.8  &    9.0 &  15.9 &   17.0 &  3.4 &  2.8 &   25.1 \\        
        &   1 Mar 97 &b&  11.9  &   17.7 &  15.7 &   25.5 &  3.3 &  9.0 &   35.8 \\        
        &            &f&  32.6  &   47.9 &  50.7 &   47.9 &  9.9 & 13.0 &   68.4 \\[1.0ex] 
\hline\\[-1ex]                                                                             
\end{tabular}\\[-1.0ex]
\label{tab:w}
\end{flushleft}

\end{table}

\begin{figure*}[t]
\includegraphics[height=88.0mm,angle=270,clip]{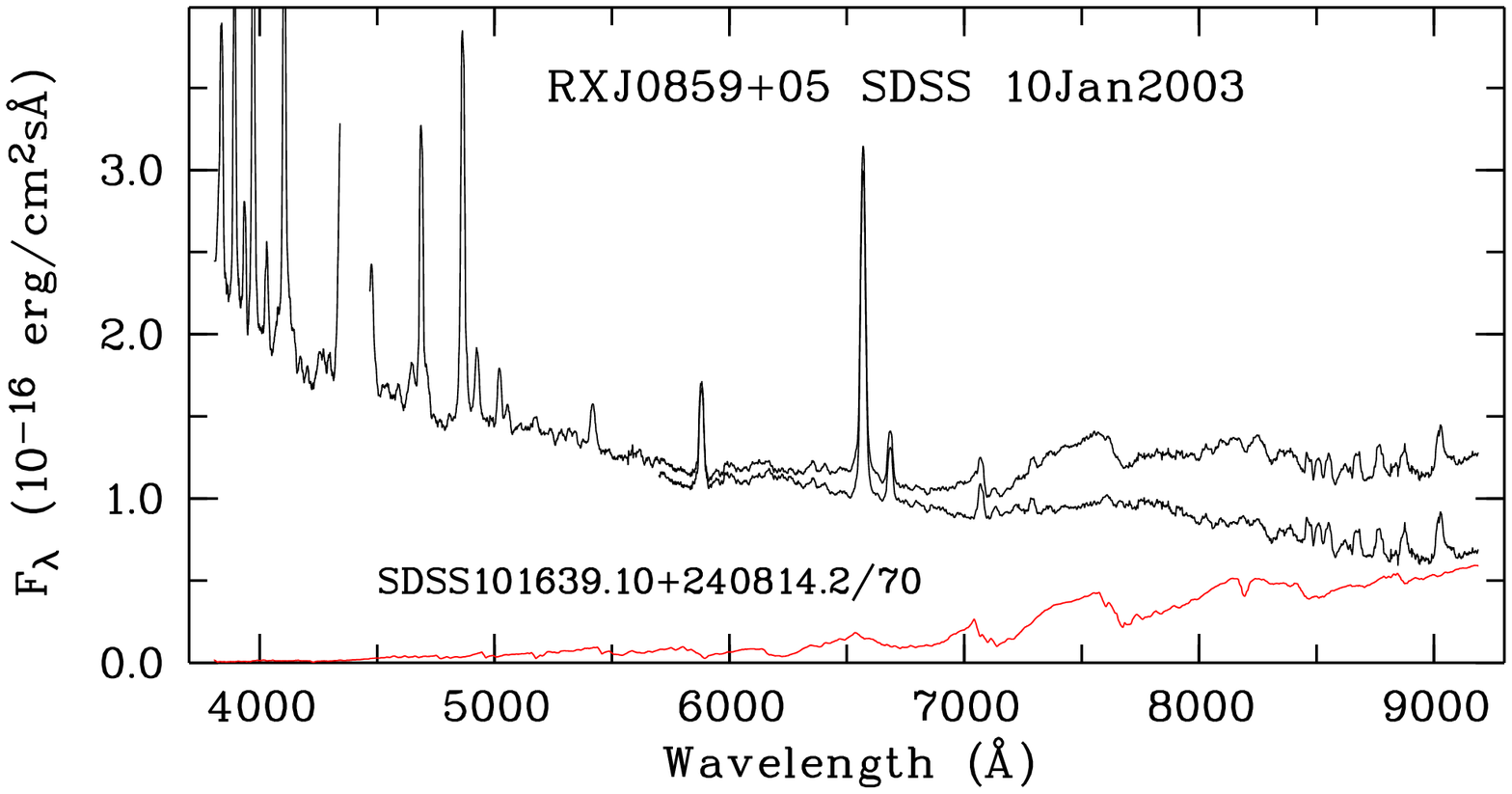}
\hfill
\includegraphics[height=89.0mm,angle=270,clip]{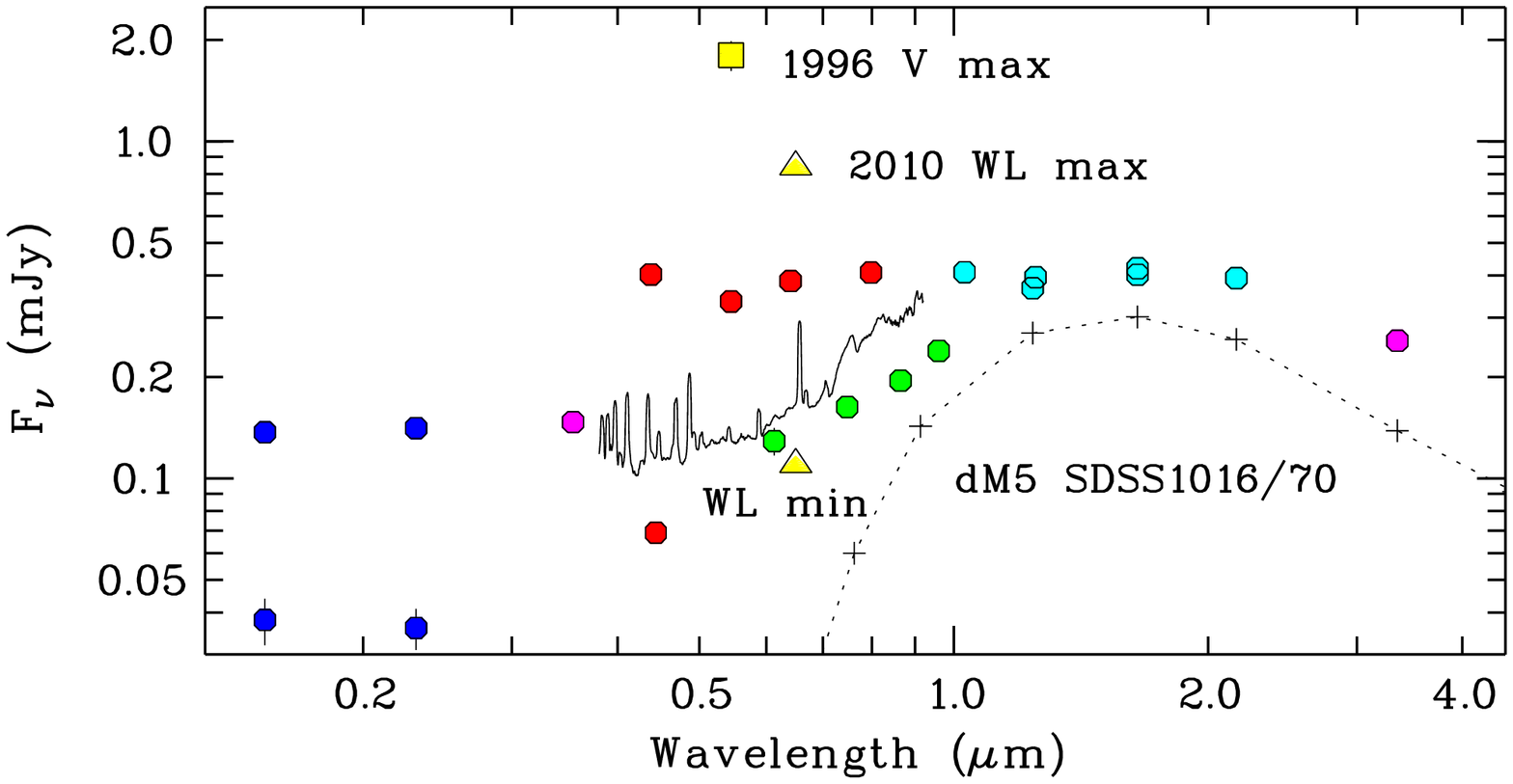}

\medskip
\begin{minipage}[t]{90mm}
\includegraphics[height=89.0mm,angle=270,clip]{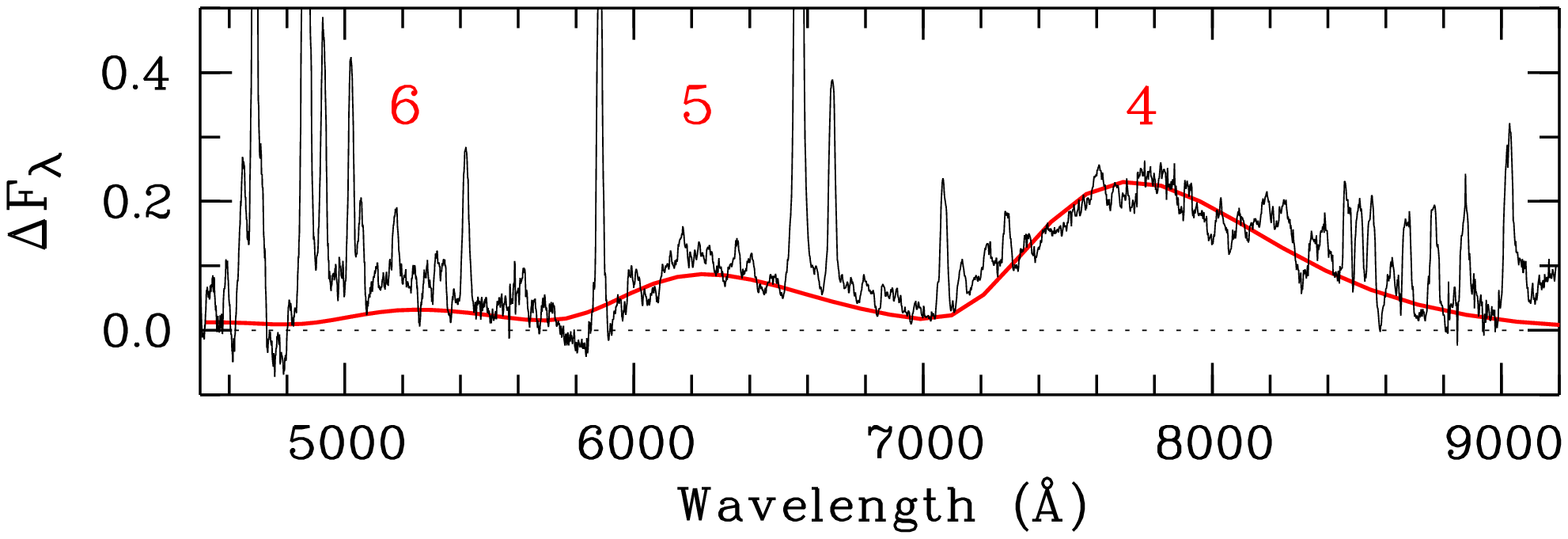}

\medskip
\includegraphics[height=89.0mm,angle=270,clip]{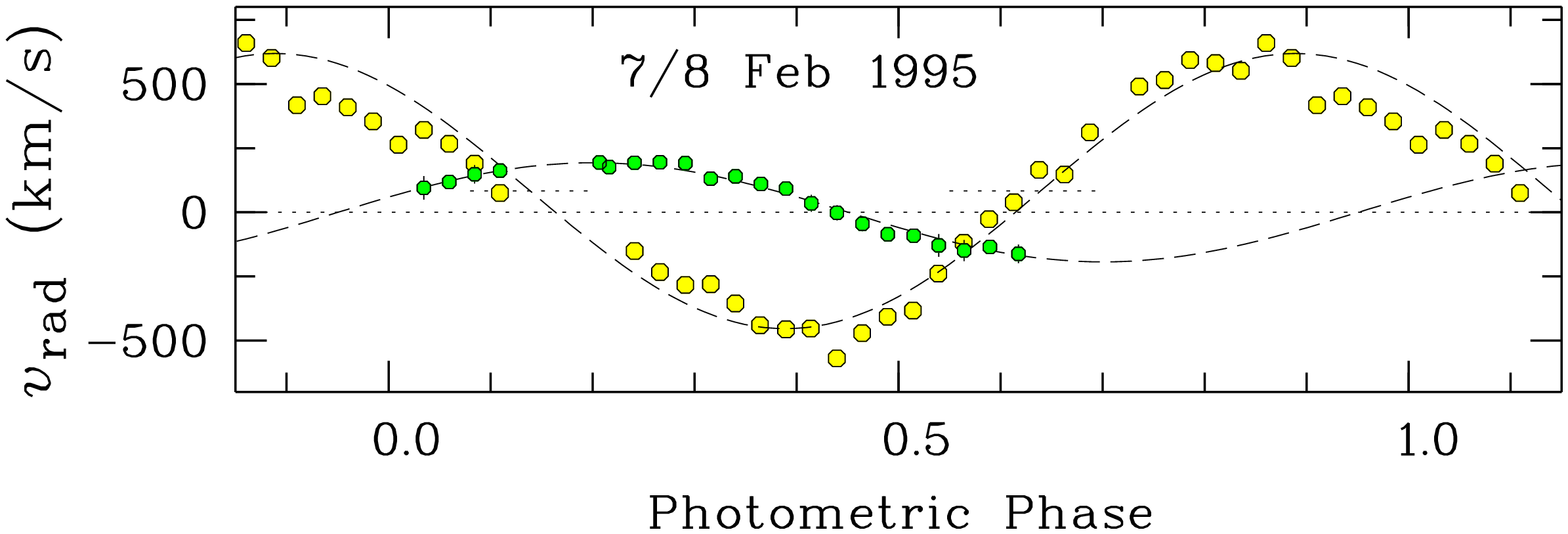}
\end{minipage}

\hspace{94mm}
\begin{minipage}[t]{90mm}
\vspace*{-63.0mm}
\includegraphics[height=89.0mm,angle=270,clip]{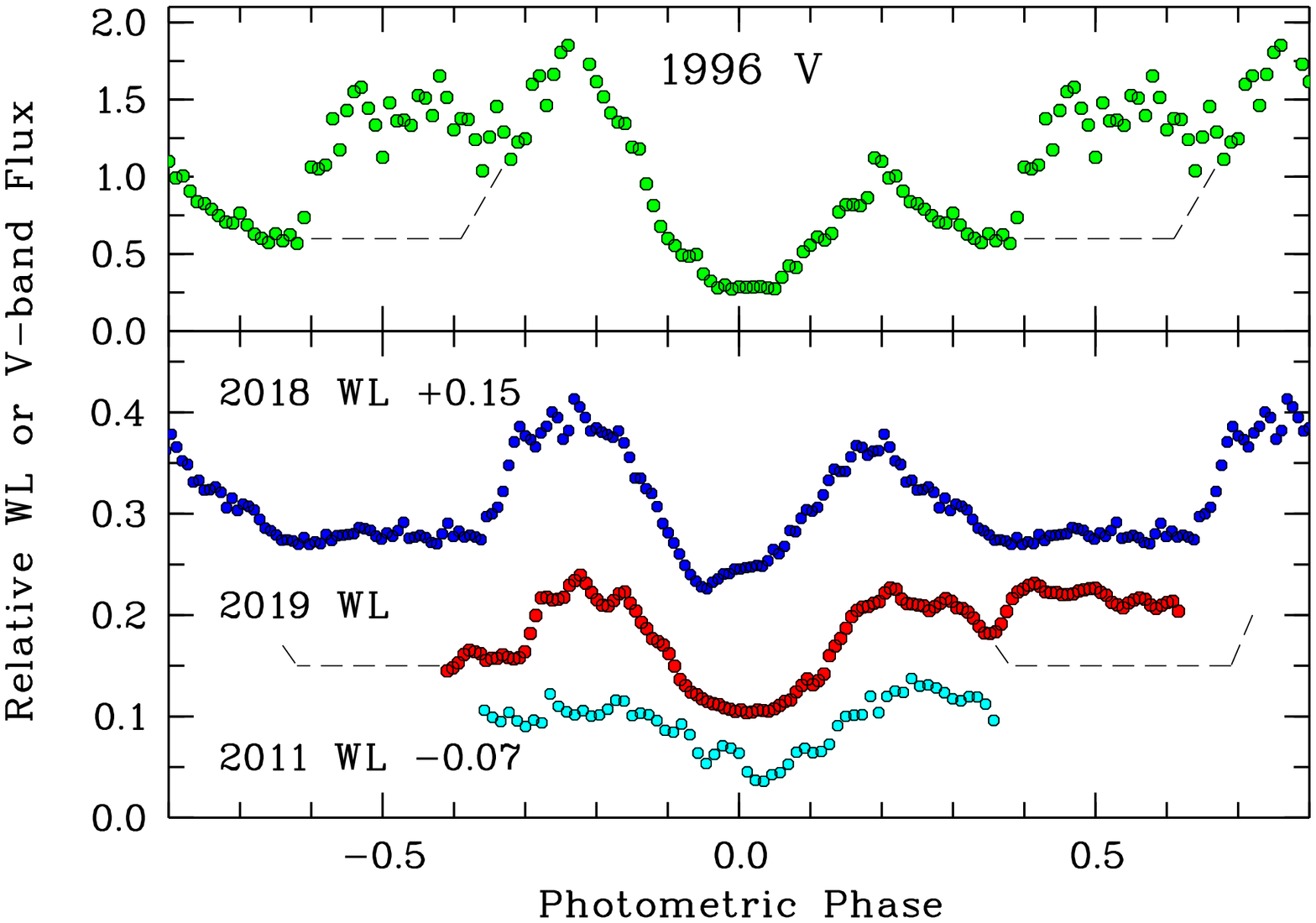}
\end{minipage}

\vspace{-1mm}
\includegraphics[height=89.0mm,angle=270,clip]{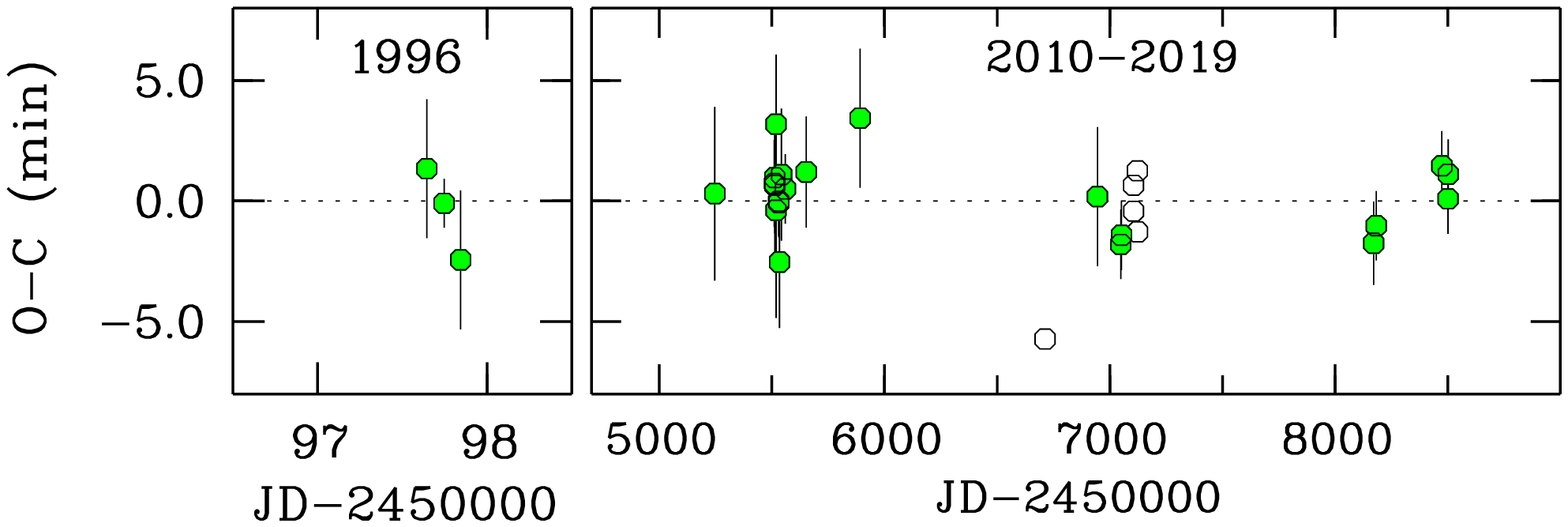}
\hfill
\includegraphics[height=89.0mm,angle=270,clip]{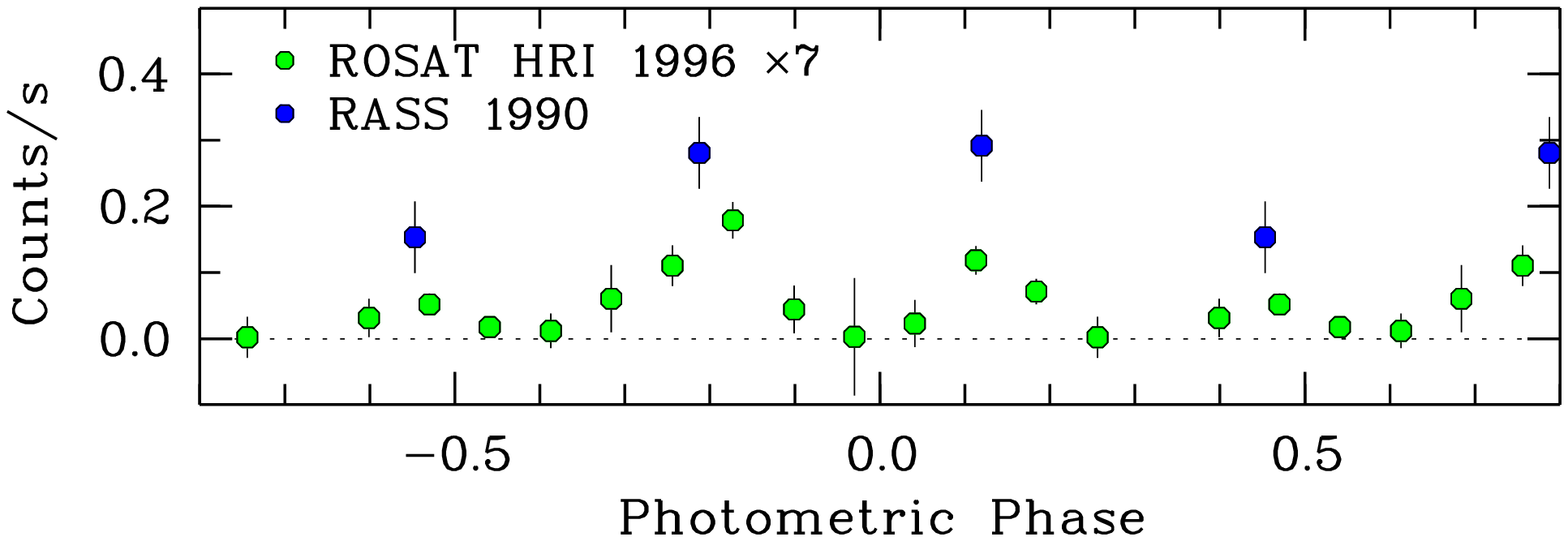}

\caption[chart]{\rxjeight. {\emph{Left, top:}} SDSS spectra of
  RXJ0859+05 (black curve) and the dM5 star SDSS\,J101639.10+240814.2
  divided by 70 to match the TiO structures in the polar spectrum
  (red). The difference spectrum (black) shows faint cyclotron
  lines. \emph{Second from top:} Model cyclotron spectrum (red) fit
  to the observed difference spectrum (black). \emph{Third from top:}
  Mean radial-velocity curves of the NEL (green) and the combined BBC
  and HCV components (yellow) of \hbet, \hgam, and \heii, measured
  from the high- and medium-resolution trailed spectra. \emph{Bottom: }
  $O\!-\!C$ diagram for the times of the primary minimum. Green dots
  show this work, and open circles show results from \citet{joshietal20}
  (see text). {\emph{Right, top:}} Overall spectral energy
  distribution, built from nonsimultaneous data (see
  text). \emph{Center panels:} V-band light curve of 15 January 1996 and
  WL light curves of on 4 March 2018, 17 January 2019, and 1 April 2011. The
  dashed lines schematically separate the emissions of the primary and
  the secondary pole. \emph{Bottom panel:} Rosat X-ray light curves
  taken 1990 with the PSPC and 1996 with the HRI. Phases are from
  Eq.~\ref{eq:0859ephem}.}
\label{fig:0859}
\end{figure*}

\section{\rxjeight\ (= \jeight) in Hydra}
\label{sec:0859}

\jeight\ was discovered 1990 in the RASS as a very soft X-ray source
and was spectroscopically identified by us as a polar. It is listed as such
in \citet{beuermannetal99}. Its orbital period of 143.9\,min, which places
it at the lower edge of the remnant period gap of polars, as defined
by \citet{bellonietal20} and \citet{schwopeetal20}.

\subsection{X-ray observations}

\jeight\ was detected in the RASS with a mean PSPC count rate of
$0.27\!\pm\!0.03$\,\cps and a hardness ratio
$HR1\,=\!-0.86\!\pm\!0.06$ (Table\,\ref{tab:xray}). Because the binary
period of 144 min (Table~\ref{tab:basic}) is commensurable with the 96
min ROSAT period, all satellite visits fall in three narrow orbital
phase intervals.  In a 32.8\,ks HRI observation from 24 April to 12 May
1996, the source was detected with 0.010\,HRI\,\cps, equivalent to
about 0.07\,PSPC\,\cps\ (Table A.1 in Appendix~\ref{sec:A}). This
observation provided nearly complete phase coverage. The lower
right-hand panel in Fig.~\ref{fig:0859} shows the two light curves,
with the HRI count rate multiplied by a factor of seven. The X-ray and
optical bright phases coincide and the central X-ray dip, near phase
zero on the ephemeris of Eq.~\ref{eq:0859ephem}, probably marks the
instance at which the line of sight to the WD passes through the
magnetically guided part of the accretion stream. We define the
soft X-ray flux in the bright phase by the two higher of
the three RASS points with a mean of 0.30\,PSPC\,\cps. This value is
entered into Col. (5) of Table~\ref{tab:xray} and is taken to
represent the high state of \jeight.  The HRI observation of 1996 has
a bright-phase count rate of 0.020\,HRI\,\cps, which converts into
possibly as much as 0.14\,PSPC\,\cps\ or half the RASS count rate,
representing an intermediate state.

The only X-ray spectral information available is that of the RASS.  A
blackbody fit yields $T_\mathrm{bb1}\!\simeq\!38$\,eV and
$N_\mathrm{H}\!\simeq\!2.8\!\times\!10^{20}$\,\atoms. The extinction
in front of \jeight\ is \ebv$\,\simeq\!0.033\!\pm\!0.008$
\citep{lallementetal18}, about 3/4 of the total galactic
extinction. The corresponding column density of cold matter is
\nh$\,=\!(3.1\!\pm\!0.8)\times\!10^{20}$ \atoms\
\citep{nguyenetal18}. The column densities from the RASS and from
\ebv\ are compatible. We fit the adjusted RASS spectrum (not shown)
with the X-ray spectral model described in Sect.~\ref{sec:31},
accepting the $N_\mathrm{H}$ value of the fit. The X-ray flux, the
luminosity, and the derived accretion rate in Table~\ref{tab:xray} are
as expected for a short-period polar \citep{townsleygaensicke09}.

\subsection{Orbital period and ephemeris}

The optical counterpart of \jeight\ was identified as a polar by a
low-resolution spectrum taken on 13 December 1993. We determined the
orbital period in 6.1\,h of continuous $V$-band photometry, using the
ESO-Dutch 90cm telescope at La Silla on 13 January 1996. The trail was
picked up using the MONET/N and MONET/S telescopes in 20 nights
between 2010 and 2019. Photometry was performed relative to the star
SDSS085908.57+053513.8, which is located 9\arcsec\ W and 101\arcsec\ S
of the target and has Sloan $r\!=\!15.83$ and $V\!\simeq\!16.17$. All
light curves possess a photometric primary minimum with a width of
about 20\,min. \jeight\ exhibits substantial variability, which is
illustrated by the light curves of 1996, 2011, 2018, and 2019 (center
right-hand panel of Fig.~\ref{fig:0859}).  The primary minimum results
from cyclotron beaming and marks the phase in which the line of sight
approaches the accretion funnel most directly. The emission from the
primary pole is visible for $\Delta \phi\!\simeq\!0.65$, indicating a
location in the upper (near) hemisphere of the WD. The dashed lines
added to the 1996 and 2019 light curves indicate the surmised emission
from the primary pole. Excess emission between $\phi\!=\!0.35$ and
0.65 likely originates from a second accretion spot in the far
hemisphere (compare the blue model light curve in
Fig.~\ref{fig:GK}). The system reached a peak brightness of
$V\!=\!15.50$ on 15~January 1996, when it was up to an order of magnitude
brighter than in later years. A linear fit to the times of 24 minima
in Table~\ref{tab:0859} in Appendix C gives the alias-free ephemeris
\begin{equation}
T_\mathrm{min}\!=\!\mathrm{BJD(TDB)}~2455246.83728(26) + 0.099951323(10)\,E.~~~
\label{eq:0859ephem}
\end{equation}
The $O\!-\!C$ diagram is displayed in the bottom left panel of
Fig.~\ref{fig:0859} (green dots). A different period was published by
\citet{joshietal20}. Their ephemeris is based on seven minimum times
added as open circles to our $O\!-\!C$ diagram, of which four timings
of 2015 agree perfectly with our data, while their three 2014 timings
are 6\,min and more than 10\,min early. Their published timings still
yield a most probable period that agrees with that of
Eq.~\ref{eq:0859ephem} within the errors, but their published period
is a less likely alias that involves a cycle count error of one orbit
over one year.

\subsection{Trailed spectra and SDSS spectrum}

We obtained trailed medium- and high-resolution optical spectra of
\jeight\ between 5 and 8 February 1995 with 6\,\AA\ and 1.6\,\AA\ FWHM
resolution, respectively, using the blue and red arms of the TWIN
spectrograph of the 3.5 m telescope on Calar Alto, Spain
(Table~\ref{tab:spec}). With exposure times of 60 min, these spectra
extend over a sizeable part of the CCD chip. As a consequence, the
Meinel OH bands in the red arm do not subtract well, complicating the
flux calibration. The Balmer and helium emission lines in the blue
spectra show well-defined narrow NEL and BBC+HCV components, with
examples shown as gray plots in Fig.~\ref{fig:vrad}. We measured their
radial velocities and show the mean velocities of \hbet, \hgam,\ and
\heii\ in Fig.~\ref{fig:0859} (third left-hand panel from the
top). The narrow component has a velocity amplitude
$K_2'\!=\!193\,\pm\!5$\,\kms. The blue-to-red zero crossing occurs at
photometric phase $\phi_\mathrm{br}\!=\!-0.06\,\pm\!0.01$ and defines
spectroscopic phase as $\phi_\mathrm{sp}\!=\!\phi_\mathrm{ph}+0.06$.
It has its zero point \emph{\textup{bona fide}} at inferior conjunction of the
secondary star and represents the true binary phase. The broad
(BBC+HVC) component has a velocity amplitude
$K_\mathrm{broad}\!\sim\!540$\,\kms\ with a $\gamma\!\sim\!83$\,\kms.
Maximum positive radial velocity is attained at photometric phase
$\phi_\mathrm{broad}\!\sim\!0.90$, roughly consistent with the
expected streaming motion of the BBC matter toward the primary pole.

Our mean blue and red spectra agree reasonably well with the better
calibrated SDSS spectrum of \jeight\ of 10 January 2003, which we display
instead in the top left panel of Fig.~\ref{fig:0859}. It features
similar equivalent widths as our mean trailed spectrum, with $W$
around 30\AA\ for the Balmer lines and \heii\ and
$W$(\heii)/$W$(\hbet)$\,\simeq\!0.60$ (Table~\ref{tab:w}).
We also show in Fig.~\ref{fig:0859} the SDSS spectrum of the dM5 star
SDSS\,J101639.10+240814.2, adjusted by a factor of $70\,\pm\!10$ to
fit the strength of the TiO bands of \jeight. It defines the $i$-band
magnitude and flux of the secondary star as
\mbox{$i\!=\!19.45\!\pm\!0.15$} and $0.060\,\pm\!0.009$\,mJy,
respectively. Subtracting the adjusted dM5 spectrum reveals waves in
the difference spectrum that we interpret as cyclotron harmonics. The
second left-hand panel from the top in Fig.~\ref{fig:0859} shows the
cyclotron line spectrum obtained by subtracting, in addition, a smooth
representation of the summed continua of the WD, the stream, and the
cyclotron component. The cyclotron line model (red curve) was
calculated with the theory of \citet{chanmugamdulk81} for a field
strength of 36~MG, an angle $\theta\!=\!70$\degr\ between the line of
sight and the field, a plasma temperature of 10\,keV, and a thickness
parameter log\,$\Lambda\!=\!2$. The fit identifies the observed humps
as the emission in the $\text{fourth}^\mathrm{}$ to $\text{sixth}^\mathrm{}$ cyclotron
harmonic. A dip centered at 5800\,\AA\  could be the \halp\
$\sigma^{-}$ Zeeman absorption trough in a field of 34\,MG, which may
correspond to the mean field in an accretion halo. Spectral structure
in the continuum shortward of 5000\,\AA\ is likely due to the Zeeman
absorption components of the higher Balmer lines.  We do not confirm
the occurrence of cyclotron emission lines in this part of the
spectrum suggested by \citet{joshietal20}.

\subsection{Spectral energy distribution}

The top right-hand panel of Fig.~\ref{fig:0859} shows the overall SED
of \jeight. It includes the SDSS spectrum (black curve) and the SDSS
$u$-band flux (magenta dot).  The Pan-STARRS data points (green), the
NOMAD and PPXML data (red), and the GALEX data of two epochs (blue)
indicate the variability of the system, as do the two yellow triangles
that describe the full range of the MONET WL measurements. The 1996
brightening (open square, peak value) may represent a rare event. An
optical flux level of $\sim\!0.4$\,mJy that possibly extends into the
near-UV seems to represent a mean high state. For
$B\!=\!36$\,MG, a plasma temperature of 10\,keV, a large viewing
angle, and a thickness parameter log$\Lambda\,\ga\,6$, the theory
predicts an optically thick cyclotron spectrum that extends into the
near-UV. At the lower end of the flux scale, the SED of the
dM5 star adjusted in the $i$ band is shown by the dotted curve. A dM4
star adjusted in the same way would have slightly lower IR
fluxes.

\subsection{System parameters}

For a CV with $P_\mathrm{orb}\!=\!144$\,min, the evolutionary model of
\citet{kniggeetal11} assumes that it entered the period gap from
longer orbital periods, causing its bloated secondary star to return
to thermal equilibrium ($f_3\!=\!1.0$) and mass transfer to cease. In
their model, $M_2\!=\!0.20$\,\msun, $R_2\!=\!0.223$\,\rsun, and the
spectral type is dM4.0. The $i$-band surface brightness is
$S_\mathrm{i}\!=\!7.6$ for a spectral type dM4 and 8.2 for dM5. At the
Gaia distance of 437\,pc, the expected magnitude is $i\!=\!19.1$ for a
dM4 star and 19.7 for dM5, compatible with the measured magnitude of
19.45 quoted above.  Our dynamical model permits any primary mass
between 0.50\,\msun\ and the mass limit with inclinations between
75\degr\ and 30\degr\ and either an unbloated secondary star of
0.21\,\msun\ ($f_3\!=\!1.0$) or a moderately bloated one with
0.17\,\msun\ ($f_3\!=\!1.10$). We can limit the inclination using the
duration of the self-eclipse of the accretion spot, $\Delta
\phi\,\simeq\,0.30$, in Eq.~\ref{eq:visi} of Sect.~\ref{sec:NEL}. The
deep primary minimum suggests that it is preferentially shaped by
cyclotron beaming, which leads to $i\!\sim\!50\,-\,60\,$\degr , and
with the measured $K_2'$, to $M_\mathrm{1}\!=\!0.56-0.72$\,\msun.  If
the wide X-ray dip at $\phi\!=\!0$ represents absorption in the
accretion stream, the lower inclinations and higher masses are
excluded, restricting the results to $i$ to $55\!-\!60\,$\degr\ and $M_1$ to
$0.56\!-\!0.66$\,\msun. The primary mass of 0.75\,\msun\ preferred by
\citet{kniggeetal11} in their evolutionary sequence requires
$i\!\simeq\!48$\degr. The derived inclination is consistent with the
visibility of the NEL for about half an orbit around superior
conjunction of the secondary star. There are ways to improve on $i$
and $M_1$. In addition to a more accurate measurement of the variation in
NEL, phase-dependent cyclotron spectroscopy or spectropolarimetry can
provide information on~$i$. Finally, a better X-ray light curve may
confirm or disprove the existence of the absorption dip and limit
$i$. Alternatively, it should be feasible to measure the temperature
and radius of the WD spectroscopically in a low state and thereby
infer its mass.

The accretion rate obtained from the X-ray fluxes in
Table~\ref{tab:xray} for $M_1\!=\!0.61$\,\msun\  and $\dot
M_\mathrm{x}\!=\!5.4\!\times\!10^{-11}$\,\msunyr \ is typical of
short-period polars, confirming that the RASS observation represented
a high state. We estimated the cyclotron flux in Table~\ref{tab:xray}
from the NOMAD and PPXML optical fluxes (red dots), which represent a
moderate high state as well. Including this component yields $\dot
M_\mathrm{x+cyc}\!=\!6.9\,\times\!10^{-11}$\,\msunyr. Interpreted as
the long-term mean accretion rate, the WD would have an equilibrium
temperature of 10500\,K and a spectral flux of 0.031\,mJy at
4600\,\AA, consistent with the lower pair of GALEX points in
Fig.~\ref{fig:0859} (upper right panel) representing the WD.

\begin{figure*}[t]
\includegraphics[height=89.0mm,angle=270,clip]{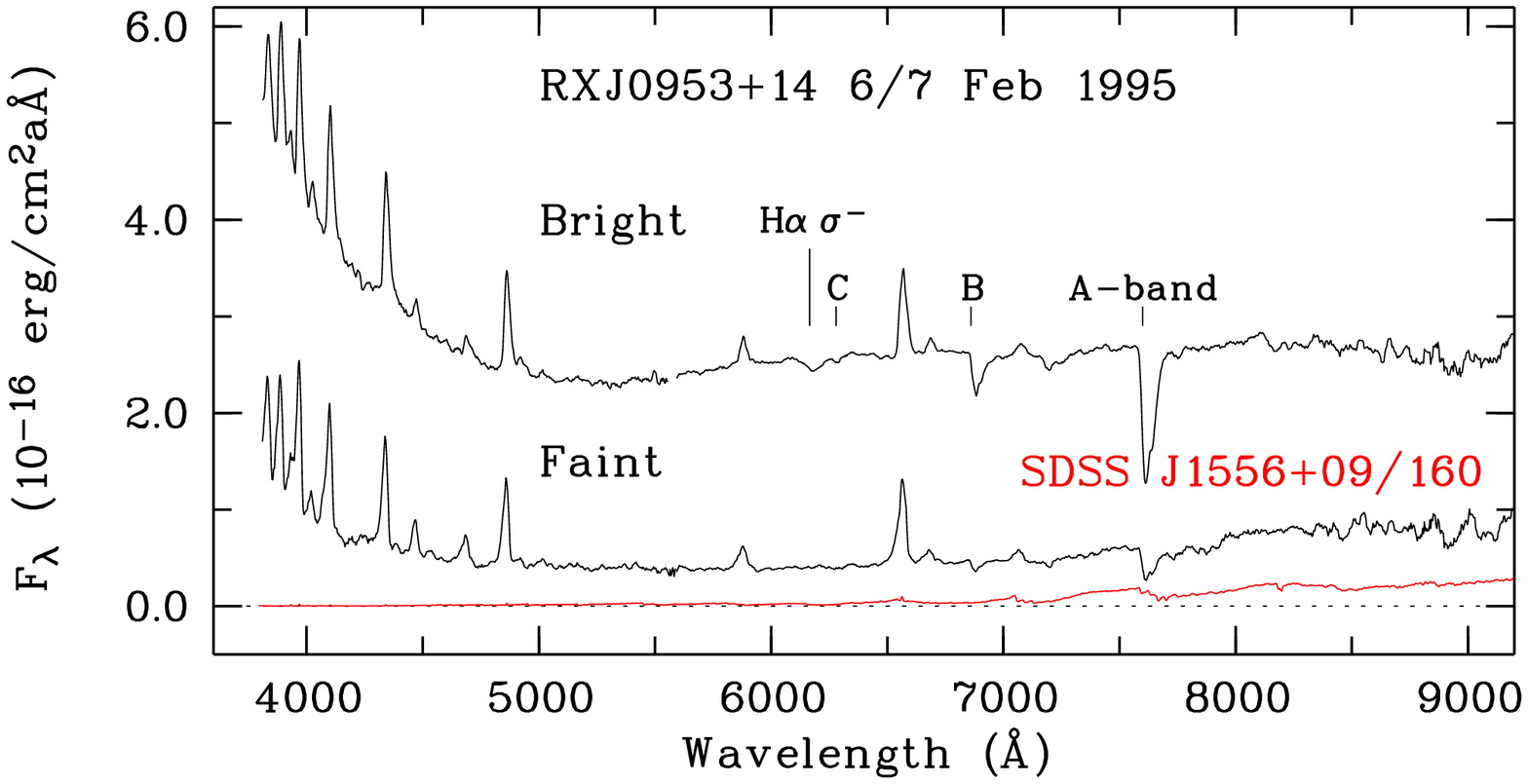}
\hfill
\includegraphics[height=89.0mm,angle=270,clip]{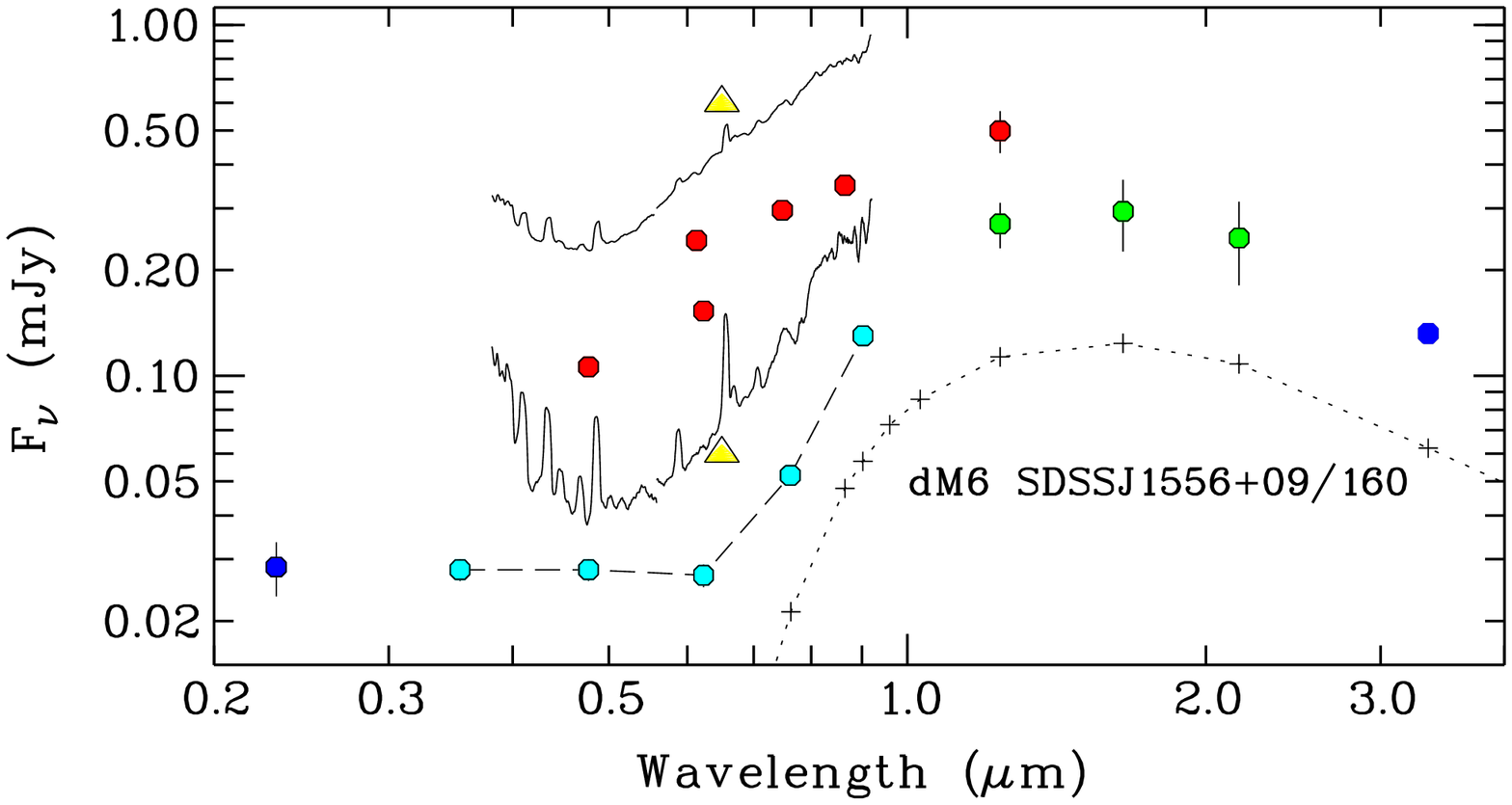}

\medskip
\begin{minipage}[t]{90mm}
\includegraphics[height=89.0mm,angle=270,clip]{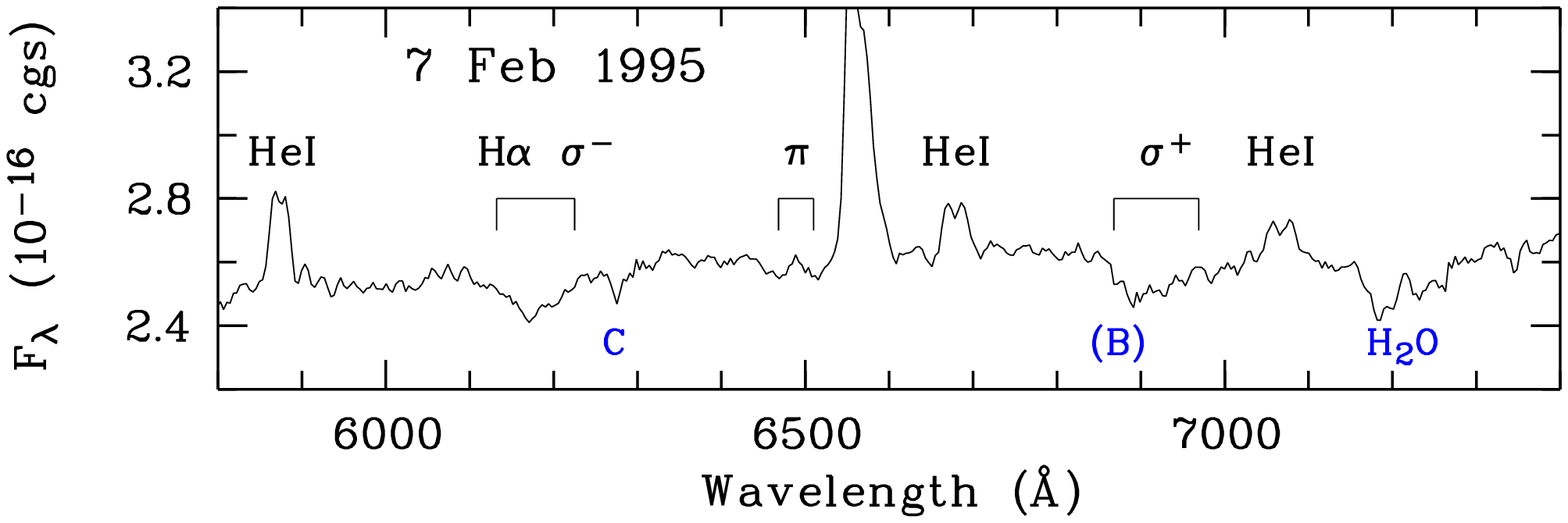}

\vspace{0.7mm}
\includegraphics[height=89.0mm,angle=270,clip]{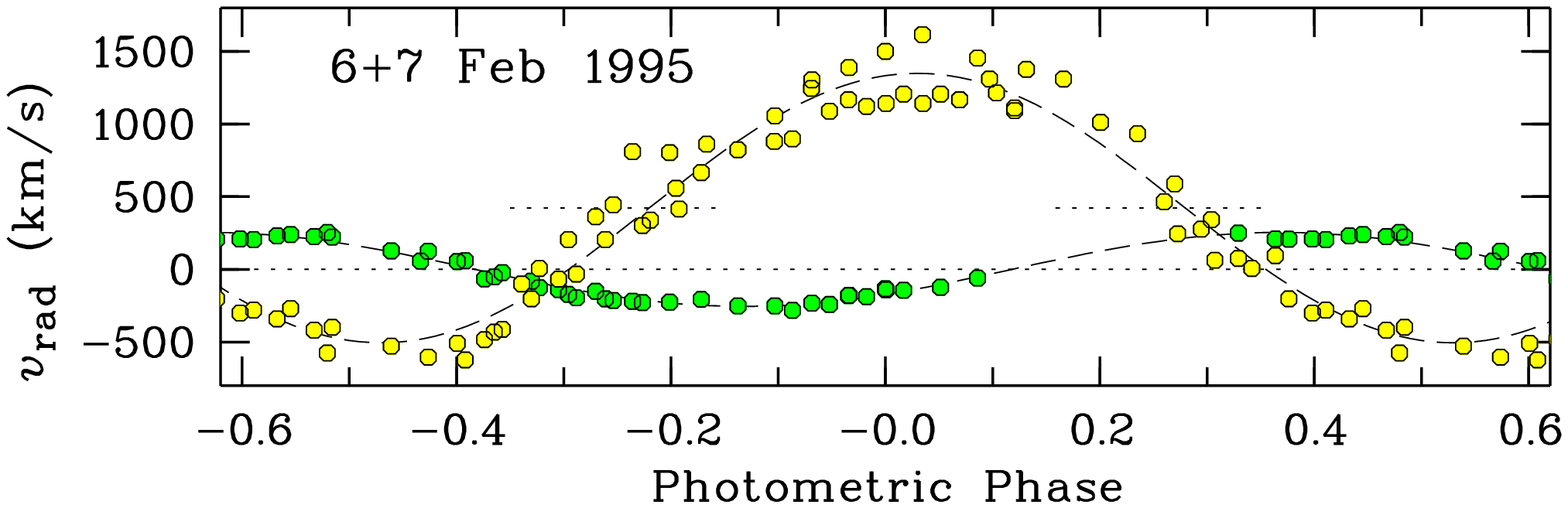}

\vspace{1.7mm}
\includegraphics[height=89.0mm,angle=270,clip]{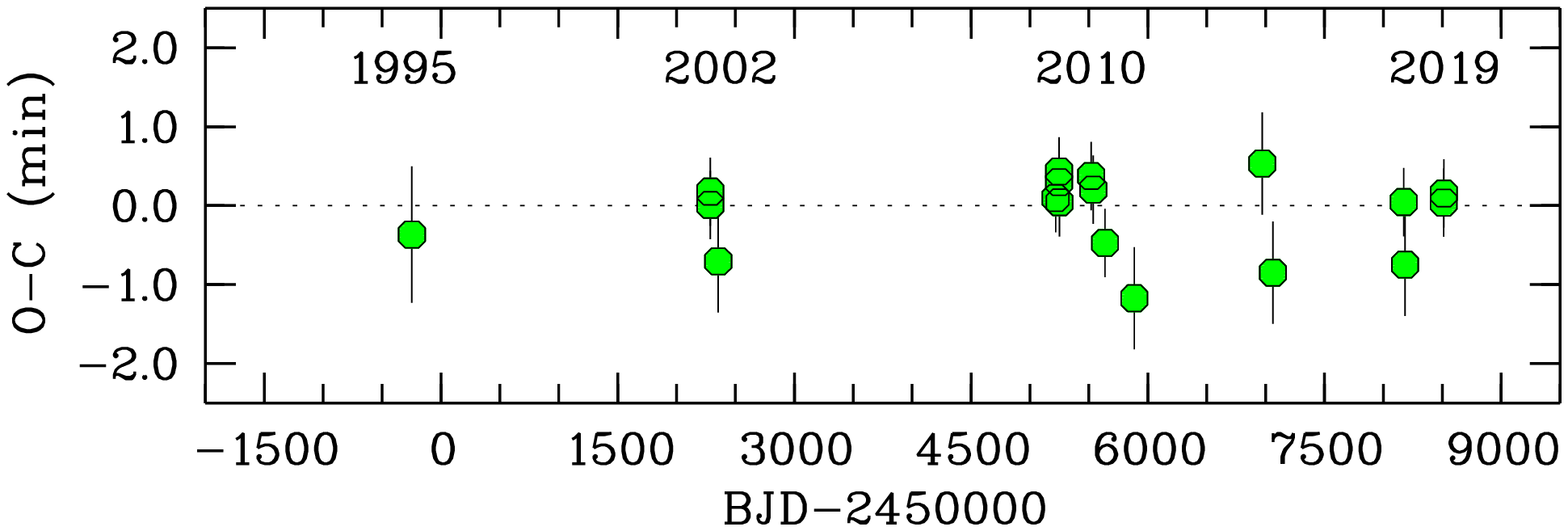}
\end{minipage}

\hspace {94mm}
\begin{minipage}[t]{90mm}

\vspace*{-90mm}
\includegraphics[height=89.0mm,angle=270,clip]{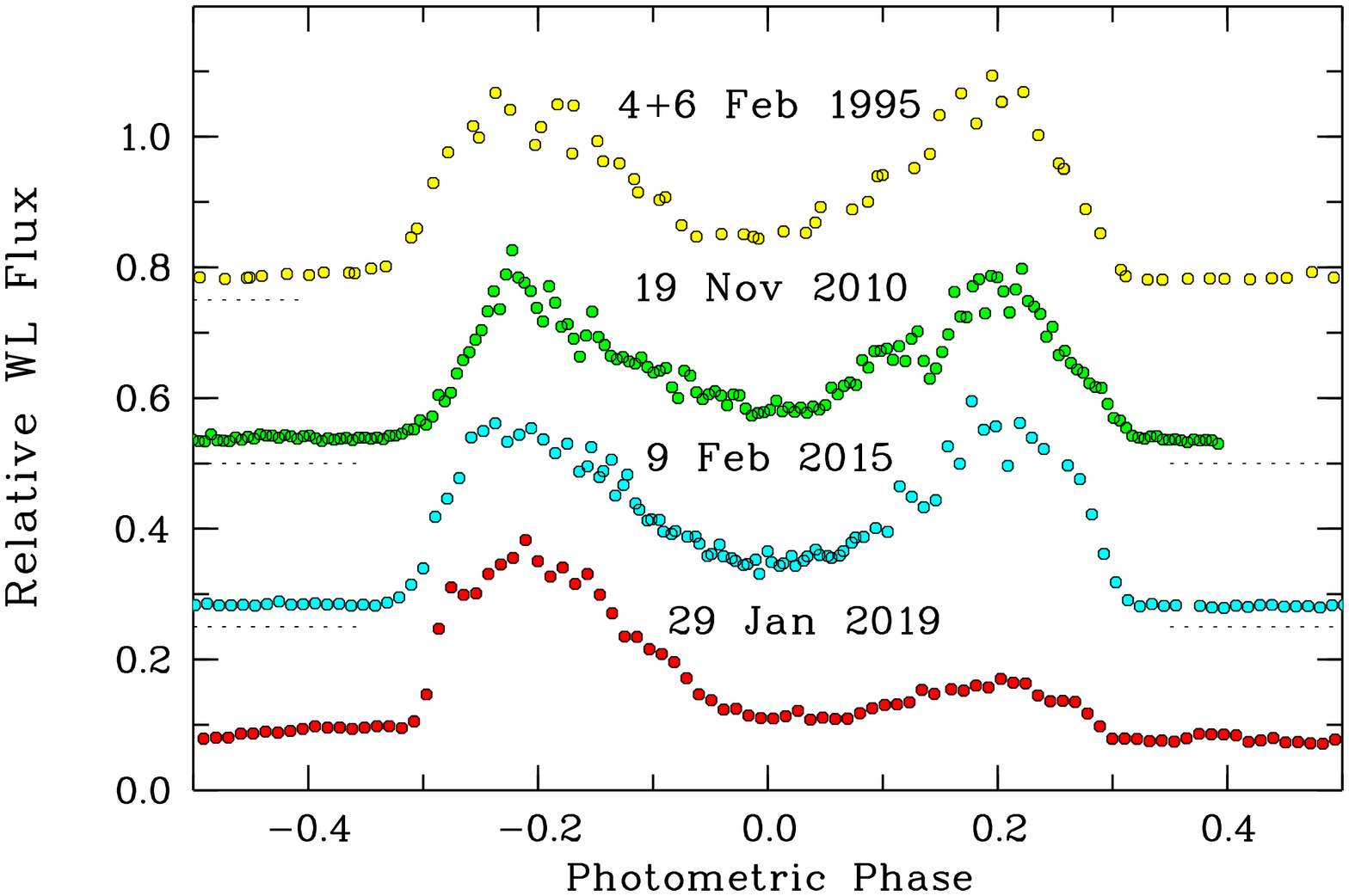}

\vspace{1.7mm}
\includegraphics[height=89.0mm,angle=270,clip]{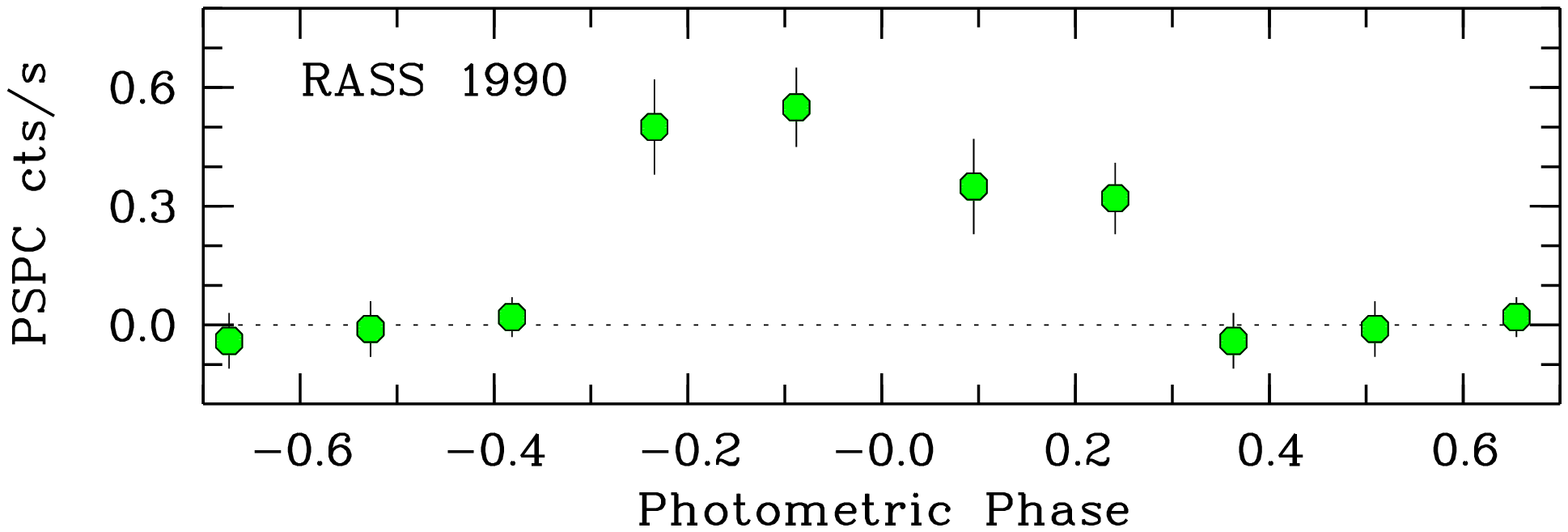}
\end{minipage}

\caption[chart]{\rxjnine. \emph{Left, top:} Combined blue and red
  flux-calibrated medium-resolution spectra at orbital maximum and
  minimum. \emph{Second from top:} Zoom into a difference spectrum
  emphasizing the \halp\ Zeeman features. \emph{Third from top:} Mean
  radial-velocity curves of the narrow and broad emission-line
  components of \halp. \emph{Bottom:} $O-C$ diagram for the center of
  the bright phase of the optical light curves. \emph{Right, top:}
  Overall spectral energy distribution, involving nonsimultaneous
  data (see text). \emph{Center:} Optical light curves taken in WL,
  taken in 1995, 2010, 2015, and 2019, shifted by multiples of 0.25
  units in the ordinate.  \emph{Bottom:} PSPC X-ray light curve taken in
  the RASS. The photometric phase is from Eq.~\ref{eq:0953mid}.}
\label{fig:0953}
\end{figure*}

\section{\rxjnine\ (= \jnine) in Leo}
\label{sec:0953}

\jnine\ was discovered in the RASS as a soft X-ray source,
spectroscopically identified by us as a polar, and is listed as such in
\citet{beuermannburwitz95} and \citet{beuermannetal99}.  The system
shows comparatively weak \heii\ line emission and was reclassified by
\citet{oliveiraetal20} as an intermediate polar. We do not share their
interpretation and confirm \jnine\ as a synchronized polar.

\subsection{X-ray observations}

In the RASS, the star was detected in November 1990 with a mean PSPC count
rate of $0.24\,\pm\!0.03$\,\cps\ and a hardness ratio
$HR1\!=\!-0.72\!\pm\!0.07$ (Table\,\ref{tab:basic}).  The RASS light
curve \citep{bolleretal16} in Fig.~\ref{fig:0953}, lower right-hand
panel, revealed a well-defined bright phase with a mean count rate of
$0.43\!\pm\!0.03$\,\cps. The phase convention refers to the center of
the optical bright phase (Eq.~\ref{eq:0953mid}). Because the flux goes
to zero during the faint phase, the orbital mean RASS spectrum is
readily scaled upward to that of the bright phase.  The extinction in
front of the source is \ebv$\,=\!0.024\!\pm\!0.006$
\citep{lallementetal18} and corresponds to
\nh$\,=\!(2.2\!\pm\!0.6)\,\times\,10^{20}$\,\atoms\
\citep{nguyenetal18}. When this value is adopted, the spectral fit with the
model of Sect.~\ref{sec:31} yields the bolometric X-ray fluxes and
luminosities listed in Table~\ref{tab:xray}.  In spite of the rather
low equivalent width of \heii\ (Table~\ref{tab:w}), this appears to be
the high state of \jnine.

\subsection{Orbital period and ephemeris}

The orbital period of \jnine\ was measured on 4--6 February 1995 by
phase-resolved $V$-band photometry with the ESO-Dutch 90 cm telescope
that extended over four consecutive orbital periods.
Photometry was performed relative to SDSS\,J095309.26+150118.8, which
is located 15\arcsec\ E and 162\arcsec\ N of the target and has AB
magnitudes $g\!=\!16.19$, $r\!=\!15.55$, and $i\!=\!15.33$.  \jnine\
was observed in WL by \citet{kanbachetal08} in 2002 and repeatedly by us between 1995 and 2019 (Table~\ref{tab:phot}). The light
curves of \jnine\ (Fig.~\ref{fig:GK} and \ref{fig:0953}) are good
examples of a single-pole accretor with strong cyclotron beaming. In
search of an appropriate fiducial mark for the period measurement, we
opted for the center of the bright phase, calculated as the mean of
the ingress and egress times. A linear fit to 18 center-bright times
yielded the alias-free ephemeris
\begin{equation}
T_\mathrm{cb}\!=\!\mathrm{BJD(TDB)}~2455217.96104(8) + 0.072022401(3)\,E.
\label{eq:0953mid}
\end{equation}
The bottom left panel of Fig.~\ref{fig:0953} shows the O-C
diagram. All measured times are listed in Table~\ref{tab:0953} in
Appendix~\ref{sec:C}. 

\subsection{Trailed spectrophotometry and Zeeman spectroscopy}

We obtained trailed optical spectra of \jnine\ on 5--8 February 1995,
using the same setup as for the previous target
(Table~\ref{tab:spec}). In this case, the correction for the Meinel
OH-bands was less problematic. We show the mean faint-phase spectrum
and the mean spectrum for intervals around the orbital maxima in
Fig.~\ref{fig:0953}, top left panel. The system reached
$r\!\simeq\!16.7$ at orbital maximum, dropping to $19.5$ in the faint
phase. The secondary is detectable in the observed faint-phase
spectrum. Its spectral type cannot securely be determined, but at the
orbital period of \jnine, the evolutionary sequence of
\citet{kniggeetal11} predicts a secondary star of spectral type about
dM5.5. We adopted the dM6 (or dM5.5) star SDSS\,J155653.99+093656.5
with $i\!=\!15.12$ as template and show its spectrum, adjusted by a
factor of $160\,\pm\,30$, in the upper left panel (red curve). This
factor yields the $i$-band magnitude and flux of the secondary star as
$20.63\!\pm\!0.21$ and $0.0203\!\pm\!0.0038$\,mJy, respectively.

The red bright-phase continuum plausibly represents cyclotron emission
from the primary pole; it extends to about 5000\,\AA\ (12th harmonic),
where the blue stream emission overtakes.  The bright-phase spectrum
contains Zeeman features of \halp, seen more clearly in the difference
spectrum of bright and faint phases (second left-hand panel from the
top). The \halp\ $\pi$ and $\sigma ^{-}$ components are undisturbed,
while the $\sigma^{+}$ component coincides with the uncorrected
atmospheric B band. These rather sharp lines with the resolved
$\pi$-components are of nonphotospheric origin and of a type usually
referred to as halo lines \citep[][their Table~2]{ferrarioetal15}. 
They occur in cool matter in the vicinity of the hot plasma emitting
the cyclotron radiation. In V834~Cen, this is the free-falling
pre-shock matter \citep{schwopebeuermann90}, and in BL~Hyi more
stationary matter at an uncertain location \citep{schwopeetal95}. In
both cases the observed lines indicate the field strength in the
vicinity of the accretion region. With 19\,MG, the derived field
strength in \jnine\  is rather low for a polar.

We measured the radial velocities of the narrow and broad components
of the Balmer lines and show the results for \halp\ in the third
left-hand panel of Fig.~\ref{fig:0953}. The narrow component has a
velocity amplitude $K_2'\!=\!254\!\pm\!7$\,\kms\ and a blue-to-red
zero crossing at photometric phase
$\phi_\mathrm{br}\!=\!0.11\,\pm\,0.01$, defining spectroscopic phase
as $\phi_\mathrm{sp}\!=\!\phi_\mathrm{ph}-0.11$, which is \emph{\textup{bona
  fide}} also the true binary phase. The broad component has a somewhat
uncertain $K_\mathrm{broad}\,\sim\,920$\,\kms\ with
$\gamma_\mathrm{broad}\!\sim\!400$\,\kms.  It reaches maximum positive
radial velocity at photometric phase
$\phi_\mathrm{ph}\!=\!0.02\,\pm\,0.02$, almost coincident with the
closest approach to the accretion funnel, when the BBC is moving away
from the observer.

The question of a possible asynchronism was raised by
\citet{oliveiraetal20}, who considered \jnine\ as an intermediate
polar. Their argument was based on a single double-peaked spectrum,
which they interpreted as originating from an accretion disk. The gray
plot in Fig.~\ref{fig:vrad} shows that \jnine\ may display a
double-peaked line profile near $\phi_\mathrm{sp}\!=\!0.5$ that
originates, however, from the superposition of NEL and the broad
component, notably in form of the HVC. We measured the spectroscopic
period by combining the data of 6 and 7 February 1995, and obtained
$P_\mathrm{n}\!=\!0\,\fd07188(15)$ and
$P_\mathrm{b}\!=\!0\,\fd07204(9)$ for the narrow and broad component,
respectively.  Both periods agree within the errors with the
photometric period of Eq.~\ref{eq:0953mid} and limit any asynchronism
to a level of $\ten{2}{-3}$. Because the simple light curve did
not change over more than 20 years, however, this places the limit at lower
than $10^{-5}$.

\subsection{Spectral energy distribution}

The overall SED is shown in the upper right panel of
Fig.~\ref{fig:0953}. The observed spectra at orbital maximum and
minium (black curves) and the MONET fluxes (yellow triangles)
delineate the range of the orbital and temporal variability over the
years. The Gaia, Pan-STARS, and a 2MASS $J$-band point (red dots) fall
within this range. The GALEX point (blue), the SDSS photometry (cyan),
part of the 2MASS data (green), and the WISE W1 point (blue) belong to
a low or an intermediate state.

The SED of the adjusted dM6 secondary star is shown by the dashed
curve. An earlier star, adjusted to the same $i$-band flux, would have
lower IR fluxes. The secondary accounts for part of the red and
IR flux observed in the SDSS (cyan), the 2MASS (green) and WISE
(blue). The remaining flux in the IR represents cyclotron
radiation or stream emission. The low fluxes measured in the SDSS
(cyan) and with GALEX (blue) can be accounted for by a WD of
0.75\,\msun\ with radius $\ten{7.6}{8}$\,cm and effective temperature
12000\,K, placed at the Gaia distance of 448\,pc (blue curve).
Measuring the temperature and radius of the WD spectroscopically
should be feasible.

\subsection{System parameters}

At the orbital period of \jnine, the evolutionary model of
\citet{kniggeetal11} predicts a moderately bloated secondary star with
a mass of 0.118\,\msun, radius of 0.161\,\rsun, and a spectral type
dM5.5 with an $i$-band surface brightness
$S_\mathrm{i}\!\simeq\!8.4$. The predicted $i$-band magnitude and flux
at the Gaia distance are 20.62 and 0.0205\,mJy, respectively,
in agreement with the observed quantities (Sect.~7.3).
For further analysis, we converted the radial-velocity amplitude
$K_2'\!=\!254$\,\kms\ into $K_2$ with our irradiation model BR08. The
well-defined length of the self-eclipse, $\Delta \phi\!\simeq\!0.40$,
places the accretion region in the upper hemisphere of the WD. The deep
cyclotron minimum and the lack of an absorption dip require
$i\,\la\,\beta\!\simeq\!\zeta$ (Sect.~\ref{sec:NEL}). Eq.~\ref{eq:visi} 
gives $i\!\simeq\!55\!-\!60$\degr\ and the measured value of $K_2'$
gives $M_1\!\simeq\!0.58\,-\,0.68$\,\msun. The standard primary mass
of \citet{kniggeetal11} of 0.75\,\msun\ would require an inclination
of 50\degr.

The accretion rate derived from the X-ray luminosity is $\dot
M_\mathrm{x}\!=\!4.6\!\times\!10^{-11}$\,\msunyr. The estimate of the
cyclotron luminosity in Table~\ref{tab:xray} is based on the observed
optical spectrophotometry, extrapolated into the near-IR. When it is included, the required accretion rate rises to $\dot
M_\mathrm{x+cyc}\!=\!5.8\!\times\!10^{-11}$\,\msunyr\ for a WD of
0.63\msun. The corresponding equilibrium temperature of the
compressionally heated WD would be 10400\,K. The predicted 4600\AA\
flux of the WD of 0.027\,mJy agrees closely with the dereddened SDSS
photometric g-band flux of $0.026\!\pm\!0.02$\,mJy
(Fig.~\ref{fig:0953}, upper right panel, cyan dots). The SDSS $ugr$
points and the GALEX flux with $0.033\!\pm\!0.03$\,mJy (blue dot)
define a flat spectrum, which likely represents the magnetic WD. The
agreement between predicted and observed spectral fluxes suggests that
the current effective temperature of the WD in \jnine\ and its equilibrium
temperature do not differ substantially.

\section{\rxjten\ (= \jten) in Hydra}
\label{sec:1002}

\jten\ was discovered in the RASS as the brightest and softest X-ray
source in our sample, spectroscopically identified by us as a polar,
and it is listed as such in \citet{beuermannthomas93},
\citet{beuermannburwitz95}, and \citet{thomasetal98}. Despite its high
degree of variability, it seems to be a tightly synchronized polar.

\subsection{X-ray observations}
\label{sec:1002x}

\begin{figure*}[t]
\includegraphics[height=89.0mm,angle=270,clip]{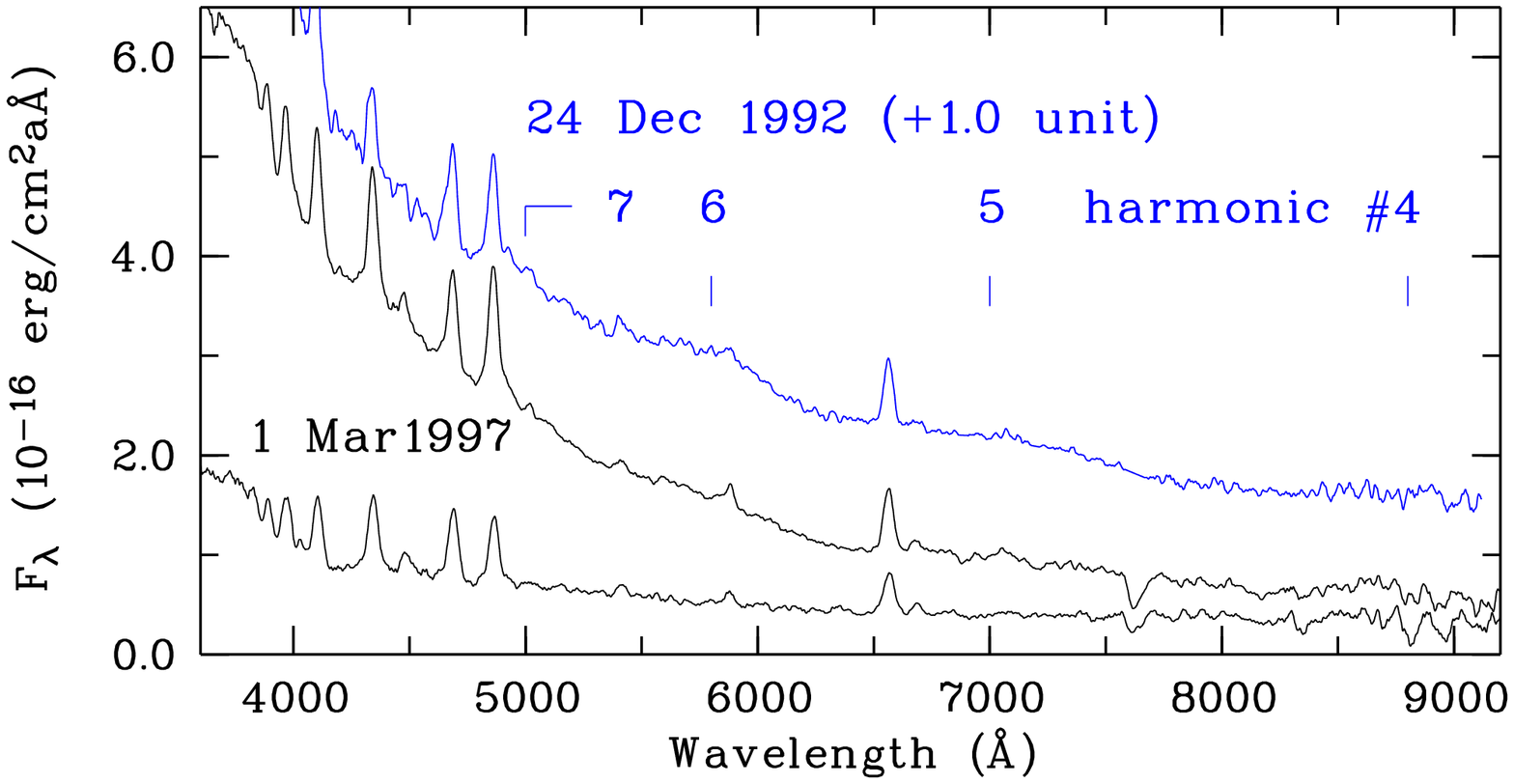}
\hfill
\includegraphics[height=88.0mm,angle=270,clip]{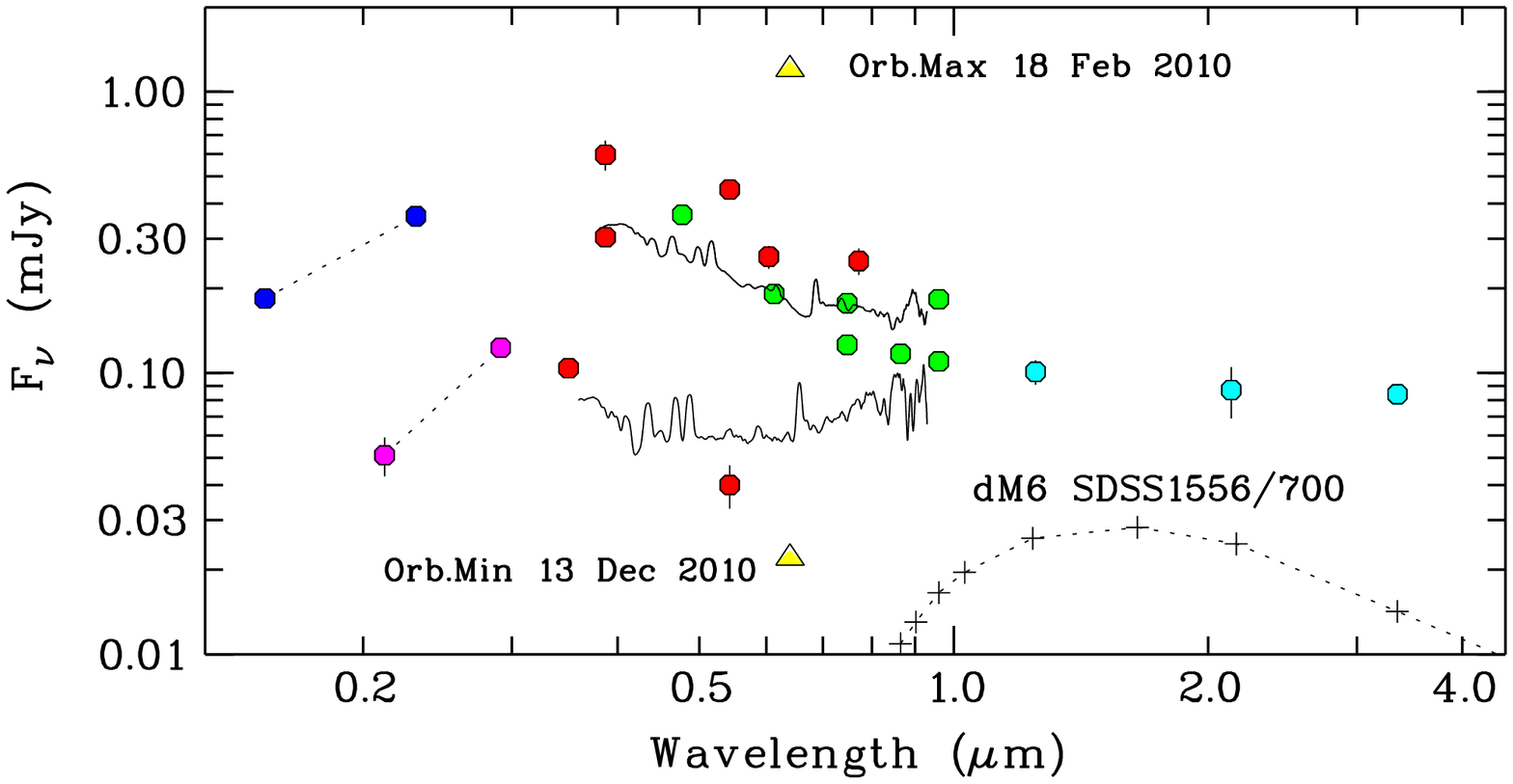}

\smallskip
\includegraphics[height=89.0mm,angle=270,clip]{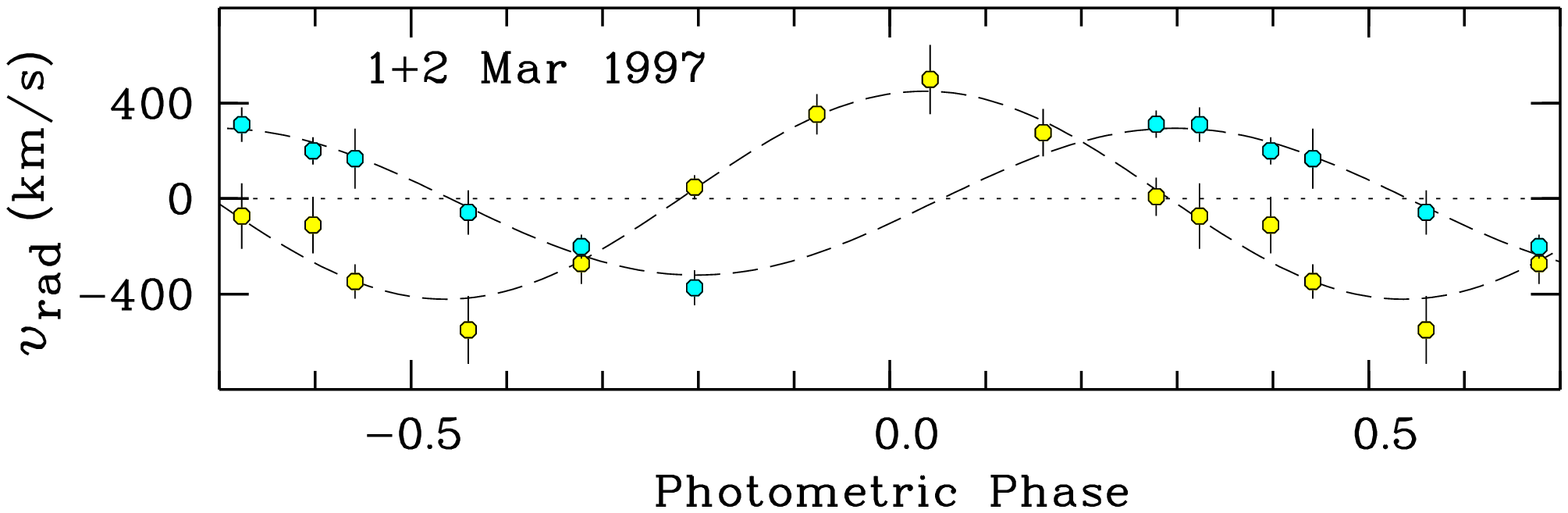}
\hfill
\includegraphics[width=28.5mm,angle=270,clip]{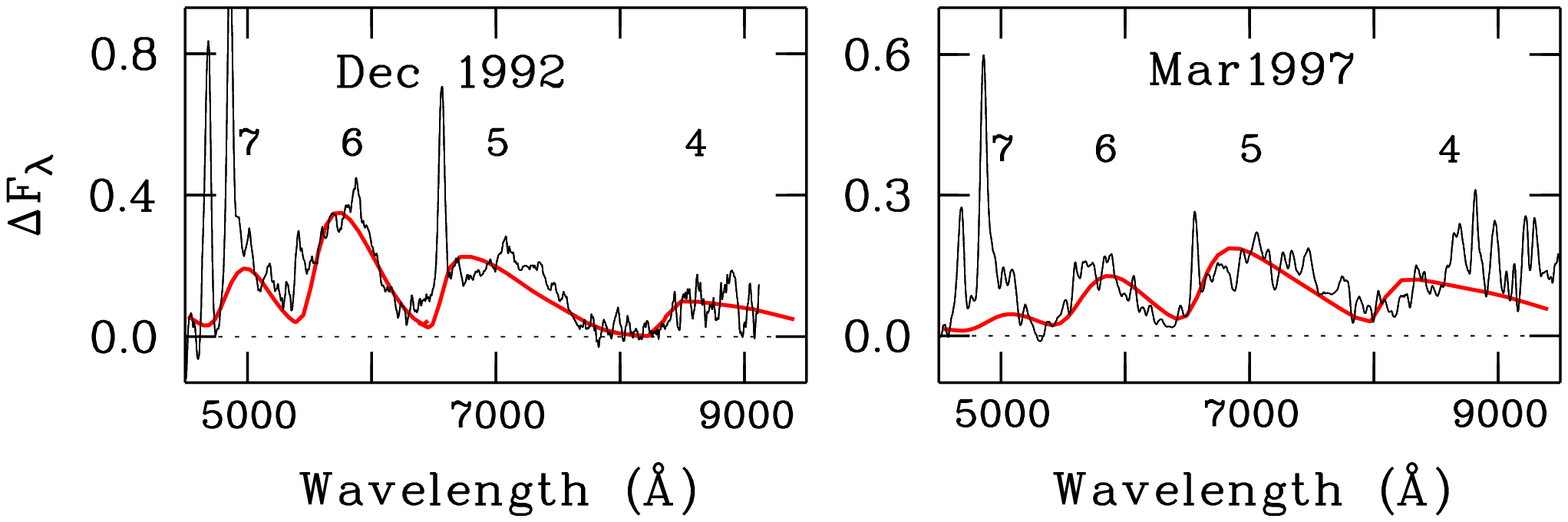}

\smallskip
\begin{minipage}[t]{90mm}
\includegraphics[height=89.0mm,angle=270,clip]{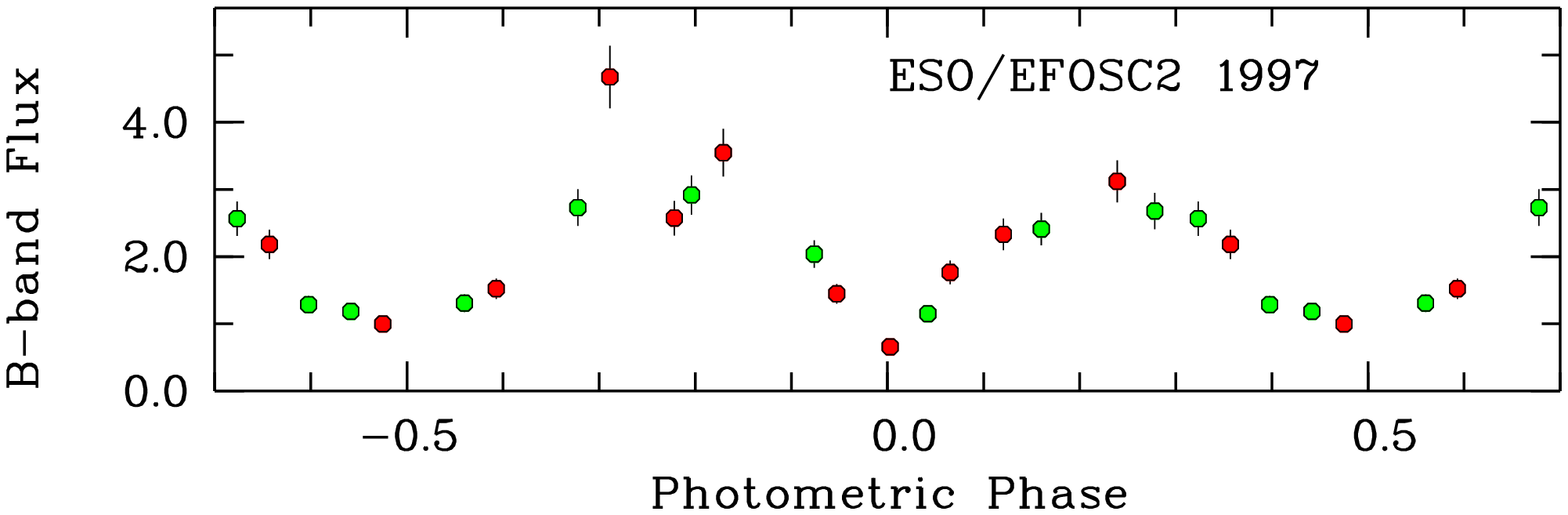}

\vspace{0.0mm} 
\includegraphics[height=89.0mm,angle=270,clip]{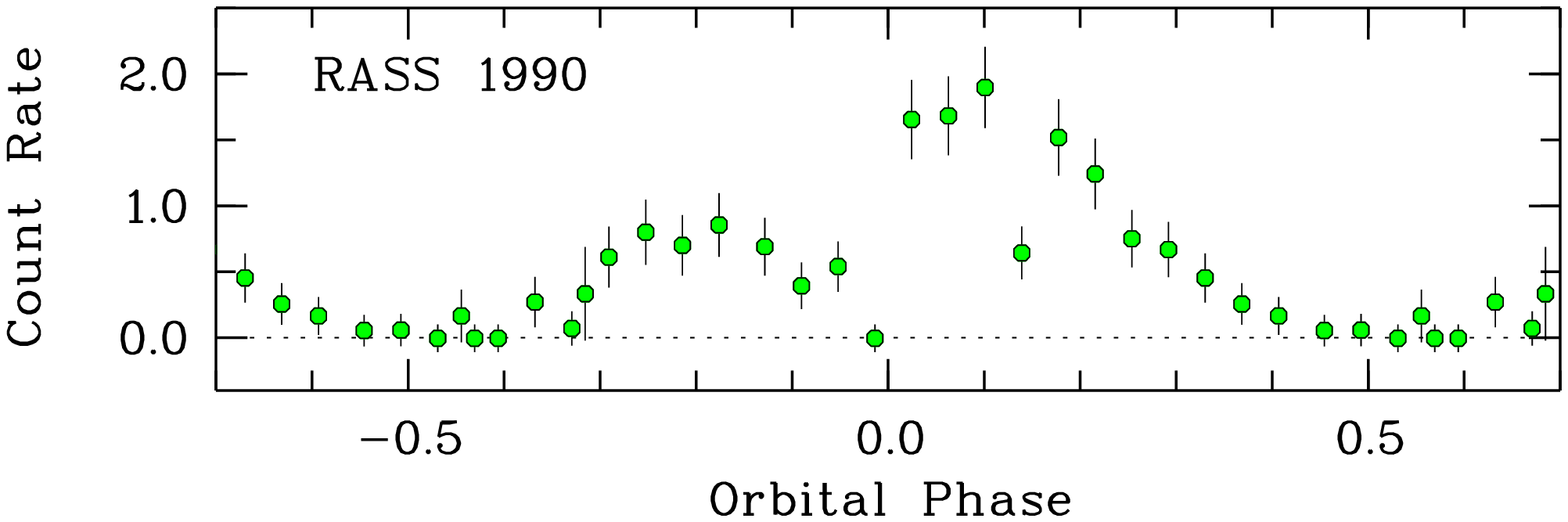}

\vspace{0.5mm}
\includegraphics[height=89.0mm,angle=270,clip]{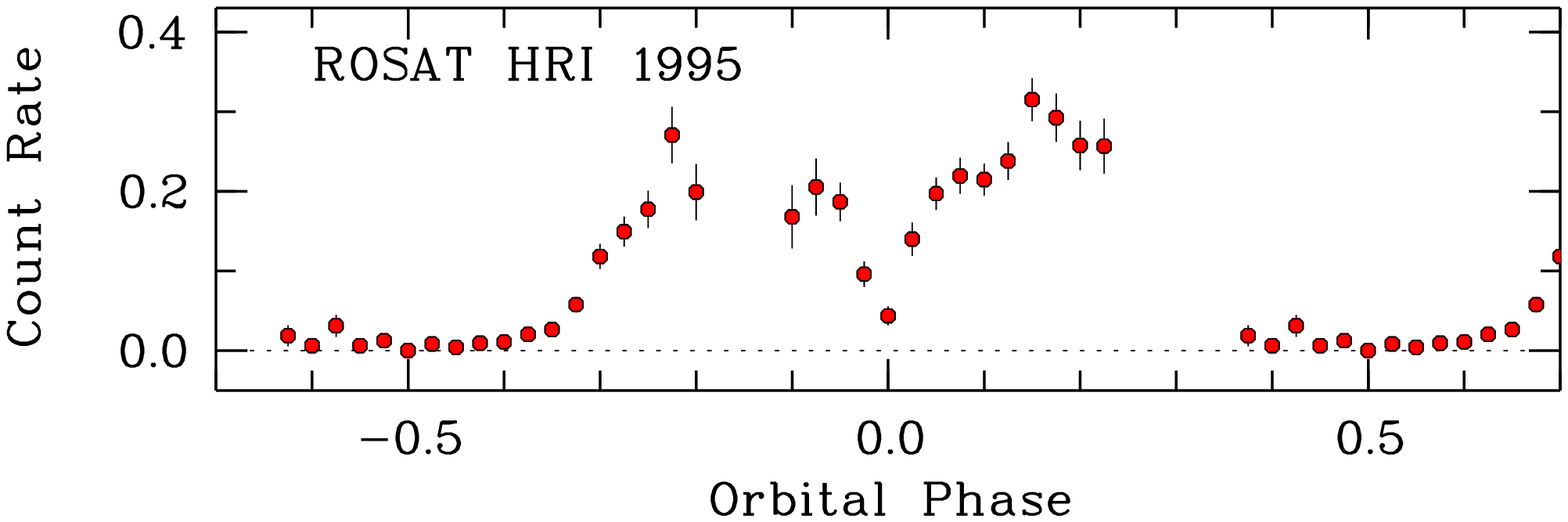}

\vspace{0.5mm}
\includegraphics[height=89.0mm,angle=270,clip]{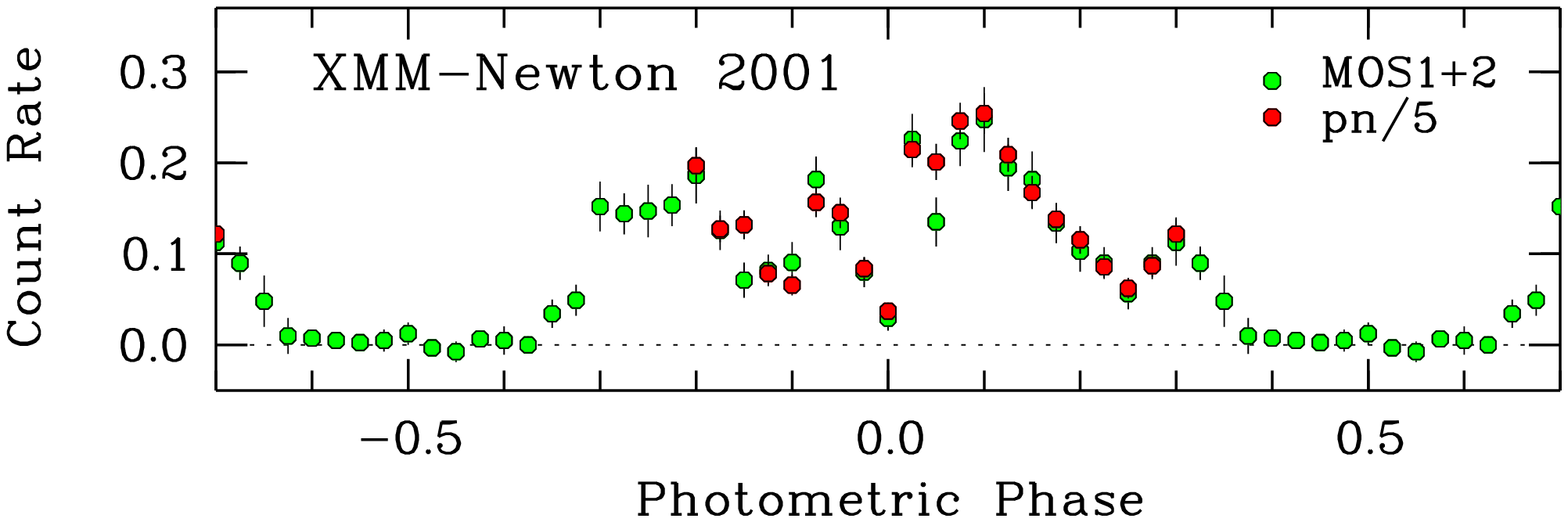}

\vspace{1.0mm}
\includegraphics[height=89.0mm,angle=270,clip]{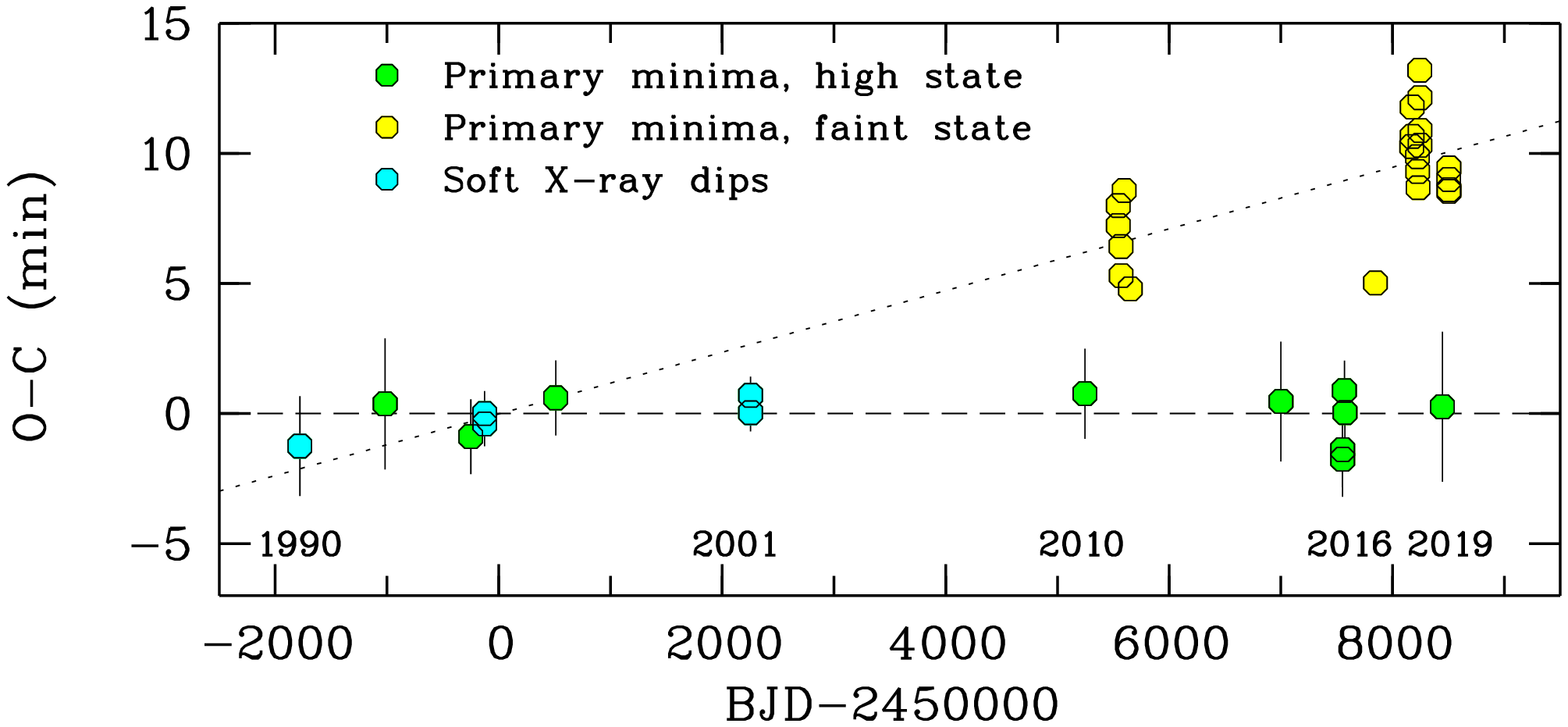}
\end{minipage}

\hspace {94mm}
\begin{minipage}[t]{90mm}

\vspace*{-141mm}  \hfill
\includegraphics[height=90.0mm,angle=270,clip]{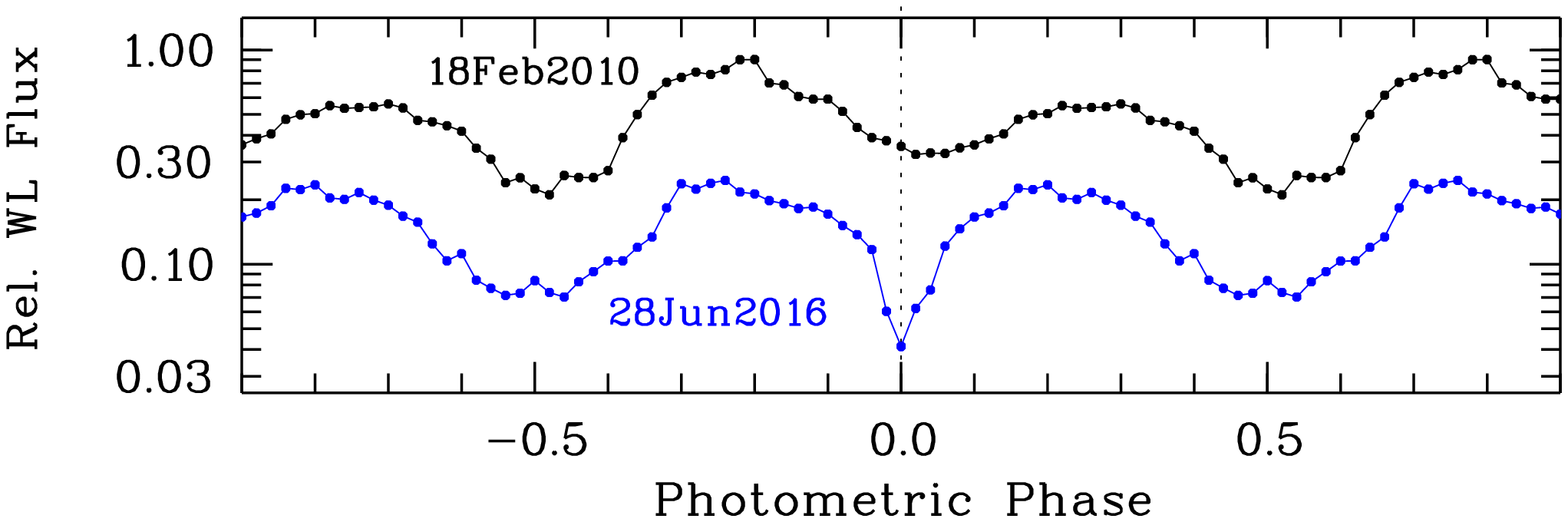}

\vspace{1.0mm}  \hfill
\includegraphics[height=90.0mm,angle=270,clip]{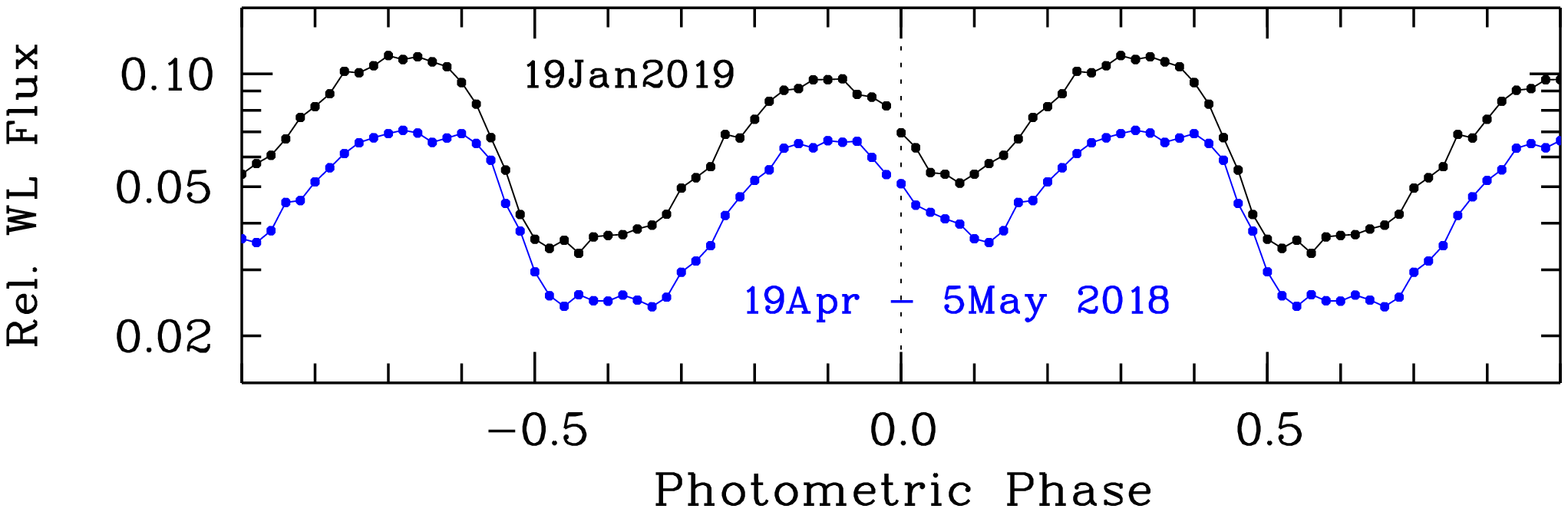}

\vspace{1.0mm} \hfill
\includegraphics[height=90.0mm,angle=270,clip]{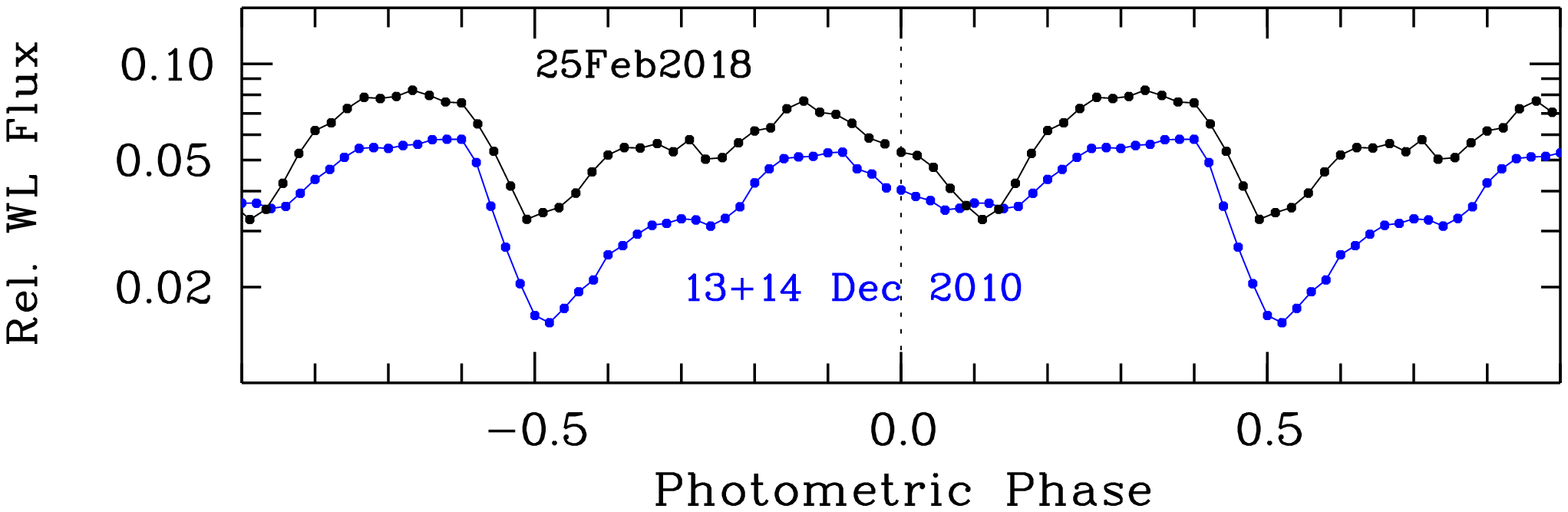}

\vspace{1.0mm} \hfill
\includegraphics[height=90.0mm,angle=270,clip]{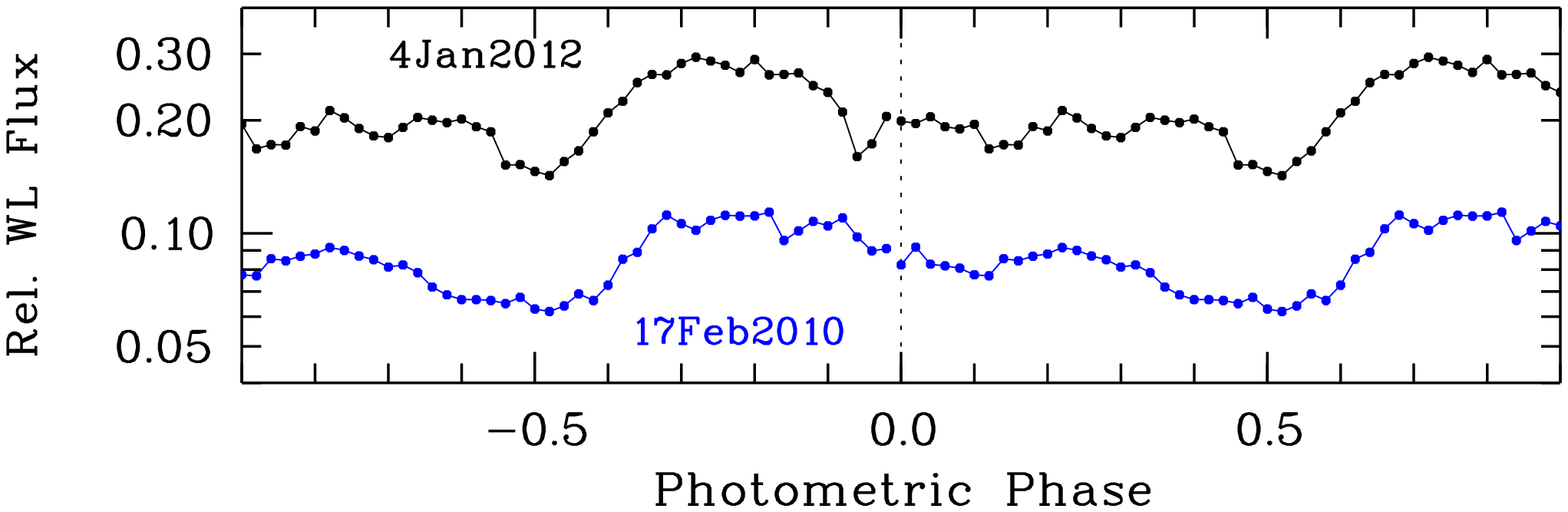}

\vspace{1.2mm} \hfill
\includegraphics[height=89.0mm,angle=270,clip]{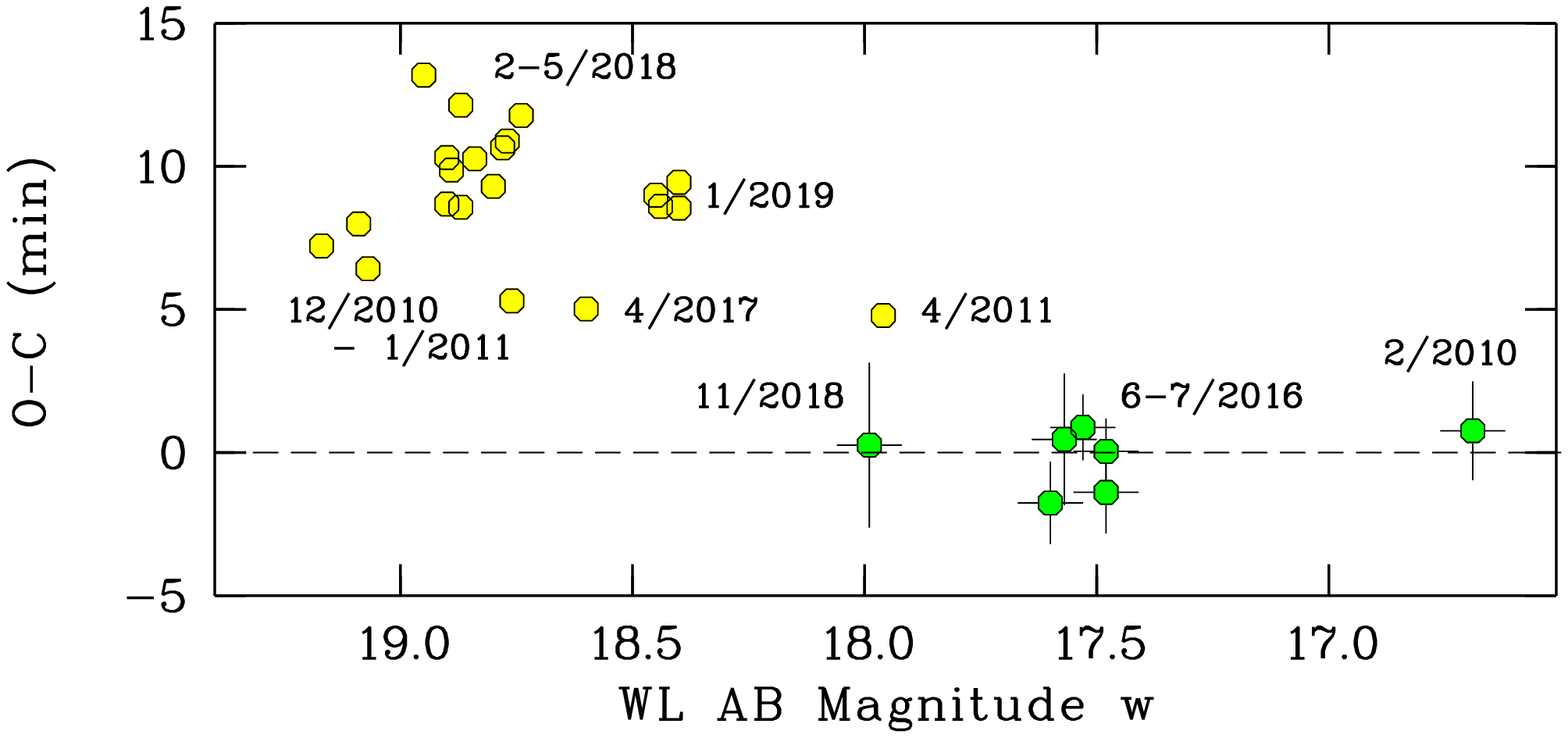}
\end{minipage}

\caption[chart]{\rxjten. \emph{Left, top: } Flux-calibrated
  low-resolution spectra on 24 December 1992 and at orbital maximum and
  minimum on 1 March 1997. Cyclotron harmonics are
  indicated. \emph{Left, second from top: } Balmer-line radial
  velocities of the narrow and broad components from spectra of 1 and
  2 March 1997.
  \emph{Left, third from top:} Spectral flux in the $B$ band of 1 and
  2 March 1997 in units is $10^{-16}$~ergs\,cm$^{-2}$s$^{-1}$\AA$^{-1}$.
  \emph{Left, next three panels: } Soft X-ray light curves of 1990,
  1995, and 2001. \emph{Right, top: } Overall nonsimultaneous
  SED. \emph{Right, second from top:} Cyclotron spectra on 24 December
  1992 and 1 March 1997 with models for a field strength of 33
  MG. \emph{Right, next four panels:} Samples of optical light curves
  taken in WL, illustrating the variability of primary and secondary
  minima. \emph{Left, bottom: } $O-C$ diagram for the primary minimum,
  from light curves with peak orbital WL AB magnitude $w\!\la\!18$.
  \emph{Right, bottom:} \oc\ for the primary minimum vs. $w$. ~~The
  photometric phase is from Eq.~\ref{eq:1002eph1}. }
\label{fig:1002}
\end{figure*}

\jten\ was detected in the RASS with a mean count rate of
$0.69\!\pm\!0.04$\,\cps\ and a hardness ratio
$HR1\!=\!-0.97\!\pm\!0.03$, implying that 98\% of the photons had
energies below the carbon edge at 0.28\,keV
(Table~\ref{tab:xray}). \jten\ was reobserved with ROSAT and the PSPC
in 1992 and 1993 and with the HRI in 1995. In 1992, it was in a low
state, with $-0.0002\pm0.0024$\,PSPC\,\cps, and in 1993 again in a
high state, with 0.71 PSPC \cps\ and 97\%\ soft
photons. \citet{ramsaycropper03} observed it with XMM-Newton in 2001
in an intermediate state. We reanalyzed their observation for the
present purpose. The lower left-hand panels of Fig.~\ref{fig:1002}
show the orbital light curves of the RASS PSPC, the 1995 ROSAT HRI,
and the 2001 XMM-Newton EPIC pn and MOS12 observations placed on the
ephemeris of Eq.~\ref{eq:1002eph1}. Common properties are a bright
phase that lasts for $\sim\!$75\% of the orbital period and a narrow
absorption dip near its center. This repetitive feature marks the
instance when the line of sight to the WD passes through the
magnetically guided part of the accretion stream.

\begin{table}[b]
\begin{flushleft}
  \caption{Fit parameters for the XMM-Newton pn and the ROSAT PSPC
    bright-phase X-ray spectra of \jten. The letter ``f'' denotes a
    frozen parameter. The quantity
    $f_\mathrm{sx,bol}\,=\!c_\mathrm{sx}f_\mathrm{bb1,bol}$ with
    $c_\mathrm{sx}\,=\!3$ and $f_\mathrm{bb1,bol}$ the bolometric flux
    of the single-blackbody fit. }
\begin{tabular}{@{\hspace{0.0mm}}c@{\hspace{1.0mm}}c@{\hspace{2.0mm}}c@{\hspace{1.0mm}}c@{\hspace{2.0mm}}c@{\hspace{3.0mm}}c@{\hspace{3.0mm}}c@{\hspace{-8.0mm}}c@{\hspace{-6.0mm}}c}\\[-2ex]
\hline\hline \\[-1.5ex]
Fit& Detector &   \nh\ & \nhint &$f_\mathrm{pc}$& \ktbbi\  &\fsxbol\ & \fthbol  & $\chi^2$ (dof)\\
& &\multicolumn{2}{c}{($10^{21}$\,cm$^{-2}$)}&&(eV)&\multicolumn{2}{c}{\hspace{-2.5mm}($10^{-11}$erg/cm$^2$s)} &      \\[0.5ex]
\hline\\[-1ex]                                             
1  & pn   & 2.15~   &     &       & 38.1 & 2.05 & 0.11 & 81.9~~(56)\\ 
2  & pn   & 0.01~   & 544 & 0.69  & 53.8 & 0.53 & 0.26 & 45.6~~(53)\\ 
3  & pn   & 1.00\,f~& 536 & 0.69  & 50.1 & 0.86 & 0.27 & 46.4~~(53)\\ 
4  & pn   & 2.90\,f~& 520 & 0.69  & 43.2 & 2.34 & 0.29 & 49.1~~(54)\\ 
5  & PSPC& 1.00\,f~&~100\,f&1.0\,f& 50.0 & 1.78 & 0.22 & 33.7~~(39)\\[1.0ex]
\hline\\
\end{tabular}\\[-2.0ex]
\label{tab:xmm}
\end{flushleft}

\vspace{-2mm}
\end{table}

The XMM-Newton EPIC pn observation and the 1993 ROSAT PSPC observation
both cover exclusively the bright phase. The PSPC spectrum is not
shown. A graph of the EPIC pn spectrum can be found in Fig.~7 of
\citet{ramsaycropper03}. We fit both spectra with the model of
Sect.~\ref{sec:31}. Table~\ref{tab:xmm} summarizes the results.  The
model in line~1 is a moderately successful fit to the pn spectrum with
\nhint$\,=\!0$. With the model in line~2, we confirm the findings of
\citet{ramsaycropper03} that the fit (i) prefers an interstellar
column density close to zero and (ii) benefits from the inclusion of
an internal absorber. A very low value of \nh\ is unrealistic,
however, given the Gaia distance of 797\,pc.  The total galactic
column density at the position of \jten\ is
$N_\mathrm{H,gal}\!=\!\ten{3.96}{20}$\,\atoms\ \citep{hi4pi16} and the
total extinction is \ebv$\,=\!0.0351$. A sizeable fraction seems to
arise in front of \jten, \ebv$\,\simeq\!0.031\!\pm\!0.009$
\citep{lallementetal18} or
\nh$\,=\!(2.9\,\pm\,0.9)\!\times\!10^{20}$\,\atoms\
\citep{nguyenetal18}. Because of tradeoffs between the parameters
\nh\ and \ktbbi, reasonably good fits are obtained for any column
density up to \nhgal\ (lines 2 to 4). Fitting the 1993 PSPC and the
2001 pn spectrum with the same \nh--\ktbbi\ combination requires
\nh$\,=\!\ten{1.0}{20}$\,\atoms\ and \ktbbi$\,=\!50$\,eV (lines 3 and
5). The blackbody fluxes for the two observations differ by about a
factor of two, which is a measure of the different brightness levels
during the two runs. We adopted these fits, but consider the fluxes of
lines 3 and 5 in Table~\ref{tab:xmm} as approximate lower limits and
marked them as such in Table~\ref{tab:xray}.

\subsection{Orbital ephemeris}
\label{sec:1002ephem}

The RASS observation has suggested a periodicity with
$P_\mathrm{orb}\!=\!0\,\fd0697(4)$, which was refined by the
separation of the narrow soft X-ray absorption dips in the 1995 HRI
observation to $P_\mathrm{orb}\!=\!0\,\fd0694(2)$. The five X-ray
dips, one in the RASS, two in the ROSAT HRI, and two in the XMM-Newton
observation (pn and MOS), did not suffice for an alias-free ephemeris,
however. Phase-resolved optical photometry and spectrophotometry of
\jten\ was performed in 1992, 1995, and 1997.  We added WL photometry
with the MONET telescopes in 27 nights between 2010 and 2019
(Tables~\ref{tab:spec}, \ref{tab:phot}, and \ref{tab:1002}). In the
photometric runs, the brightness of the target was measured relative
to a comparison star C1 located at
RA(2000)\,=\,$10^\mathrm{h}02^\mathrm{m}11\fs4$, DEC(2000)\,=
$-19\degr26\arcmin12\arcsec$ or 5\arcsec\,W and 35\arcsec\,S of the
target (Fig~\ref{fig:fc1002}). It has an AB magnitude
$r\!=\!15.95$\,\footnote{https://panstarrs.stsci.edu} and colors
$g-r\!=\!0.66$ and $r-i\!=\!0.24$, giving $w\!\simeq\!16.0$ for unity
relative to the WL flux.

Improving the X-ray derived period by optical data proved tedious
because the light curves show a pronounced variability, display phase
shifts in the positions of the minima, and possess two minima per
orbit, which annoyingly flipped the role of the more prominent one
over the years. The spectrophotometric light curve of 1997 in the left
part of Fig.~\ref{fig:1002} (third panel from the top) and selected WL
light curves of 2010--2019 on the right side (panels three to six from
the top) provide an overview. All light curves are phased on the
ephemeris of Eq.~\ref{eq:1002eph1} used as a 
\linebreak \clearpage \noindent 
common reference. They
show that (i) the cyclotron-dominated bright phase and the X-ray
bright phase coincide, (ii) the primary optical minimum at
$\phi\!\simeq\!0$ is produced by cyclotron beaming and occurs close in
time to the X-ray absorption dip (see the discussion in
Sect.~\ref{sec:NEL}), and (iii) the secondary minimum occurs when the
accretion spot dips behind the horizon at $\phi\!\simeq\!0.5$. The
light curves in the right-hand panels are highly variable. The cases
of 18 February 2010 and 28 June 2016 suggest that either the primary spot
wanders in latitude or the minimum is filled up by the emission of an
independent second accretion region. Repeatedly over the years, we
observed light curves in which the separation of the primary and the
secondary minimum differed from half an orbital period (e.g.,  13+14
December 2010 and 25 February 2018). This apparent shift is probably caused by a
second emission region that appears near $\phi\!=\!0.6$ (compare the
similar blue model light curve in the right panel of
Fig.~\ref{fig:GK}). On other occasions, both minima appear displaced
by up to +0.12 in phase, requiring a longitudinal shift of the spot by
up to 40\degr (e.g., April--May 2018). The light curves of 4 January 2012 and
17 February 2010 lack any trace of the primary minimum.

Our search for a long-term ephemeris is based on a total of five X-ray
dips and 70 times of optical minima, not all measured from complete
orbital light curves. Because there is no unique way to distinguish
primary and secondary minima observationally, we started our search
for a long-term ephemeris by considering the mixed bag of primary and
secondary minima and calculating a periodogram in the vicinity of
\po/2. This procedure yielded a unique (alias-free) ephemeris. We
assigned cycle number $E\!=\!0$ to the X-ray dip in the XMM-Newton pn
light curve as our key primary minimum. This allowed us to identify
all minima with even cycle numbers on the \po$/2$ fit as primary
ones. We proceeded by calculating an ephemeris on a \po\ basis for the
subset of all primary minima with redefined cycle numbers. We kept the
definition of $E\!=\!0$.

We show the \oc\ values of all primary minima based on the ephemeris
of Eq.~\ref{eq:1002eph1} in the bottom left panel of
Fig.~\ref{fig:1002}.  This ephemeris satisfies all minima of
1990--2001 and also selected minima of 2010, 2014, 2016, and 2018
(green dots, shown with error bars), but fails to meet other groups of
minima of 2010, 2017, 2018, and 2019 (yellow dots, shown without error
bars to avoid clutter). The likely physical cause of the discrepant
\oc\ values becomes clear from the bottom right panel, where we show
the same data plotted versus the brightness of the system, measured by
the peak orbital WL magnitude $w$. In high states with $w\!<\!18$,
\oc\ averages zero, while in intermediate or low states with
$w\!\ga\!18$, \oc\ reaches up to $\sim\!12$\,min. Obviously, no
linear ephemeris can describe the up and down of \oc.

Selecting the early data (five X-ray dips and four primary optical
minima) and the seven primary minima from high-state light curves with
$w\!<\!18$ defines the alias-free ephemeris
\begin{equation}
  T_\mathrm{min}\!=\!\mathrm{BJD(TDB)}~2452254.38167(21) + 0.069427553(5)\,E,~~~
\label{eq:1002eph1}                              
\end{equation}
which served as our reference and is represented by the dashed line in
the bottom left panel of Fig.~\ref{fig:1002}. The X-ray data represent
high states (ROSAT) or a moderately high state (XMM-Newton). The
spectrophotometry of 1992 and 1997 is characterized by strong \heii\
emission lines (Table~\ref{tab:w}), which indicates the presence of an
intense XUV component and marks them as 'high state' as well. Fitting the
early data alone yields the same ephemeris as Eq.~\ref{eq:1002eph1}
within the uncertainties. This shows that the early X-ray and optical data and
the later high-state data are entirely compatible with a common linear
ephemeris.

Selecting instead the mutually exclusive subset of primary minima of
2010--2019 with $w\!>\!18$ (yellow dots in the bottom panels of
Fig.~\ref{fig:1002}), supplemented by the seven timings of 1990--1997,
but excluding the now discrepant 2001 XMM timings, we obtain the
longest period compatible with part of the data,
\begin{equation}
T_\mathrm{min}\!=\!\mathrm{BJD(TDB)}~2452254.38353(25) + 0.069427610(4)\,E.~~~
\label{eq:1002eph2}                              
\end{equation}
This ephemeris has an orbital period 4.9\,ms longer than that of
Eq.~\ref{eq:1002eph1} and is represented by the dotted line in the
bottom left panel of Fig.~\ref{fig:1002}. Both ephemerides assign the
same cycle numbers to all minima of the 30 yr covered by
our data. Despite the remaining uncertainty, our ephemerides are
therefore alias-free. The times of all primary minima used for Eq.~\ref{eq:1002eph1} or
Eq.~\ref{eq:1002eph2} are listed in Table~\ref{tab:1002} in
Appendix~\ref{sec:C}. The high-state spot position is stable over at least about
1.5 mag in $w,$ and we argue that the period in Eq.~\ref{eq:1002eph1}
more likely represents the true binary period.

The \oc\ variations in \jten\ are unlike anything observed in
synchronized polars. Does \jten\ lack synchronism? It showed no
evidence for a shorter, intermediate-polar like periodicity.  Between
1990 and 2001, X-ray period, optical photometric period, and
spectroscopic period agreed, severely limiting the permitted degree of
asynchronism.  The data of 26 December 1992 to 1 January 1993 and of $1\!-\!2$
March 1997 yielded spectroscopic periods of 0\,\fd069400(50) and
0\,\fd069422(110), which deviate from the photometric period of
Eq.~\ref{eq:1002eph1} by $-2.4\,\pm\,4.3$\,s and $-0.5\,\pm\,9.5$\,s,
respectively. The two epochs combined lead to a spectroscopic
period of 0\,\fd06942755(19), which perfectly agrees with the
photometric period within its error of 17\,ms. This excludes a
sizeable asynchronism as in the post-nova V1500\,Cyg, but not
necessarily the minute level detected in DP\,Leo and thought to
reflect the slow oscillation of the two stellar components about their
magnetostatic equilibrium position \citep{beuermannetal14}. However,
even if active in \jten, this process provides no explanation for the
observed up and down of \oc. We are therefore left with the
classical explanations: (i)~At a reduced accretion rate, the stream
penetrates less deeply into the magnetosphere, the spot moves closer to
the line connecting the two stars, and the closest approach to the
spot occurs later. (ii)~At a reduced accretion rate, the stream
switches from a ballistic trajectory in the orbital plane to a
magnetically guided path starting from the secondary star, a
possibility considered in Papers I and II. (iii)~In a complex field
geometry, the stream may be directed to different positions on the WD
surface depending on the accretion rate and the ram pressure it
exerts.  All polars experience drastic variations in the accretion
rate, yet none has so far displayed apparent or real spot movements
similar to those observed in \jten.

\subsection{Spectrophotometry and cyclotron spectroscopy}
\label{sec:1002field}

Low- and medium-resolution phase-resolved spectrophotometry was
performed in December 1992 and March 1997 (Table~\ref{tab:spec}).  The top
left panel in Fig.~\ref{fig:1002} shows the single spectrum of 1992
(blue) and the 1997 spectra at orbital maximum and minimum
(black). They are characterized by a blue continuum, strong Balmer and
\heii\ emission lines, the Balmer jump in emission, and the absence of
spectral features of the stellar components. The equivalent widths of
\heii\ and \hbet\ (Table~\ref{tab:w}) fall in the general area
populated by polars \citep[][their Fig.\,2]{oliveiraetal20}.
Furthermore, the bright-phase line ratio near unity indicates that
\jten\ was in a high state in December 1992 and March 1997. The Balmer lines
in the medium-resolution spectra of both years are fairly wide, as
noted already by \citet{oliveiraetal17}. Their FWHM varies between
1100 and 2300\,\kms\ over the orbit. Balmer and helium lines consist
of a broad component, representing a mixture of the BBC and HVC, and a
well-defined narrow component, representing the NEL. The second left
panel in Fig.~\ref{fig:1002} shows the radial-velocity curves of the
narrow component, measured from the 2 March 1997 spectra, and of the
broad component, obtained from 1 and 2 March.  The narrow component has
a radial-velocity amplitude $K_2'\!=\!307\pm30$\,\kms, with a
blue-to-red zero crossing at photometric phase
$\phi_\mathrm{br,n}\!=\!0.055\!\pm\!0.024$.  The broad component has a
radial-velocity amplitude $K_\mathrm{broad}\!=\!436\pm47$\,\kms,
reaching maximum positive radial velocity at
$\phi_\mathrm{broad}\!=\!0.044\!\pm\!0.013$.  If the narrow line
tracks the motion of the secondary star, spectroscopic (or binary) and
photometric phases are related by
$\phi_\mathrm{sp}\!=\!\phi_\mathrm{ph}-0.055$, with
$\phi_\mathrm{ph}\!=\!0$ some $20\degr\pm\,9\degr$ in azimuth before
inferior conjunction.  Maximum positive broad-line radial velocity
occurs $4\degr\!\pm\!5\degr$ before inferior conjunction, but primary
minima that are 10 min late occur at binary phase
$\phi_\mathrm{sp}\!\simeq\!0.05$ or 5 min past inferior
conjunction. Very similar numbers were obtained for 1992. Whether they
still applied between 2010 and 2019, when the pronounced \oc\
variations occurred, remains uncertain.

The spectrum of 24 December 1992 displays weak cyclotron lines superposed
on the blue optically thick cyclotron continuum and the optically thin
stream emission. Lines at the same positions are also detected in the
difference between the 1997 orbital maximum and minimum spectra. They
are displayed in Fig.~\ref{fig:1002} (second right-hand panels from
the top) with an estimated continuum interactively subtracted. They
were fit by constant-temperature models calculated with the theory
of \citet{chanmugamdulk81} (red curves) for a field strength of 33\,MG
with k$T_\mathrm{e}\!\simeq\!7\,$\,keV, an angle against the field
line $\theta\!=\!70-80^{\circ}$, and a thickness parameter
log\,$\Lambda\!\simeq\!3.8$.  For these parameters, the lines
represent the fourth to seventh harmonic.  We cannot entirely exclude that
the observed lines are harmonics 3\,--\,6 in a field of 40\,MG.

\subsection{Spectral energy distribution}

The top right-hand panel of Fig.~\ref{fig:1002} shows the overall SED
of \jten\ that includes the 1 March 1997 bright and faint-phase spectra
(black solid curves), a summary of nonsimultaneous photometry, and a
representation of the secondary star from our dynamical model
presented below. The photometry is from GALEX (blue), the XMM-Newton
UV monitor (magenta), Gaia, SkyMapper, and the XMM-Newton V monitor
(red), PanSTARRS (green), VISTA (cyan), and ALLWISE W1 (magenta). The
full range of the brightness variations of our extensive MONET WL
observations (yellow triangles) spans a factor of 50. None of the
observations represents a low state, although such drops in brightness
exist, as evidenced by the 1992 ROSAT PSPC observation and the
long-term light curve of the Catalina Sky Survey
\citep{drakeetal09}\footnote{http://crts.caltech.edu/}, which shows a
drop to $>\!20$\,mag at the end of 2007 from a \mbox{general level of
  $18\pm1$ mag}.

\subsection{System parameters}
\label{sec:1002system}

At the orbital period of \jten, the evolutionary model of
\citet{kniggeetal11} predicts a mildly bloated secondary star with a
mass of 0.108\,\msun, a radius of 0.153\,\rsun, and a spectral type
dM5.6. With an $i$-band surface brightness
$S_\mathrm{i}\!\simeq\!8.5$, the $i$-band magnitude and flux at the
Gaia distance are 22.08 and 0.0053\,mJy. It is not surprising that the
secondary star is not detected in the observed spectrum. The
occurrence of the soft X-ray absorption dip requires that the
inclination $i$ of the system exceeds the inclination of the accreting
field line $\zeta$ in the accretion spot. In the optical and X-ray
light curves, the primary accretion spot is visible for 3/4 of the
orbital period. It is self-occulted for a phase interval $\Delta
\phi\!\simeq\!0.25$, implying $i\!>\!50$\degr\ (Eq.~\ref{eq:visi} in
Sect.~\ref{sec:NEL}).  On the other hand, if the deep minima in some
light curves are produced by cyclotron beaming, $i$ cannot be too
large, suggesting $i\!\simeq\!55-60$\degr. When we convert the observed NEL
radial-velocity amplitude of $K_2'\!=\!307$\,\kms\ into $K_2$ with our
irradiation model BR08, the corresponding mass range is
$M_1\!\simeq\!0.75\,-\,0.89$\,\msun. For $i\!=\!75$\degr, just
avoiding an eclipse, the minimum primary mass consistent with the
measured $K_2'$ is $M_1\!\simeq\!0.63$.

The accretion rate based on the 1993 high-state PSPC observation is
$\dot M_\mathrm{x}\!\ga\!\ten{6.3}{-11}$\,\msunyr. When the
cyclotron flux from the high photometric data points in the SED of
Fig.~5 (see Table~\ref{tab:xray} column 13) is included, the required
accretion rate rises to $\dot
M_\mathrm{x+cyc}\!\ga\!\ten{8.7}{-11}$\,\msunyr. Both $\dot M$ values
are quoted as lower limits because they are based on an X-ray fit with a
lower-than-standard interstellar absorbing column density
(Sect. \ref{sec:1002x}).  Correspondingly, for the XMM-Newton
observation in a moderately high or intermediate state complemented by
an estimate of the cyclotron luminosity from the orbital mean of our
spectrophotometry, $\dot M_\mathrm{x+cyc}\!\ga\!\ten{4.7}{-11}$\,\msunyr.
If either one of these rates equals the secular mean, the
corresponding equilibrium temperatures of the WD from compressional
heating would fall between $\ga\!12800$\,K and $\ga\!15000$\,K for a
WD of 0.82\,\msun. The predicted 4600\AA\ flux of the WD in the high
state is 0.014\,mJy, about a factor of two below the lowest MONET flux
(yellow triangle). A spectroscopic temperature measurement and mass
estimate of the WD in a low state appears feasible.

\section{Discussion}
\label{sec:disc}

In this last paper of a series of three, we report results on five
ROSAT-discovered polars collected over three decades.  Papers I and II
\citep{beuermannetal17,beuermannetal20} contained in-depth analyses of
V358\,Aqr and the eclipsing polar HY\,Eri. The results on the present
five objects are less complete, but they are accompanied by accurate
linear ephemerides, which allow the correct phasing of past and future
observations. There is no evidence of a variation in orbital
period or for an asynchronism in any of the five targets, and we
consider them as \emph{\textup{bona fide}} synchronous rotators.

Two of our targets, \jone\ and \jeight, belong to the league of bright
polars that reach 15 or 16 mag. At the other end of the brightness
scale, the distant object \jsix\ resides near the bounce period and
does not appear to exceed 19 mag. \jone\ is the second polar after
VY~For \citep{beuermannetal89}, in which the main active pole appears to
be permanently hidden behind the WD. \citet{cropper97} advocated
polarimetry of VY~For to investigate its accretion geometry. Studying
the brighter system \jone\ may be more profitable.

The evolution of polars differs from that of nonmagnetic CVs.
The concept of reduced magnetic braking of
\citet{lietal94} was devised to explain the effective disappearance of
the period gap for polars and was successfully employed by
\citet{bellonietal20} in their binary stellar evolution code. Polars
show at most a remnant gap, and systems with \po$\,\la\!150$\,min
behave effectively as short-period polars \citep{schwopeetal20}. In
the current sample, this applies to \jeight. There is agreement that
all short period CVs suffer larger angular momentum losses (AML) than
predicted by gravitational radiation alone. \citet{kniggeetal11}
included the additional AML as a numerical scaling factor, while
\citet{bellonietal20} implemented the empirical consequential angular
momentum loss eCAML of \citet{schreiberetal16} in their code. CAML
describes an accretion-related process, for instance, the time-averaged
Bondi-Hoyle-type frictional energy loss that the secondary star
experiences in the nova shells that are expelled at intervals by the WD. Both
authors effectively raised the AML and thereby $\dot M$ in an attempt
to better match observed quantities as the bounce period or the space
density of CVs.

The concept of an 'observed' accretion rate was considered
questionable for a long time, but was placed on firmer ground by (i)
the advent of the Gaia trigonometric distances
\citep{bailerjonesetal18} and (ii) the progress in constructing a
3D extinction map for the solar neighborhood
\citep{lallementetal18}, which helps to derive reliable optical and
X-ray luminosities.  Calculating the accretion rate requires knowledge
of the WD mass. For three of our targets, we measured the radial-velocity amplitude $K_2'$ of the narrow component of the Balmer lines,
which is thought to originate on the irradiated surface of the secondary star.
Converting $K_2'$ into the amplitude $K_2$ of the center of mass of the
secondary star requires knowledge of the inclination $i$ of the
system and an irradiation model for the emission line in question. In
this pilot study, we have applied our irradiation model BR08
\citep{beuermannreinsch08,beuermannetal17} although it was not
optimized for the Balmer lines and \heii. Ways to improve $i$ have
been noted in the relevant subsections. The derived primary masses
depend only weakly on $M_2$ because the stellar models and the
moderate bloating in short-period polars leave little freedom.  All
five systems, however, are open for a more direct determination of
$M_1$ by a spectroscopic measurement of the temperature and radius of
the WD in a low state of the respective system.

For five of the seven targets in Papers I-III of this series, we
measured the field strength in the accretion region spectroscopically.
The mean field strength for the present sample of 30\,MG does not
differ much from the 33.4\,MG of the complete sample of our 27 ROSAT
soft and hard X-ray discovered polars and the 38\,MG of all polars in
Table~2 of \citet{ferrarioetal15}. Measuring the WD temperature and
thereby its radius requires models of magnetic atmospheres. Such
models were calculated by \citet{jordan92}, but significant
uncertainties related to Stark broadening in the presence of
a magnetic field remain and we emphasize again the need for calculations of the
shifts of individual Stark components (see Paper II).

The polars discussed in this paper feature X-ray spectra that consist
of a soft quasi-blackbody and a hard thermal X-ray component. Hard
X-rays originate primarily in the cooling flow of the accretion
stream that develops downstream of the strong shock that is set up above the
stellar surface. Its temperature distribution is determined by the
competition between bremsstrahlung and cyclotron cooling and the
associated radiative transfer
\citep{woelkbeuermann96,fischerbeuermann01}. Soft X-rays originate
from the heated stellar atmosphere of the spot and its
surroundings. Insight into the temperature distribution of the
optically thick soft X-ray emission was obtained from the
high-resolution spectrum of the prototype polar AM~Her measured down
to 92~eV (135\,\AA) with the Low Energy Transmission Grating
Spectrometer (LETGS) on board Chandra \citep{beuermannetal12}. The
analysis revealed the expected spread in temperature and demonstrated
that modeling the soft X-ray component by a single blackbody is a
severe approximation. In the case of AM Her, the single-blackbody
model underestimated the bolometric flux of the XUV and soft X-ray
component by a factor of $3.7\!\pm\!0.7$. We approximately accounted
for this effect by raising the bolometric energy flux obtained for a
single-blackbody fit by a factor $c_\mathrm{sx}\!=\!3$.

\citet{townsleygaensicke09} and \citet{palaetal20} reported reliable
effective temperatures of nine polars and of 42 non-mCVs with
\po$\!<\!2.1$\,h. The mean temperatures and periods are
$12850\!\pm\!570$\,K and 103\,min for the polars and
$14350\!\pm\!370$\,K and 92\,min for the non-mCVs.  According to the theory of
compressional heating of the WD (Eq.~\ref{eq:temp} in
Sect.~\ref{sec:33}), these temperatures translate into long-term mean
accretion rates of $\langle\dot
M_\mathrm{polar}\rangle\!=\!\ten{6.7}{-11}$\msunyr\ and $\langle\dot
M_\mathrm{non-mCV}\rangle\!=\!\ten{1.05}{-10}$\msunyr, respectively,
assuming a WD mass of 0.75\,\msun. The mean accretion rate of the five
polars in Table~\ref{tab:xray} with a mean WD mass of 0.72\,\msun\ and
a mean period of 103\,min is $\langle\dot
M_\mathrm{x+cyc}\rangle\!=\!\ten{6.8}{-11}$\msunyr\ in their normal
high states. We consider the close agreement with $\langle\dot
M_\mathrm{polar}\rangle$ coincidental, but the general
agreement supports our upward correction of the
single-blackbody flux by the factor $c_\mathrm{sx}\!=\!3$.  The mean
temperature of single-blackbody fits for the five targets in this
paper is 39\,eV, which is similar to that of AM~Her,
k$T\mathrm{bb}\!=\!32.3$\,eV. Using a common value of
$c_\mathrm{sx}$ therefore appears plausible, but in general, employing a
multitemperature model is preferable.

Different mean accretion rates for short-period polars and non-mCVs
are currently not predicted by the evolutionary
models. \citet{kniggeetal11} do not distinguish between the subtypes
of CVs. In their best-fit model, they predict $\dot
M\!=\!\ten{7.0}{-11}$\msunyr\ at 100\,min for $M_1\!=\!0.75$\,\msun,
which corresponds rather to the temperature-derived value for polars
than to that for non-mCVs. \citet{bellonietal20} explicitly accounted for the
reduced magnetic braking in polars, but at short orbital periods or
$M_2\!<\!0.20$\,\msun, the fraction $\Phi$ of open field lines of the
secondary star vanishes for a CV with $M_1\!=\!0.75$\,\msun\ and
$B\!=\!30$\,MG \citep[][their Fig.~2]{bellonietal20}.  From that point
on, their code assigns the same AML to polars and non-mCVs and
predicts $\dot M\!\simeq\!\ten{1.0}{-10}$\msunyr\ at 100\,min, which
is correct for non-mCVs, but exceeds $\langle\dot
M_\mathrm{polar}\rangle$. Our targets with known field strengths,
\jeight, \jnine, and \jten,\ lie in the $\Phi\!=\!0$ regime. Hence,
there is evidence that short-period polars and non-mCVs evolve
differently, but the physical cause remains elusive so far.

The question of the energy balance between the soft and hard X-ray
emission of polars was intensely debated at the time when the ROSAT
and XMM-Newton soft X-ray measurements became available
\citep[e.g.,][]{ramsayetal94,beuermannburwitz95,ramsaycropper04a}. The
background was the prediction by \citet{kinglasota79} and
\citet{lambmasters79} that one-half of the accretion energy escaped as
hard X-rays and the other half was intercepted by the photosphere of the WD
and reprocessed into soft X-rays. The high sensitivity of ROSAT for
soft X-rays and its very limited hard X-ray response led to the notion
that intense soft X-ray emission was one of the hallmarks of polars,
and the luminosity ratio of soft to hard X-rays seemed to exceed
unity by a large factor. The theoretical description of shocks buried
in the photosphere provided an understanding of the dominance of soft
X-rays \citep{kuijperspringle82}. The internal absorption of
hard X-rays \citep{ramsaycropper04a} showed that the hard X-ray flux
had been severely underestimated and that cyclotron radiation needs to be
included in the energy balance of the post-shock cooling
flow. Furthermore, the case of AM~Her showed that the soft X-ray
luminosity had been underestimated as well. Taking all these caveats into
account, the best estimate for AM~Her is
$L_\mathrm{sx}/(L_\mathrm{hx}+L_\mathrm{cyc})\!=\!4.3\!\pm\!2.0$
\citep{beuermannetal12}, and the consensus is that the luminosity ratio
exceeds unity only in a small number of polars in their high-accretion
states \citep{ramsaycropper04b}. The results for the present five
sources in column 16 of Table~\ref{tab:xray} support this conclusion.
When the high-state values are selected, the mean luminosity ratio is
$L_\mathrm{sx}/(L_\mathrm{hx}+L_\mathrm{cyc})\!=\!2.2\!\pm\!1.1$. Furthermore, at
the reduced accretion rate in intermediate and low states the temperature of the heated surrounding of the spot
drops and the emission moves out of the soft X-ray band, causing the
system to turn into a (apparently) hard X-ray source
\citep{ramsayetal04,beuermannetal08,schwopeetal20}.

We have analyzed five previously neglected polars and
derived accretion rates based primarily on ROSAT PSPC data.
A modern instrument such as eRosita \citep{predehletal20} will
provide a more adequate data base for an up-to-date study of a larger
population of CVs.

\begin{acknowledgements}
    
  We thank the anonymous referee for a constructive and helpful report
  that improved the presentation. Our fifth author, Hans-Christoph
  Thomas, analyzed a large part of the early data before his untimely
  death.  Most of the more recent photometric data were collected with
  the MONET telescopes of the Monitoring Network of Telescopes, funded
  by the Alfried Krupp von Bohlen und Halbach Foundation, Essen, and
  operated by the Georg-August-Universit\"at G\"ottingen, the McDonald
  Observatory of the University of Texas at Austin, and the South
  African Astronomical Observatory.  The spectroscopic and part of the
  photometric observations were made at the European Southern
  Observatory La Silla, Chile, and the Calar Alto Observatory, Spain.
  We made use of the Sloan Digital Sky Survey (SDSS), the Two Micron
  All Sky Survey (2MASS), the Wide-field Infrared Survey Explorer
  (WISE), the Galaxy Evolution explorer (GALEX), the PanSTARRS data
  base, and further sources accessed via the VizieR Photometric viewer
  operated at CDS, Strasbourg, France. We thank Jan Kurpas of the
  Astronomisches Institut Potsdam for producing the finding charts.

\end{acknowledgements}

\bibliographystyle{aa}

\clearpage

\appendix
\section{ROSAT count rates and energy fluxes} 
\label{sec:A}

\vspace{-28mm}
The polars discussed in this paper were discovered as soft high
galactic latitude X-ray sources in the ROSAT All Sky Survey
(RASS). The survey was performed with the position-sensitive
proportional counter (PSPC) as detector, which possesses moderate
energy resolution with an FWHM of 0.4\,keV at 1\,keV photon energy.
Subsequent pointed ROSAT observations were made with either the PSPC
or the High Resolution Imager (HRI). The HRI lacked energy resolution.
For quick reference, examples of an integrated PSPC count rate of
1.0\,\cps\ converted into an HRI count rate are given in
Table~\ref{tab:crrat} for selected incident blackbody spectra and
thermal APEC spectra absorbed by a column density \nh\ of neutral
matter of solar composition. They were calculated with the NASA HEASARC tool
PIMMS\footnote{https://heasarc.gsfc.nasa.gov/cgi-bin/Tools/w3pimms/w3pimms.pl}. 
For the blackbodies absorbed by
\nh$\,\la\!3\!\times\!10^{20}$\,\atoms, the PSPC is typically more
sensitive than the HRI by a factor \mbox{$7\!\pm\!2$} and for the APEC
spectra by about a factor of three.
 
\vspace{-28mm}
In Tabel~\ref{tab:flux} we list the energy fluxes for incident
blackbody spectra of three temperatures and column densities each. The
first line quotes the energy flux of the absorbed spectrum incident on
the detector in the ROSAT band of $0.10\!-\!2.0$\,keV for 1.0~PSPC
\cps. The unabsorbed flux is quoted in the second and the bolometric
flux in the third line. For a temperature \ktbbi$\,=\!60$\,eV and a low
column density, the PSPC measures an energy flux that is not far from
the bolometric flux. For 20\,eV, however, it recovers only a small
fraction of the flux and the corrections become very large.

\pagebreak

\begin{table}[t]
\begin{flushleft}
  \caption{HRI count rates equivalent to 1 PSPC\,\cps\ for blackbody
    and thermal APEC spectra with the quoted temperatures and neutral
    absorbers \nh\ in \ergs for 1 PSPC\,\cps.}
\begin{tabular}{@{\hspace{0.0mm}}c@{\hspace{3.0mm}}c@{\hspace{3.0mm}}c@{\hspace{3.0mm}}c@{\hspace{5.0mm}}c@{\hspace{1.0mm}}c@{\hspace{1.0mm}}c@{\hspace{1.0mm}}c}\\[-5ex]
  \hline\hline \\[-1.5ex]
  \nh     & \multicolumn{3}{c}{Blackbody} & \multicolumn{4}{c}{APEC} \\[0.5ex]
  (cm$^{-2}$)& 20\,eV& 40\,eV& 60\,eV& 1.0\,keV & 3.1\,keV & 9.7\,keV & 27.3\,keV \\[0.5ex]
  \hline\\[-1ex]
  1e19      & 0.112 & 0.127 & 0.143 & 0.319 & 0.242 & 0.270 & 0.271 \\
  1e20      & 0.117 & 0.138 & 0.162 & 0.362 & 0.315 & 0.310 & 0.313 \\
  3e20      & 0.129 & 0.163 & 0.203 & 0.392 & 0.360 & 0.352 & 0.352 \\
  1e21      & 0.147 & 0.280 & 0.379 & 0.407 & 0.385 & 0.375 & 0.373 \\
  1e22      & 0.494 & 0.350 & 0.377 & 0.388 & 0.362 & 0.346 & 0.346 \\[1.0ex]
  \hline\\
\end{tabular}\\[-1.0ex]
\label{tab:crrat}
\end{flushleft}

\vspace{-5mm}
\end{table}

\begin{table}[t]
\begin{flushleft}
  \caption{Blackbody energy fluxes in units of $10^{-12}\,$\ergs\ for
    three values each of \ktbbi\ and neutral absorber \nh\ in \atoms\
    normalized to produce 1\,PSPC \cps. The three lines refer to the
    absorbed and unabsorbed fluxes in the interval $0.1\!-\!2$\,keV
    and the bolometric flux.}
\begin{tabular}{@{\hspace{1mm}}l@{\hspace{3mm}}r@{\hspace{1.5mm}}r@{\hspace{1.5mm}}r@{\hspace{2mm}}r@{\hspace{1.5mm}}r@{\hspace{1.5mm}}r@{\hspace{2mm}}r@{\hspace{1.5mm}}r@{\hspace{1.5mm}}r}\\[-2ex]
  \hline\hline \\[-1.5ex]
Energy&\multicolumn{3}{c}{\ktbbi$\,=\,$20\,eV}&\multicolumn{3}{c}{\ktbb$\,=\,$40\,eV}&\multicolumn{3}{c}{\ktbb$\,=\,$60\,eV}\\[0.5ex]
Range &1e19&1e20&3e20&1e19&1e20&3e20&1e19&1e20&3e20\\
  \hline\\[-1ex]
Absorbed    &  3.9 &   2.4 &   2.2 &  2.7 &  2.6 &  3.4 &  3.3 &  3.8 &  6.1 \\
Unabsorbed  &  5.3 &  24.3 &  88.2 &  3.3 &  8.1 & 32.0 &  3.6 &  7.6 & 24.1 \\
Bolometric  & 21.4 &  97.4 & 217.0 &  4.6 & 11.3 & 44.6 &  4.1 &  8.9 & 27.6 \\[1.0ex]
\hline\\[-1.0ex]
\end{tabular}
\label{tab:flux}
\end{flushleft}

\vspace{-20mm}
\end{table}

\section{Finding charts in Sloan g} 
\label{sec:B}

\begin{figure}[h]
\includegraphics[height=80.0mm,angle=0,clip]{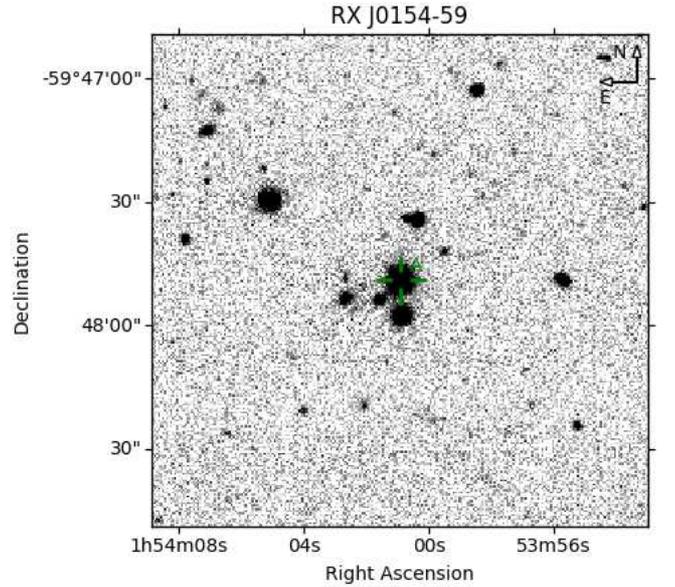}
\caption[chart]{Finding chart for \rxjone. Size is
  2\arcmin $\times $2\arcmin. The comparison star is 2\arcsec\ E and
  256\arcsec\ N of the target and outside the image.}
\label{fig:fc0154}
\vspace{-5mm}
\end{figure}

\begin{figure*}[bht]
\includegraphics[height=80.0mm,angle=0,clip]{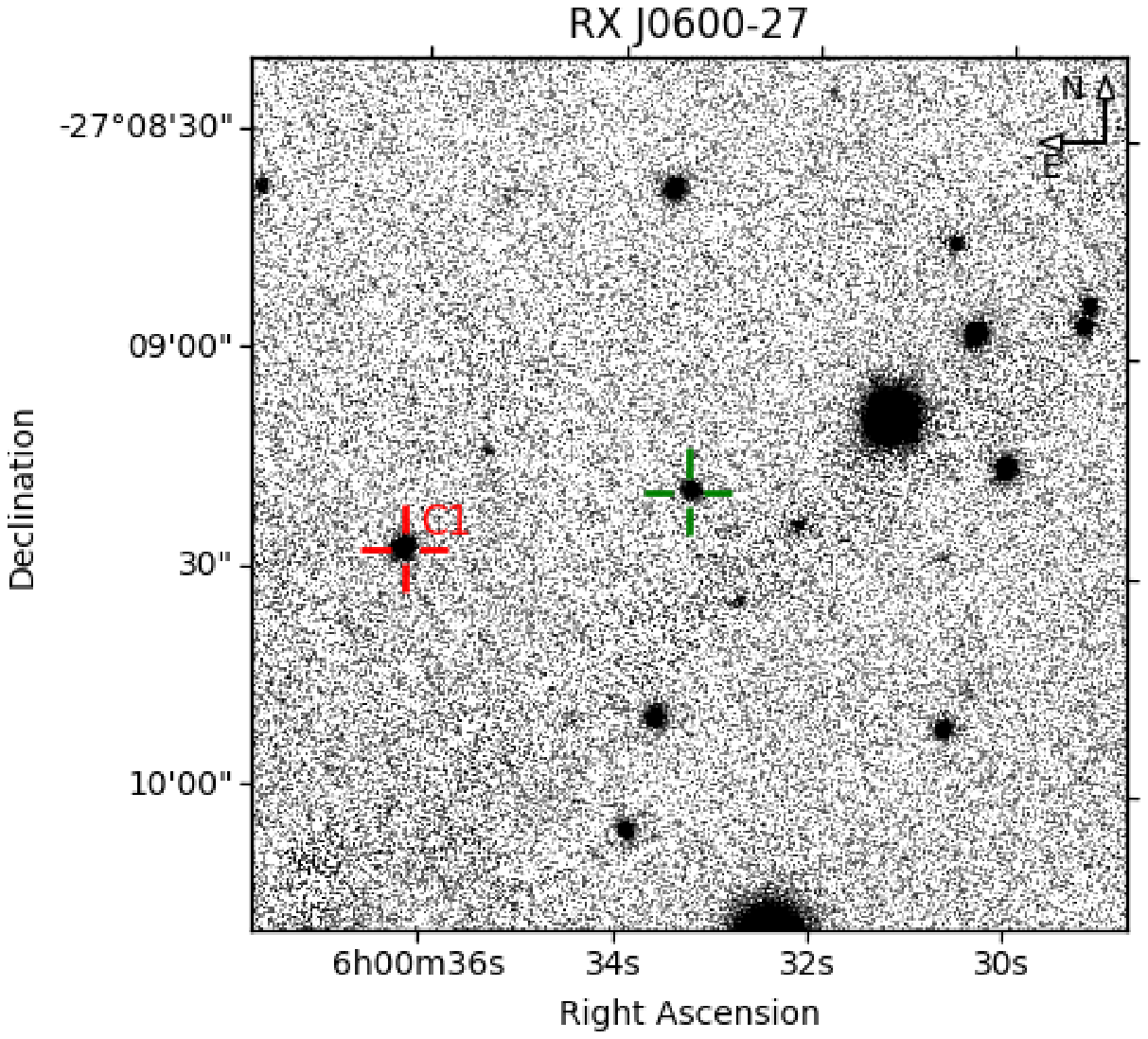}
\hspace{3mm}
\includegraphics[height=80.0mm,angle=0,clip]{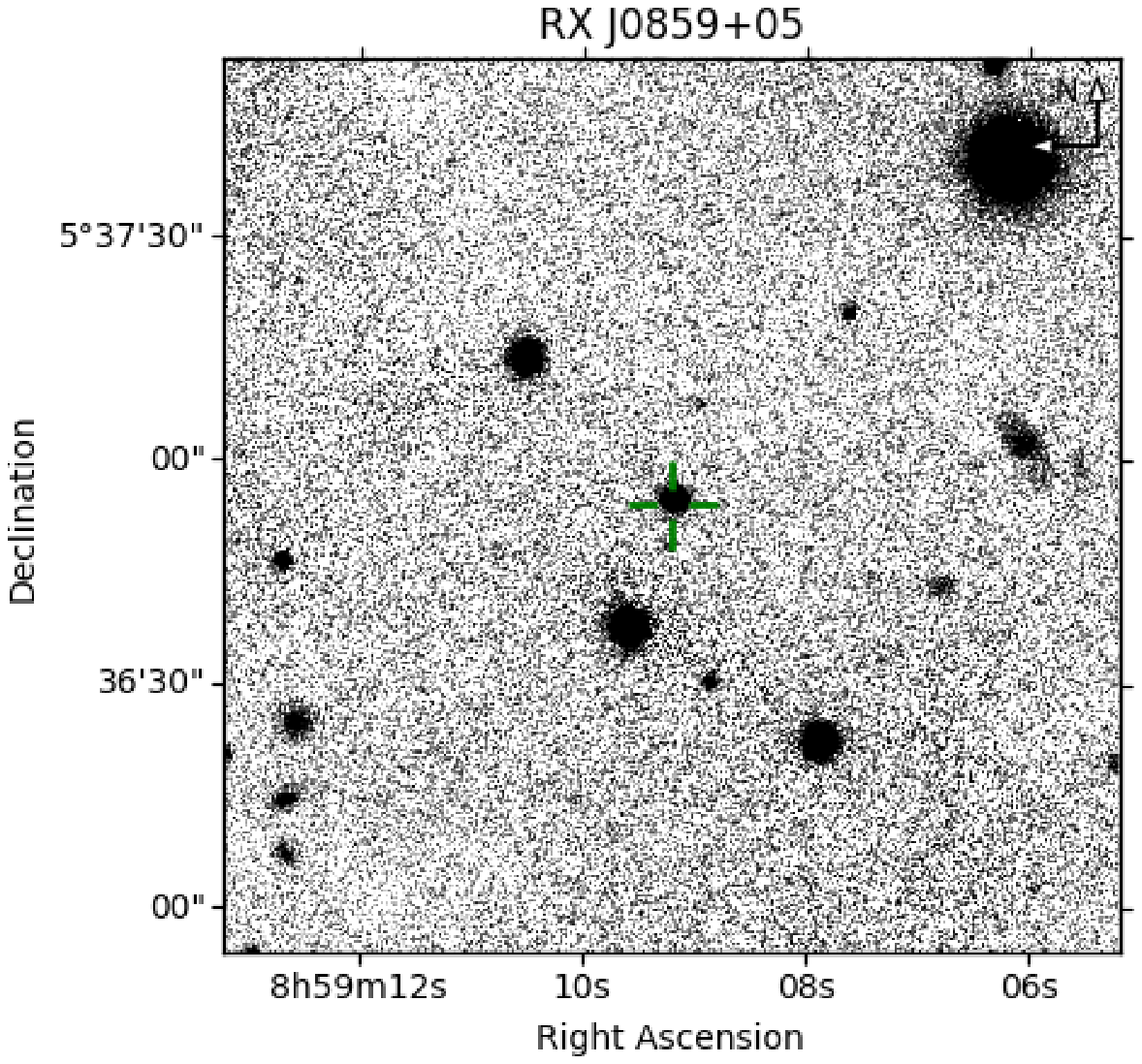}
\caption[chart]{{\emph Left: }Finding chart for \rxjsix. Size is
  2\arcmin $\times $2\arcmin. The comparison star is 40\arcsec\ E and
  8\arcsec\ S of the target and marked C1. \mbox{{\emph Right: } Finding
  chart for \rxjeight. Size is 2\arcmin $\times $2\arcmin. The
  comparison star is 15\arcsec\ E and 162\arcsec\ N of the target and
  outside the image.}}
\label{fig:fc0600}
\end{figure*}

\begin{figure*}[t]
\includegraphics[height=80.0mm,angle=0,clip]{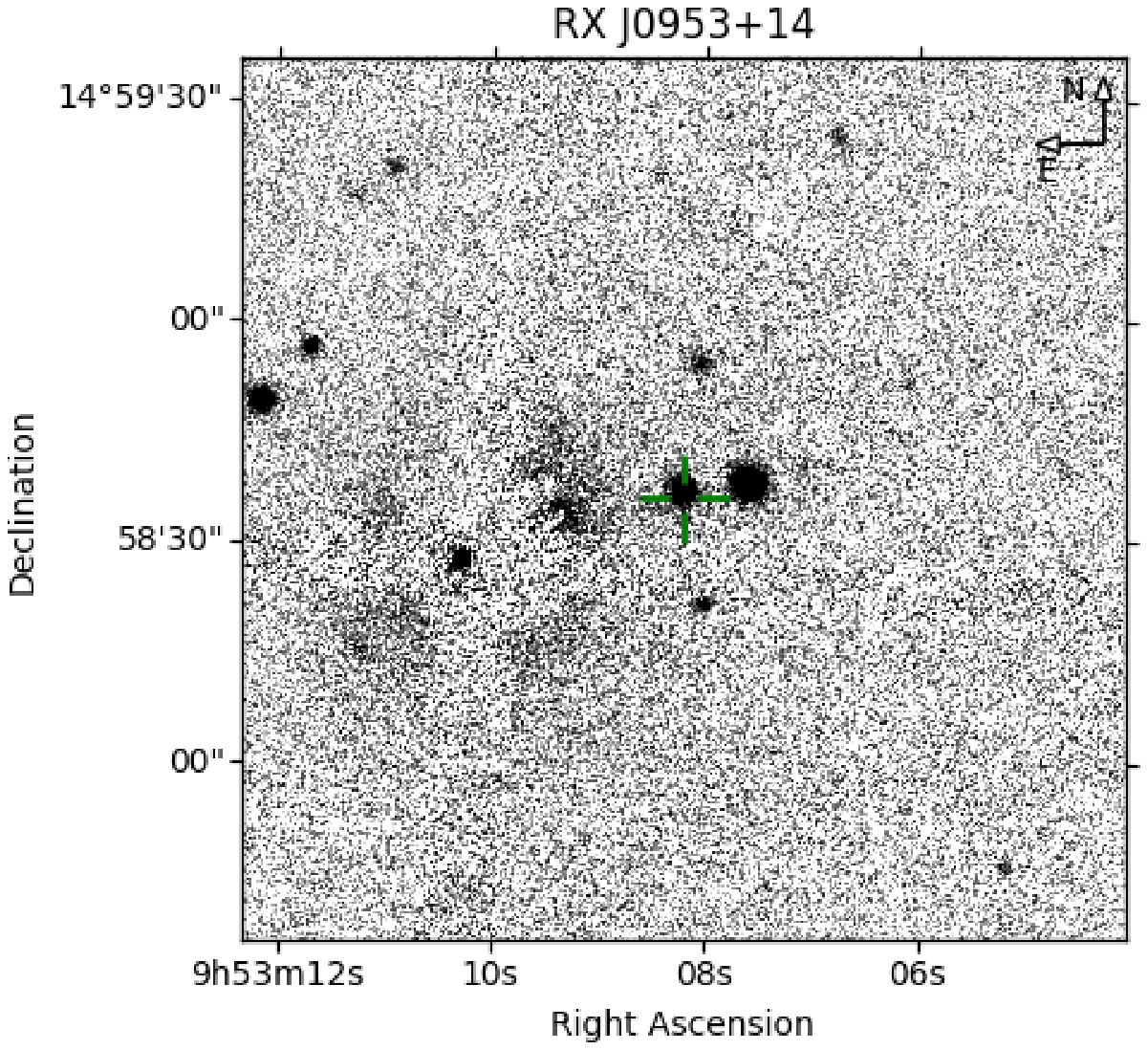}
\hspace{3mm}
\includegraphics[height=80.0mm,angle=0,clip]{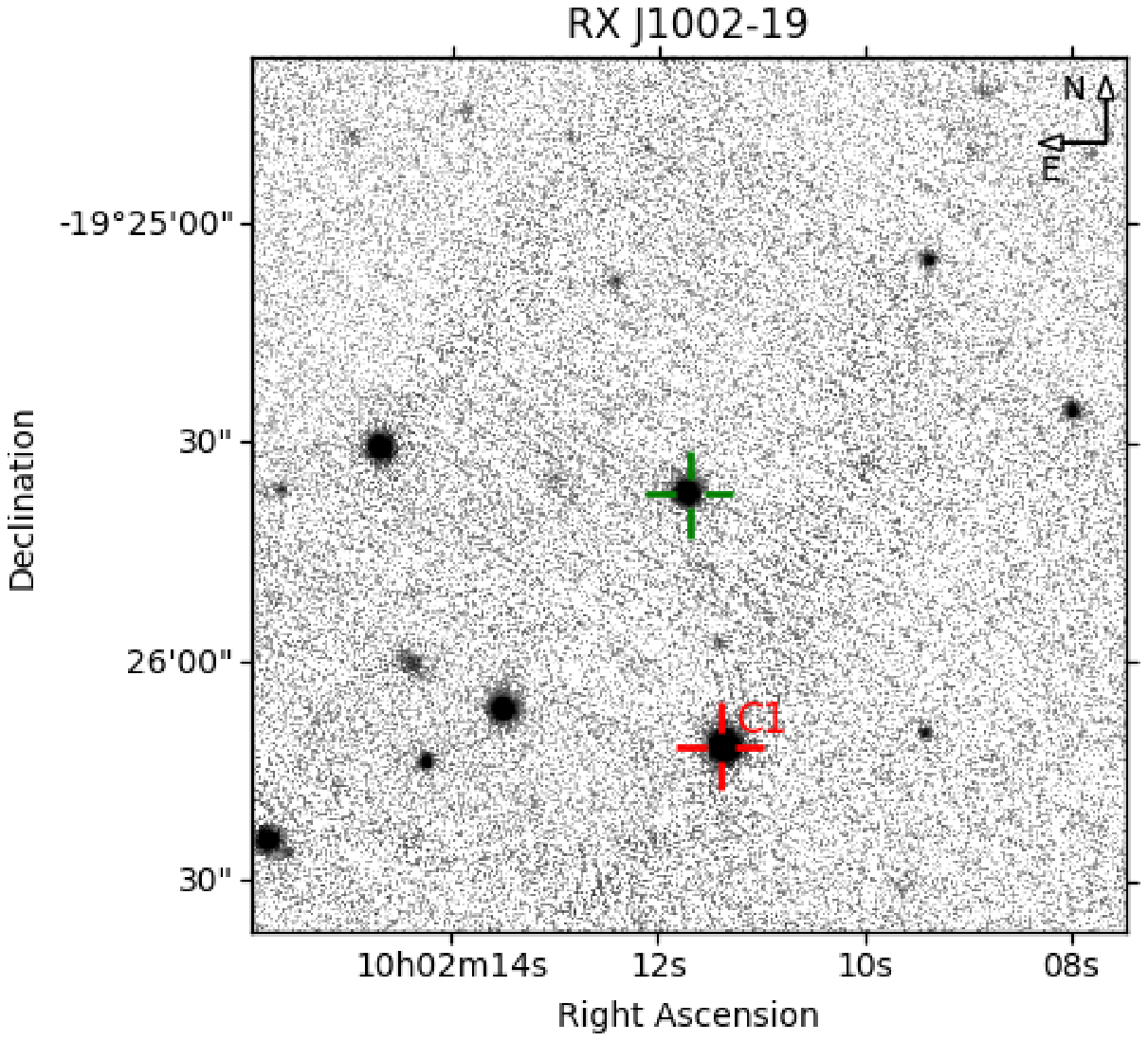}
\caption[chart]{{\emph Left: } Finding chart for \rxjnine. Size is
  2\arcmin $\times $2\arcmin. The comparison star is 9\arcsec\ W and
  101\arcsec\ S of the target and outside the image. {\emph Right: }
  Finding chart for \rxjten. Size is 2\arcmin $\times $2\arcmin. The
  comparison star is 5\arcsec\ W and 36\arcsec\ S of the target and
  marked C1.}
\label{fig:fc1002}
\end{figure*}

\pagebreak

\section{Observed times in BJD (TDB)} 
\label{sec:C}

\begin{table}[hb]
\begin{flushleft}
  \caption{Observed times of optical minima of \rxjone\ transformed from UTC to BJD (TDB).}
\begin{tabular}{@{\hspace{2.0mm}}r@{\hspace{2.0mm}}c@{\hspace{4.0mm}}c@{\hspace{2.0mm}}r@{\hspace{2.0mm}}r@{\hspace{3.0mm}}c@{\hspace{3.0mm}}c@{\hspace{3.0mm}}c}\\[-3ex]
\hline\hline \\[-1.5ex]
 Cycle  & BJD(TDB)     &Error&$O\!-\!C$&Expos& Band & Instr.\\
        & 2400000+     &(min)& (min)&(min)&         &     \\[0.5ex]    
\hline\\[-1ex]                                                               
-129778 & 49247.672100 &  1.4 &  1.8& 2.5 & V       & (1) \\
-129777 & 49247.732600 &  1.4 &  0.0& 2.5 & V       & (1) \\
-129776 & 49247.795000 &  1.4 &  0.9& 2.5 & V       & (1) \\
-128306 & 49338.592000 &  4.3 & -0.0& 8.0 & Spec    & (2) \\
-119106 & 49906.855400 &  1.4 & -1.1& 2.0 & V       & (3) \\
-119105 & 49906.917700 &  1.4 & -0.3& 2.0 & V       & (3) \\
-119090 & 49907.844570 &  1.4 &  0.2& 2.0 & V       & (3) \\
-119089 & 49907.906600 &  1.4 &  0.6& 2.0 & V       & (3) \\
-116841 & 50046.760000 &  1.4 & -0.1& 5.0 & Spec    & (2) \\
      0 & 57263.764246 &  0.7 & -0.7& 2.0 & Sloan r & (4) \\
      1 & 57263.825894 &  0.7 & -0.9& 2.0 & Sloan r & (4) \\
      2 & 57263.888474 &  0.7 &  0.3& 2.0 & Sloan r & (4) \\
     32 & 57265.740385 &  0.7 & -1.3& 1.5 & Sloan r & (4) \\
     49 & 57266.790959 &  0.7 & -0.6& 1.5 & Sloan r & (4) \\
     50 & 57266.852630 &  0.7 & -0.7& 1.5 & Sloan r & (4) \\
     82 & 57268.830126 &  0.7 &  0.6& 1.5 & Sloan r & (4) \\
     83 & 57268.892077 &  0.7 &  0.9& 1.5 & Sloan r & (4) \\
   4434 & 57537.642297 &  0.7 & -0.9& 1.0 & U       & (5) \\
   4514 & 57542.584319 &  0.7 & -0.0& 1.0 & Sloan g & (5) \\
   4515 & 57542.646022 &  0.7 & -0.1& 1.0 & Sloan g & (5) \\
   4580 & 57546.661802 &  1.0 &  1.2& 1.0 & WL      & (5) \\
   4595 & 57547.586244 &  0.9 & -1.8& 1.0 & WL      & (5) \\
   4596 & 57547.648947 &  0.9 & -0.5& 1.0 & WL      & (5) \\
   4675 & 57552.529255 &  0.9 &  0.5& 1.0 & WL      & (5) \\
   4676 & 57552.590257 &  0.9 & -0.6& 1.0 & WL      & (5) \\
   4774 & 57558.644130 &  0.9 &  0.3& 1.0 & WL      & (5) \\
  11909 & 57999.356810 &  1.0 &  0.1& 1.0 & Sloan g & (5) \\
  12105 & 58011.463446 &  1.0 &  0.3& 1.0 & WL      & (5) \\
  12168 & 58015.354383 &  1.0 & -0.3& 1.0 & WL      & (5) \\
  12169 & 58015.416603 &  1.0 &  0.3& 1.0 & WL      & (5) \\
  12170 & 58015.477993 &  1.0 & -0.2& 1.0 & WL      & (5) \\
  13544 & 58100.347313 &  0.7 &  0.4& 1.0 & WL      & (5) \\
  13546 & 58100.470187 &  0.7 & -0.5& 1.0 & WL      & (5) \\
  17515 & 58345.626916 &  1.0 &  0.3& 1.0 & WL      & (5) \\
  18273 & 58392.447349 &  0.9 &  1.0& 1.0 & WL      & (5) \\
  18321 & 58395.411922 &  0.9 &  0.6& 1.0 & WL      & (5) \\
  18336 & 58396.338846 &  0.9 &  1.2& 1.0 & WL      & (5) \\
  18337 & 58396.399895 &  0.9 &  0.1& 1.0 & WL      & (5) \\
  18352 & 58397.327088 &  0.6 &  1.1& 1.0 & WL      & (5) \\ [1.0ex]
 \hline\\                                                                       
\end{tabular}\\[-1.0ex]
\footnotesize{(1) ESO/Dutch 0.9\,m, (2) MPI/ESO\,2.2m, EFOSC\,2, (3) ESO/Danish 2.5\,m, 
  (4) ESO\,2.2\,m, GROND, (5) SAAO MONET/S 1.2\,m.}\\[-3.0ex]
\label{tab:0154}
\end{flushleft}

\vspace{2.5mm}
\end{table}

\vspace{0mm}
\begin{table}[b]
\begin{flushleft}
  \caption{Observed times of optical maxima ($m\!=\!+1$) and minima
    ($m\!=\!-1$) for \rxjsix\ transformed from UTC to BJD(TDB). A
    maximum and the following minimum have been assigned the same
    cycle number. The maxima define $\phi\!=\!0$ in
    Eq.~\ref{eq:0600ephem}, the minima occur on average at
    $\phi\!=\!0.55$. There may be a cycle count error by one orbit at
    $E\!=\!-151194$.}
\begin{tabular}{@{\hspace{2.0mm}}r@{\hspace{2.0mm}}c@{\hspace{2.0mm}}c@{\hspace{2.0mm}}r@{\hspace{2.0mm}}r@{\hspace{3.0mm}}c@{\hspace{3.0mm}}c@{\hspace{3.0mm}}c}\\[-5ex]
\hline\hline \\[-1.5ex]
 Cycle  & BJD(TDB)        & $m$ &Error &$O\!-\!C$& Expos   & Band    & Instr.\\
        & 2400000+        &     &(min)  & (min)  & (min)   &         &     \\[0.5ex]    
\hline\\[-1ex]                                                               
-151194 & 49753.5883336 & $+1$ &   3.0 &  -0.5 &    5.0  & V       & (1) \\
-151194 & 49753.6201522 & $-1$ &   3.0 &   1.6 &    5.0  & V       & (1) \\
      0 & 58015.5526565 & $+1$ &   3.5 &  -1.0 &    1.0  & WL      & (2) \\
      0 & 58015.5805583 & $-1$ &   3.5 &  -4.5 &    1.0  & WL      & (2) \\
      1 & 58015.6067600 & $+1$ &   3.5 &  -1.8 &    1.0  & WL      & (2) \\[1.0ex]
\hline\\[-5ex]
\end{tabular}
\end{flushleft}
\end{table}

\begin{table}[thb]
\begin{flushleft}
\vspace{14.5mm}
\begin{tabular}{@{\hspace{2.0mm}}r@{\hspace{2.0mm}}c@{\hspace{2.0mm}}c@{\hspace{2.0mm}}r@{\hspace{2.0mm}}r@{\hspace{3.0mm}}c@{\hspace{3.0mm}}c@{\hspace{3.0mm}}c}\\[-5ex]
\hline\hline \\[-1.5ex]
 Cycle  & BJD(TDB)        & $m$ &Error &$O\!-\!C$& Expos   & Band    & Instr.\\
        & 2400000+        &     &(min)  & (min)  & (min)   &         &     \\[0.5ex]    
\hline\\[-1ex]                                                               
    145 & 58023.4810547 & $+1$ &   3.8 &   6.0 &    1.0  & WL      & (2) \\
    237 & 58028.5352205 & $-1$ &   3.5 &   1.0 &    1.0  & WL      & (2) \\
    238 & 58028.5556718 & $+1$ &   3.5 &  -4.5 &    1.0  & WL      & (2) \\
    238 & 58028.5917742 & $-1$ &   3.5 &   3.8 &    1.0  & WL      & (2) \\
    239 & 58028.6126754 & $+1$ &   3.5 &  -1.1 &    1.0  & WL      & (2) \\
    457 & 58040.5578307 & $-1$ &   4.0 &   2.1 &    1.0  & WL      & (2) \\
    458 & 58040.5808020 & $+1$ &   3.5 &   0.2 &    1.0  & WL      & (2) \\
    529 & 58044.4636276 & $+1$ &   5.8 &   4.6 &    1.0  & WL      & (2) \\
   1335 & 58088.5041883 & $+1$ &   5.8 &   0.0 &    1.0  & WL      & (2) \\
   1353 & 58089.5192909 & $-1$ &   2.8 &   1.7 &    1.0  & WL      & (2) \\
   1591 & 58102.5223240 & $-1$ &   2.3 &  -1.8 &    1.0  & WL      & (2) \\
   1592 & 58102.5501542 & $+1$ &   2.3 &   3.3 &    1.0  & WL      & (2) \\
   1609 & 58103.5055411 & $-1$ &   2.3 &  -2.4 &    1.0  & WL      & (2) \\
   1610 & 58103.5334112 & $+1$ &   3.5 &   2.8 &    1.0  & WL      & (2) \\
   1682 & 58107.4641785 & $+1$ &   2.3 &  -2.5 &    1.0  & WL      & (2) \\
   1682 & 58107.4964786 & $-1$ &   2.3 &   0.3 &    1.0  & WL      & (2) \\
   1683 & 58107.5199587 & $+1$ &   2.3 &  -0.8 &    1.0  & WL      & (2) \\
   1683 & 58107.5511187 & $-1$ &   2.3 &   0.3 &    1.0  & WL      & (2) \\
   1684 & 58107.5745987 & $+1$ &   2.3 &  -0.9 &    1.0  & WL      & (2) \\
   1734 & 58110.3364302 & $-1$ &   2.3 &  -1.9 &    1.0  & WL      & (2) \\
   1735 & 58110.3623902 & $+1$ &   2.3 &   0.5 &    1.0  & WL      & (2) \\
   1735 & 58110.3945801 & $-1$ &   2.3 &   3.1 &    1.0  & WL      & (2) \\
   1736 & 58110.4444301 & $-1$ &   3.5 &  -3.8 &    1.0  & WL      & (2) \\
   1737 & 58110.4682301 & $+1$ &   2.3 &  -4.5 &    1.0  & WL      & (2) \\
   1737 & 58110.4993300 & $-1$ &   2.3 &  -3.4 &    1.0  & WL      & (2) \\
   1738 & 58110.5268300 & $+1$ &   2.3 &   1.2 &    1.0  & WL      & (2) \\
   1755 & 58111.4844283 & $-1$ &   2.3 &  -1.3 &    1.0  & WL      & (2) \\
   1810 & 58114.4585864 & $+1$ &   2.3 &  -2.7 &    1.0  & WL      & (2) \\
   1810 & 58114.4918062 & $-1$ &   2.3 &   1.5 &    1.0  & WL      & (2) \\
   1811 & 58114.5129561 & $+1$ &   3.5 &  -3.1 &    1.0  & WL      & (2) \\
   1811 & 58114.5445258 & $-1$ &   2.3 &  -1.3 &    1.0  & WL      & (2) \\
   1812 & 58114.5679957 & $+1$ &   2.3 &  -2.5 &    1.0  & WL      & (2) \\
   2814 & 58169.3232900 & $+1$ &   2.3 &  -0.7 &    1.0  & WL      & (2) \\
   2816 & 58169.4332140 & $+1$ &   2.3 &   0.2 &    1.0  & WL      & (2) \\
   3929 & 58230.2839183 & $-1$ &   2.3 &   1.2 &    1.0  & WL      & (2) \\
   3947 & 58231.2688652 & $-1$ &   2.3 &   3.1 &    1.0  & WL      & (2) \\
   6058 & 58346.5885790 & $+1$ &   3.5 &  -4.2 &    1.0  & WL      & (2) \\
   6058 & 58346.6193800 & $-1$ &   3.5 &  -3.6 &    1.0  & WL      & (2) \\
   7483 & 58424.4891501 & $-1$ &   1.7 &  -2.2 &    1.0  & WL      & (2) \\
   7484 & 58424.5162715 & $+1$ &   1.7 &   1.9 &    1.0  & WL      & (2) \\
   7631 & 58432.5453950 & $+1$ &   1.7 &  -3.4 &    1.0  & WL      & (2) \\
   7631 & 58432.5775760 & $-1$ &   1.7 &  -0.8 &    1.0  & WL      & (2) \\
   7722 & 58437.5236975 & $+1$ &   2.3 &   4.7 &    1.0  & WL      & (2) \\
   7722 & 58437.5493075 & $-1$ &   2.3 &  -2.1 &    1.0  & WL      & (2) \\
   8523 & 58481.3225700 & $-1$ &   1.2 &   1.9 &    1.0  & WL      & (2) \\
   8524 & 58481.3450900 & $+1$ &   1.2 &  -0.7 &    1.0  & WL      & (2) \\
   8524 & 58481.3774100 & $-1$ &   1.2 &   2.2 &    1.0  & WL      & (2) \\
   8596 & 58485.3102370 & $-1$ &   1.2 &  -0.1 &    1.0  & WL      & (2) \\
   8597 & 58485.3346220 & $+1$ &   1.2 &  -0.0 &    1.0  & WL      & (2) \\
   8597 & 58485.3655300 & $-1$ &   1.2 &   0.8 &    1.0  & WL      & (2) \\
   8765 & 58494.5160935 & $+1$ &   1.2 &   1.6 &    1.0  & WL      & (2) \\
   8765 & 58494.5440108 & $-1$ &   1.2 &  -1.9 &    1.0  & WL      & (2) \\
   8766 & 58494.5707000 & $+1$ &   1.2 &   1.6 &    1.0  & WL      & (2) \\
   8819 & 58497.4970370 & $-1$ &   1.2 &   1.3 &    1.0  & WL      & (2) \\
   8820 & 58497.5206460 & $+1$ &   1.2 &   0.3 &    1.0  & WL      & (2) \\
   8820 & 58497.5517050 & $-1$ &   1.2 &   1.3 &    1.0  & WL      & (2) \\
   9127 & 58514.3250900 & $-1$ &   1.7 &  -2.4 &    1.0  & WL      & (2) \\
   9128 & 58514.3493900 & $+1$ &   1.7 &  -2.4 &    1.0  & WL      & (2) \\
   9128 & 58514.3799800 & $-1$ &   1.7 &  -2.0 &    1.0  & WL      & (2) \\[1.0ex]
 \hline\\                                                                       
\end{tabular}\\[-1.0ex]
\footnotesize{(1) ESO/Dutch 0.9\,m, (2) SAAO MONET/S 1.2\,m}\\[-3.0ex]
\label{tab:0600}
\end{flushleft}

\vspace{-0mm}
\end{table}

\begin{table}[b]
\begin{flushleft}
  \caption{Observed times of optical minima for \rxjeight\ transformed from UTC
    to BJD(TDB).}
\begin{tabular}{@{\hspace{2.0mm}}r@{\hspace{2.0mm}}c@{\hspace{2.0mm}}c@{\hspace{2.0mm}}r@{\hspace{2.0mm}}r@{\hspace{3.0mm}}c@{\hspace{3.0mm}}c@{\hspace{3.0mm}}c}\\[-3ex]
\hline\hline \\[-1.5ex]
 Cycle & BJD(TDB)   & Error &$O\!-\!C$& Expos &Band & Instr.\\
       & 2400000+   & (min) & (min)   & (min) &     &       \\[0.5ex]    
\hline\\[-1ex]
-51517 & 50097.645900 &   2.9 &   1.3 &  0.5  & V   & (1) \\
-51516 & 50097.744854 &   1.0 &  -0.1 &  0.5  & V   & (1) \\
-51515 & 50097.843176 &   2.9 &  -2.4 &  0.5  & V   & (1) \\
     0 & 55246.837490 &   3.6 &   0.3 &  1.0  & WL  & (2) \\
  2643 & 55511.009128 &   1.8 &   0.7 &  1.0  & WL  & (2) \\
  2662 & 55512.908370 &   2.3 &   1.0 &  1.0  & WL  & (2) \\
  2663 & 55513.008100 &   2.7 &   0.7 &  1.0  & WL  & (2) \\
  2722 & 55518.904510 &   4.5 &  -0.4 &  1.0  & WL  & (2) \\
  2723 & 55519.006940 &   2.9 &   3.2 &  1.0  & WL  & (2) \\
  2841 & 55530.798980 &   1.4 &  -0.0 &  1.0  & WL  & (2) \\
  2842 & 55530.898890 &   1.4 &  -0.1 &  1.0  & WL  & (2) \\
  2872 & 55533.895720 &   2.7 &  -2.5 &  1.0  & WL  & (2) \\
  2962 & 55542.893854 &   2.7 &   1.1 &  1.0  & WL  & (2) \\
  3132 & 55559.885170 &   1.4 &   0.5 &  1.0  & WL  & (2) \\
  4061 & 55652.740434 &   2.3 &   1.2 &  1.0  & WL  & (2) \\
  6454 & 55891.925500 &   2.9 &   3.4 &  1.0  & WL  & (2) \\
 16989 & 56944.910430 &   2.9 &   0.2 &  1.0  & WL  & (2) \\
 18028 & 57048.758480 &   1.4 &  -1.8 &  1.0  & WL  & (2) \\
 18058 & 57051.757280 &   1.4 &  -1.4 &  1.0  & WL  & (2) \\
 29250 & 58170.412259 &   1.7 &  -1.7 &  1.0  & WL  & (3) \\
 29369 & 58182.306969 &   1.4 &  -1.0 &  1.0  & WL  & (3) \\
 32282 & 58473.466899 &   1.4 &   1.5 &  1.0  & WL  & (3) \\
 32552 & 58500.452806 &   1.4 &   0.1 &  1.0  & WL  & (3) \\
 32562 & 58501.453029 &   1.4 &   1.1 &  1.0  & WL  & (3) \\
 \hline\\                                                                       
\end{tabular}\\[-1.0ex]
\footnotesize{(1) ESO/Dutch 0.9\,m, (2) McDonald Observatory MONET/N 1.2\,m,  (2) SAAO MONET/S 1.2\,m.}\\[-3.0ex]
\label{tab:0859}                                                               
\end{flushleft}

\vspace{5mm}
\end{table}

\begin{table}[b]
\begin{flushleft}
  \caption{Observed times of the center of the optical bright phase for
    \rxjnine\ transformed from UTC to BJD(TDB).}
\begin{tabular}{@{\hspace{2.0mm}}r@{\hspace{2.0mm}}c@{\hspace{2.0mm}}c@{\hspace{2.0mm}}r@{\hspace{2.0mm}}r@{\hspace{3.0mm}}c@{\hspace{3.0mm}}c@{\hspace{3.0mm}}c}\\[-3ex]
\hline\hline \\[-1.5ex]
 Cycle & BJD(TDB)      & Error  &$O\!-\!C$& Expos &Band & Instr.\\
       & 2400000+      & (min)  & (min)   & (min) &     &       \\[0.5ex]    
\hline\\[-1ex]
-75884 &  49752.612917 &   0.9 &  -0.4 &   4.0   & V   & (1) \\
-40714 &  52285.641017 &   0.4 &  -0.0 &   1.0   & WL  & (2) \\
-40713 &  52285.713156 &   0.4 &   0.2 &   1.0   & WL  & (2) \\
-39773 &  52353.413600 &   0.6 &  -0.7 &   3.0   & WL  & (3) \\
     0 &  55217.961107 &   0.4 &   0.1 &   1.0   & WL  & (4) \\
   388 &  55245.906033 &   0.4 &   0.4 &   1.0   & WL  & (4) \\
   389 &  55245.977962 &   0.4 &   0.3 &   1.0   & WL  & (4) \\
   444 &  55249.939017 &   0.4 &   0.0 &   1.0   & WL  & (4) \\
  4193 &  55519.951229 &   0.4 &   0.4 &   1.0   & WL  & (4) \\
  4430 &  55537.020418 &   0.4 &   0.2 &   1.0   & WL  & (4) \\
  5803 &  55635.906705 &   0.4 &  -0.5 &   1.0   & WL  & (4) \\
  9261 &  55884.959682 &   0.6 &  -1.2 &   1.0   & WL  & (4) \\
 24353 &  56971.922940 &   0.6 &   0.5 &   1.0   & WL  & (4) \\
 25616 &  57062.886273 &   0.6 &  -0.8 &   1.0   & WL  & (4) \\
 41021 &  58172.391979 &   0.4 &   0.0 &   1.0   & WL  & (5) \\
 41175 &  58183.482879 &   0.6 &  -0.7 &   1.0   & WL  & (5) \\
 45730 &  58511.545461 &   0.4 &   0.0 &   1.0   & WL  & (5) \\
 45758 &  58513.562168 &   0.4 &   0.2 &   1.0   & WL  & (5) \\
 \hline\\                                                                       
\end{tabular}\\[-1.0ex]
\footnotesize{(1) ESO/Dutch 0.9\,m, (2) Calar Alto 3.5\,m OPTIMA \citep{kanbachetal08},  (3)
  Observatorio Astron{\'o}mico de Mallorca, 30-cm, (4) McDonald Observatory MONET/N 1.2\,m,  (5) SAAO MONET/S 1.2\,m.}\\[-3.0ex]
\label{tab:0953}                                                              
\end{flushleft}

\vspace{-0mm}
\end{table}

\begin{table}[t]
\begin{flushleft}
  \caption{Observed times of the X-ray dips and the primary optical
    minima of \rxjten\ transformed from UTC to BJD(TDB). The minima
    included in the fit for Eq.~\ref{eq:1002eph1} have $i7\!=\!1$,
    those for Eq.~\ref{eq:1002eph2} have $i8\!=\!1$. The \oc\ values
    are correlated with the WL AB magnitude $w$.}
\begin{tabular}{@{\hspace{0mm}}r@{\hspace{1.0mm}}c@{\hspace{1.0mm}}c@{\hspace{1.0mm}}r@{\hspace{1.0mm}}r@{\hspace{2.0mm}}r@{\hspace{1.0mm}}r@{\hspace{1.0mm}}r@{\hspace{1.0mm}}c@{\hspace{1.0mm}}c@{\hspace{1.0mm}}c}\\[-3ex]
 \hline\hline \\[-1.5ex]
 Cycle & BJD(TDB)     & Error & i7 & $O\!-\!C$             & i8 & $O\!-\!C$             & Exp~ &Band & $w$   & Instr.\\
 & 2400000+     &       &    & Eq.~\ref{eq:1002eph1} &    & Eq.~\ref{eq:1002eph2} &       &     &       &       \\
 &              & (min) &    & (min)                 &    & (min)                 & (min) &     & (mag) &       \\[0.5ex]
 \hline\\[-1ex]
 -58114 & 48219.66788 &   1.9 & 1 &  -1.4     & 1 &   0.7  &   0.5 &  X   &        & (1) \\
 -47122 & 48982.81668 &   2.5 & 1 &   0.3     & 1 &   1.5  &  10.0 &  Sp  &        & (2) \\
 -36074 & 49749.85143 &   1.4 & 1 &  -1.0     & 1 &  -0.7  &   2.5 &  V   &        & (3) \\
 -34287 & 49873.91908 &   0.9 & 1 &  -0.1     & 1 &   0.0  &   0.9 &  X   &        & (4) \\
 -34286 & 49873.98852 &   0.9 & 1 &  -0.1     & 1 &   0.1  &   0.9 &  X   &        & (4) \\
 -25144 & 50508.69555 &   1.8 & 1 &   0.4     & 1 &  -0.2  &  10.0 &  Sp  &        & (2) \\
 -25130 & 50509.66776 &   2.5 & 1 &   0.7     & 1 &   0.1  &  10.0 &  Sp  &        & (2) \\
      0 & 52254.38165 &   0.7 & 1 &  -0.0     & 0 &  -2.7  &   0.3 &  X   &        & (5) \\
      0 & 52254.38211 &   0.7 & 1 &   0.6     & 0 &  -2.1  &   1.0 &  X   &        & (6) \\
  43087 & 55245.80718 &   1.7 & 1 &   0.7     & 0 &  -5.5  &   1.0 &  WL  &  16.69 & (7) \\
  47381 & 55543.93358 &   2.9 & 0 &   7.2     & 1 &   0.6  &   1.0 &  WL  &  19.17 & (7) \\
  47396 & 55544.97553 &   1.7 & 0 &   8.0     & 1 &   1.4  &   1.0 &  WL  &  19.09 & (7) \\
  47754 & 55569.82950 &   1.4 & 0 &   6.4     & 1 &  -0.2  &   1.0 &  WL  &  19.07 & (7) \\
  47757 & 55570.03700 &   1.4 & 0 &   5.3     & 1 &  -1.4  &   1.0 &  WL  &  18.76 & (7) \\
  48157 & 55597.81030 &   1.9 & 0 &   8.6     & 1 &   1.9  &   1.0 &  WL  &  18.87 & (7) \\
  48947 & 55652.65544 &   1.4 & 0 &   4.8     & 1 &  -2.0  &   1.0 &  WL  &  17.96 & (7) \\
  68353 & 56999.96355 &   2.3 & 1 &   0.5     & 0 &  -7.9  &   1.0 &  WL  &  17.57 & (7) \\
  76308 & 57552.25820 &   1.4 & 1 &  -1.7     & 0 & -10.8  &   1.0 &  WL  &  17.60 & (8) \\
  76309 & 57552.32788 &   1.4 & 1 &  -1.4     & 0 & -10.4  &   1.0 &  WL  &  17.48 & (8) \\
  76538 & 57568.22837 &   1.2 & 1 &   0.9     & 0 &  -8.1  &   1.0 &  WL  &  17.53 & (8) \\
  76610 & 57573.22657 &   1.2 & 1 &   0.1     & 0 &  -9.0  &   1.0 &  WL  &  17.48 & (8) \\
  80528 & 57845.24718 &   1.4 & 0 &   5.0     & 1 &  -4.3  &   1.0 &  WL  &  18.60 & (8) \\
  85210 & 58170.31063 &   1.7 & 0 &  10.3     & 1 &   0.5  &   1.0 &  WL  &  18.84 & (8) \\
  85282 & 58175.31047 &   1.7 & 0 &  11.8     & 1 &   2.1  &   1.0 &  WL  &  18.74 & (8) \\
  85283 & 58175.37911 &   1.4 & 0 &  10.7     & 1 &   0.9  &   1.0 &  WL  &  18.78 & (8) \\
  85958 & 58222.24217 &   1.7 & 0 &   9.9     & 1 &   0.1  &   1.0 &  WL  &  18.89 & (8) \\
  85987 & 58224.25517 &   1.4 & 0 &   9.3     & 1 &  -0.5  &   1.0 &  WL  &  18.80 & (8) \\
  86045 & 58228.28154 &   1.4 & 0 &   8.7     & 1 &  -1.1  &   1.0 &  WL  &  18.90 & (8) \\
  86261 & 58243.28102 &   1.4 & 0 &  13.2     & 1 &   3.4  &   1.0 &  WL  &  18.95 & (8) \\
  86277 & 58244.39114 &   1.4 & 0 &  12.2     & 1 &   2.3  &   1.0 &  WL  &  18.87 & (8) \\
  86289 & 58245.22339 &   2.0 & 0 &  10.9     & 1 &   1.1  &   1.0 &  WL  &  18.77 & (8) \\
  86290 & 58245.29243 &   2.0 & 0 &  10.3     & 1 &   0.5  &   1.0 &  WL  &  18.90 & (8) \\
  89174 & 58445.51451 &   2.9 & 1 &   0.3     & 0 &  -9.8  &   1.0 &  WL  &  17.99 & (8) \\
  89994 & 58502.45117 &   2.0 & 0 &   9.0     & 1 &  -1.1  &   1.0 &  WL  &  18.45 & (8) \\
  90008 & 58503.42288 &   2.0 & 0 &   8.6     & 1 &  -1.5  &   1.0 &  WL  &  18.44 & (8) \\
  90009 & 58503.49290 &   2.0 & 0 &   9.5     & 1 &  -0.7  &   1.0 &  WL  &  18.40 & (8) \\
  90010 & 58503.56170 &   2.0 & 0 &   8.6     & 1 &  -1.6  &   1.0 &  WL  &  18.40 & (8) \\[1.0ex]
 \hline\\
\end{tabular}\\[-1.0ex]
\footnotesize{(1) RASS PSPC, (2) MPI/ESO/ 2.2\,m EFOSC2, spectrophotometry, (3) ESO/Dutch 0.9\,m, (4) ROSAT HRI, (5) XMM-Newton EPIC pn, (6) XMM-Newton EPIC MOS12, (7) McDonald Observatory MONET/N 1.2\,m, (8) SAAO MONET/S 1.2\,m.}\\[-3.0ex]
\label{tab:1002}                                                              
\end{flushleft}
\end{table}

\end{document}